%% file: main.tex
\documentclass[sn-nature]{sn-jnl}
\usepackage{graphicx}
\usepackage{multirow}
\usepackage{lineno}
\usepackage{amsmath,amssymb,amsfonts}
\usepackage{amsthm}
\usepackage{mathrsfs}
\usepackage[title]{appendix}
\usepackage{xcolor}
\usepackage{textcomp}
\usepackage{manyfoot}
\usepackage{booktabs}
\usepackage{algorithm}
\usepackage{algorithmicx}
\usepackage{algpseudocode}
\usepackage{listings}
\usepackage{geometry}
\usepackage{titlesec}
\usepackage{tabularx}
\usepackage{xr}
\usepackage{makecell}

\titlelabel{\thetitle.\quad}

\begin{document}
\title[Article Title]{Analog fast Fourier transforms for scalable and efficient signal processing}

\author*[1]{\fnm{T. Patrick} \sur{Xiao}}\email{txiao@sandia.gov}
\author[1]{\fnm{Ben} \sur{Feinberg}}\email{bfeinbe@sandia.gov}
\author[1]{\fnm{David K.} \sur{Richardson}}\email{dkricha@sandia.gov}
\author[1]{\fnm{Matthew} \sur{Cannon}}\email{mcannon@sandia.gov}
\author[1]{\fnm{Calvin} \sur{Madsen}}\email{cfmadse@sandia.gov}
\author[2]{\fnm{Harsha} \sur{Medu}}\email{Harsha.Medu@infineon.com}
\author[2]{\fnm{Vineet} \sur{Agrawal}}\email{Vineet.Agrawal@infineon.com}
\author[3]{\fnm{Matthew J.} \sur{Marinella}}\email{m@asu.edu}
\author[4]{\fnm{Sapan} \sur{Agarwal}}\email{sagarwa@sandia.gov}
\author*[1]{\fnm{Christopher H.} \sur{Bennett}}\email{cbennet@sandia.gov}

\affil*[1]{\orgname{Sandia National Laboratories}, \orgaddress{\city{Albuquerque}, \state{New Mexico}, \postcode{87123}, \country{USA}}}
\affil[2]{\orgname{Infineon Technologies}, \orgaddress{\city{San Jose}, \state{California}, \postcode{95134}, \country{USA}}}
\affil[3]{\orgdiv{Department of Electrical, Computer, and Energy Engineering}, \orgname{Arizona State University}, \orgaddress{\city{Tempe}, \state{Arizona}, \postcode{85281}, \country{USA}}}
\affil[4]{\orgname{Sandia National Laboratories}, \orgaddress{\city{Livermore}, \state{California}, \postcode{94551}, \country{USA}}}

\abstract{
Edge devices are being deployed at increasing volumes to sense and act on information from the physical world. The discrete Fourier transform (DFT) is often necessary to make this sensed data suitable for further processing -- such as by artificial intelligence (AI) algorithms -- and for transmission over communication networks. Analog in-memory computing has been shown to be a fast, energy-efficient, and scalable solution for processing edge AI workloads, but not for Fourier transforms. This is because of the existence of the fast Fourier transform (FFT) algorithm, which enormously reduces the complexity of the DFT but has so far belonged only to digital processors. Here, we show that the FFT can be mapped to analog in-memory computing systems, enabling them to efficiently scale to arbitrarily large Fourier transforms without requiring large sizes or large numbers of non-volatile memory arrays. We experimentally demonstrate analog FFTs on 1D audio and 2D image signals, performing analog computations on up to 524K charge-trapping memory devices simultaneously, where each device has precisely tunable, low-conductance analog states. The scalability of both the new analog FFT approach and the charge-trapping memory device is leveraged to compute a 65,536-point analog DFT, a scale that is otherwise inaccessible by analog systems and which is $>$500$\times$ larger than any previous analog DFT demonstration. Analog FFT cores can provide higher energy efficiency and performance per area than specialized digital FFT processors at all FFT sizes, while also functioning as efficient matrix multiplication engines for AI workloads.
}

\newgeometry{bottom=2.5cm, left=2.5cm, right=2.5cm, top=2.5cm}

\maketitle

\newgeometry{bottom=2.5cm, left=2.5cm, right=2.5cm, top=2.5cm}

\input{intro}
\input{cooleytukey}
\input{sonos}
\input{audio}
\input{vector_radix}
\input{scaling}
\input{conclusion}

\input{main.bbl}
\input{methods}

\section*{Acknowledgements}
We thank A. Talin and W. Wahby for comments and feedback on the manuscript. This work was supported by the Laboratory-Directed Research and Development (LDRD) Programs at Sandia National Laboratories. This article has been authored by an employee of National Technology \& Engineering Solutions of Sandia, LLC under Contract No. DE-NA0003525 with the U.S. Department of Energy (DOE). The employee owns all right, title and interest in and to the article and is solely responsible for its contents. The United States Government retains and the publisher, by accepting the article for publication, acknowledges that the United States Government retains a non-exclusive, paid-up, irrevocable, world-wide license to publish or reproduce the published form of this article or allow others to do so, for United States Government purposes. The DOE will provide public access to these results of federally sponsored research in accordance with the DOE Public Access Plan https://www.energy.gov/downloads/doe-public-access-plan.

\newpage
\input{suppmat}

\end{document}

%% file: intro.tex
\section{Introduction}
\label{sec:intro}

The increasing deployment of remote and inter-connected sensors and actuators is leading to the collection of ever greater volumes of data from the physical world. Acting on this information in real time requires computation to be done where the data is sensed, and the raw sensed signals must often be processed by linear transforms to be made suitable for further processing, communication, or storage. One of the most ubiquitous transforms is the discrete Fourier transform (DFT), which converts spatial or temporal signals to their frequency representation. When used in edge or Internet-of-Things (IoT) devices, the DFT can form images of the sensed environment from radar or lidar raw data \cite{Walker1980,Jakowatz2012,Richards2005,Sun2020}, process raw audio waveforms to aid in speech recognition \cite{Amodei2016,Radford2023}, and modulate signals for wireless communication with other IoT devices \cite{Hwang2009,Cai2018,Michailow2014}. Though the need for processing at the edge is rapidly growing, the end of Dennard scaling means that there will no longer be improvements to the energy per arithmetic operation using CMOS logic \cite{Hennessy2019}. Conventional digital processors may soon prove inadequate for the large-scale data processing capabilities that are desired in future edge devices.

Analog in-memory computing (IMC) systems are a potential solution to overcome these scaling limits. These systems exploit analog circuit laws to rapidly and efficiently compute matrix-vector multiplications (MVMs) inside memory arrays, while greatly reducing the crippling energy overhead of data movement between the memory and processor in traditional von Neumann architectures \cite{Burr2017,xiao2020analog,sebastian2020memory}. These advantages enable analog IMC to process machine learning (ML) inference workloads with potentially orders-of-magnitude greater performance per watt, as explored in numerous recent works \cite{Li2019,yao2020fully,xue2021cmos,wan2022compute,huang2024memristor,ambrogio2023analog,LeGallo2023,Fick2022}. However, these benefits have yet to fully materialize for executing the large DFT operations that often come before or after the ML algorithm in edge systems, and would thus become the bottleneck. This is because while the DFT can be expressed as an MVM, it can be computed using far fewer mathematical operations by using the fast Fourier transform (FFT) algorithm \cite{Duhamel1990}. Although the FFT has been an indispensable instrument of digital signal processing (DSP) for decades, analog IMC systems have yet to capitalize on the benefits of the FFT, severely limiting their efficiency and scalability.

Analog IMC systems to date have relied exclusively on a direct MVM mapping of the DFT and its real-only counterpart, the discrete cosine transform (DCT) \cite{cai2016memristor}. With this approach, analog IMC retains its low energy-per-operation only up to DFT sizes that can fit within a single memory array. Therefore, the largest experimentally demonstrated analog DFTs and DCTs have been limited to 64 or 128 points per dimension \cite{li2018analogue,hu2018memristor,zhao2023energy,Huang2025,Wang2025}.
Relying on these prior methods would preclude analog processing from scaling to the much larger DFT sizes that are routinely needed for many practical applications.
For example, mobile or IoT devices that use 5G wireless communications standards need to support FFTs of up to 4096 points to generate orthogonal frequency-division multiplexing (OFDM) waveforms \cite{Zaidi2018}.
Large 2D FFTs with well over 1000 or 10,000 points per dimension also enable high-resolution range-velocity maps using frequency-modulated continuous wave (FMCW) lidar or radar for autonomous vehicles \cite{Kim2022_FMCW, Heo2021}, as well as high-resolution synthetic aperture radar imagery for satellite- or aircraft-based remote sensing applications \cite{Yu2019}.

In this article, we demonstrate a new, more scalable mapping of the Fourier transform onto analog IMC architectures that is based on the classic Cooley-Tukey FFT algorithm \cite{cooley1965algorithm}. This method factorizes large DFTs into smaller elementary DFTs that are computed by analog MVMs, which enables a small number of modestly sized memory arrays to efficiently process arbitrarily large DFTs. Compared to prior analog IMC approaches, the analog FFT reduces the energy and area scaling of the DFT from $O(N^2)$ to $O(N \text{log} N)$, just as the original FFT algorithm reduced the complexity of the DFT on digital processors.
The analog FFT can also yield more accurate DFT computation at any transform size, independently of the memory technology used for computing.
The superior scalability of the analog FFT extends the practical range of mathematical kernels that can be efficiently processed using analog IMC.
This enables a flexible architecture of analog IMC cores (Fig. \ref{fig:cooley_tukey}e) that can be re-programmed to accelerate different kernels, such as neural network layers and FFTs, to achieve low-power end-to-end execution of edge workloads.

As a proof-of-concept, we implemented analog FFTs on a large array of charge-trapping flash memory devices based on the silicon-oxide-nitride-oxide-silicon (SONOS) material stack, which was fabricated in a 40-nm complementary metal-oxide-semiconductor (CMOS) process. 
The subthreshold operation and precise analog programmability of this memory enables half a million devices to simultaneously participate in a single analog MVM while retaining high accuracy.
We used the SONOS array to experimentally compute the frequency spectra of 1D audio and 2D image signals.
Leveraging the scalability of both the FFT and the memory technology, we compute analog DFTs of up to 65,536 points, which is more than $500\times$ larger than any prior DFT computed using analog hardware \cite{Wang2025}.
We also show using experimentally-validated models that analog IMC hardware can provide accurate end-to-end acceleration of complex edge workloads such as automatic speech recognition, which combine FFTs with ML processing.

%% file: cooleytukey.tex
\section{The analog fast Fourier transform}
\label{sec:cooleytukey}

\begin{figure}[t]
\centering
\includegraphics[width=\textwidth]{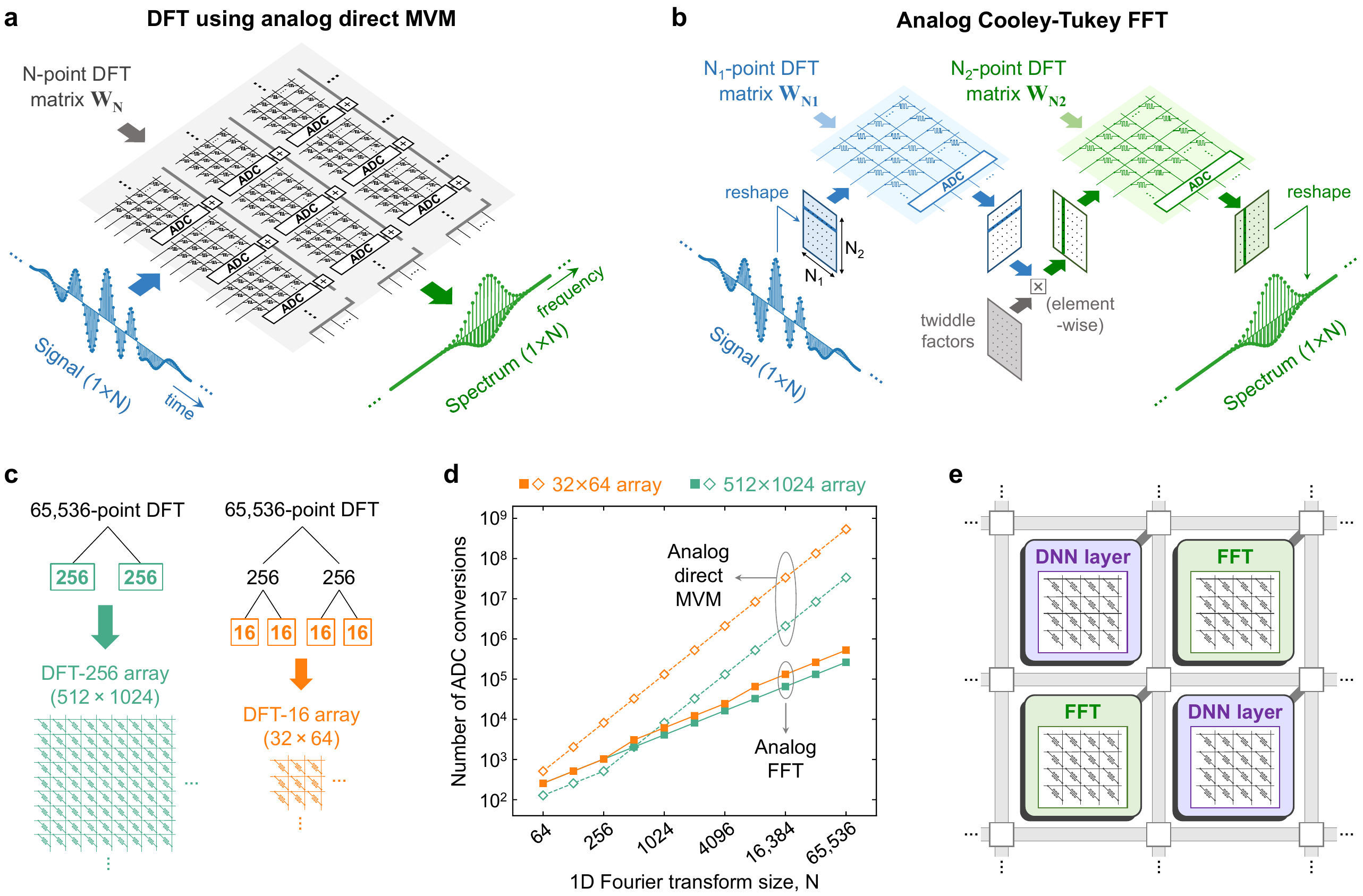}
\caption{\textbf{Processing large Fourier transforms using analog in-memory computing.} (a) The direct MVM approach requires a large DFT matrix to be split across many arrays. (b) The analog Cooley-Tukey FFT factorizes the $N$-point DFT into smaller DFTs of size $N_1$ and $N_2$. Only the real part of the temporal signal and frequency spectrum are shown for simplicity. (c) Two of many ways to factorize a 65,536-point DFT using the analog FFT. The leaves of the trees are elementary DFTs mapped to analog MVMs, and the branches are Cooley-Tukey factorizations. (d) Comparison of how the number of ADC conversions scales with DFT size for the analog direct MVM and the analog FFT. We consider analog IMC systems with a maximum single-array DFT size of 16 points ($32 \times 64$ array size), and 256 points ($512 \times 1024$ array size). (e) A mesh fabric of analog IMC cores can accelerate a diverse range of workloads. The same cores for processing FFTs can be reprogrammed to execute DNN layers and other kernels.}
\label{fig:cooley_tukey}
\end{figure}

Analog in-memory computing systems and digital processors are governed by fundamentally different scaling laws for their energy consumption. The energy associated with digital processing generally scales with the number of arithmetic operations such as multiplies and adds. Analog IMC systems follow a different scaling law, due to an essential characteristic of these architectures: though the matrix computations inside the memory array are extremely efficient, almost all the energy is consumed by peripheral circuits, particularly those involved in sensing and converting analog summed currents to digital outputs with 8-bit or similar resolution \cite{HERMES2022,wan2022compute,xiao2020analog,Yu2021,Aguirre2024}. Therefore, the total number of conversions by the analog-to-digital converter (ADC) is a useful measure that generally scales together with the total energy consumption, independent of the specific memory technology or circuit implementation.

Beyond its importance for signal processing, the discrete Fourier transform epitomizes the different ways that the two hardware paradigms can scale. The DFT converts a signal $\mathbf{x}$ with length $N$ to its frequency spectrum $\mathbf{X}$:

\begin{equation}
\label{eq:dft}
X_k = \sum_{n=0}^{N-1}x_n e^{-i2\pi nk/N}
\end{equation}
This equation can be explicitly written as a complex-valued MVM: $\mathbf{X} = \mathbf{W}_N\mathbf{x}$, where $\mathbf{W}_N$ is the $N$-point DFT matrix of size $N \times N$. The element at the ($n$, $k$) position of this matrix has the value $\left(\omega_N\right)^{nk}$, where $\omega_N = e^{-i2\pi/N}$.

In digital hardware, the DFT is generally computed using an FFT algorithm. The FFT exploits divide-and-conquer techniques to compute the DFT using far fewer operations than by computing the above MVM directly. This fundamentally reduces the DFT's computational complexity, and hence its energy scaling, from $\mathcal{O}(N^2)$ to $\mathcal{O}(N \text{log}_2 N)$ \cite{Duhamel1990}. The reduction is possible by exploiting the symmetries of the DFT matrix. For matrices without symmetry or periodicity, such as matrices of deep neural network (DNN) weights, the number of operations is irreducible without the use of approximations.

The conventional analog IMC approach to the DFT is to program the real and imaginary parts of the matrix $\mathbf{W}_N$ onto a resistive memory crossbar, then compute Equation \ref{eq:dft} directly as an analog MVM \cite{cai2016memristor,li2018analogue,zhao2023energy,Song2024}. When the DFT is computed directly using a single array, the energy scales with the number of outputs of a DFT, which is $\mathcal{O}(N)$. However, this result holds only for small $N$. As $N$ becomes large, the matrix $\mathbf{W}_N$ eventually exceeds the maximum size of a memory array. There are many factors that constrain the physical size of resistive crossbars, including the maximum current supported by the peripheral circuitry \cite{wan2022compute}, parasitic $IR$ voltage drops along the array interconnects \cite{xiao2021analysis}, accumulation of memory device conductance errors and noise \cite{OnTheAccuracy}, write disturb \cite{Burr2014}, and process yield considerations. In this regime, the direct MVM approach requires a large DFT matrix to be split across many physical arrays, with each array producing partial results that are digitized then added, as shown in Fig. \ref{fig:cooley_tukey}a. The number of ADC conversions needed to produce a single element of $\mathbf{X}$ scales as $\mathcal{O}(N)$, leading to a total DFT energy scaling of $\mathcal{O}(N^2)$. The area also scales as $\mathcal{O}(N^2)$, regardless of whether it is dominated by the peripheral circuits or the memory array. The rapid, quadratic growth of energy and area with $N$ limits the size and nature of problems that can be processed efficiently with this approach using analog hardware.

We show that analog IMC hardware can overcome this limit by implementing the FFT algorithm and inheriting its fundamental scalability advantages. Specifically, we use Cooley and Tukey's original two-factor FFT algorithm to compute any DFT whose length is a composite number: $N = N_1 \times N_2$ \cite{cooley1965algorithm,Duhamel1990}. The input $\mathbf{x}$ and output $\mathbf{X}$, which are vectors of length $N$, can be reshaped into 2D matrices with dimensions $N_1 \times N_2$: we call these $\mathbf{\tilde{x}}$ and $\mathbf{\tilde{X}}$, respectively. The indices of the new matrices are related to those of the original vectors by $\tilde{x}_{n_1,n_2} = x_{N_2n_1+n_2}$ and $\tilde{X}_{k_2,k_1} = X_{k_2+N_1k_1}$. By exploiting the periodicity of $\mathbf{W}_N$, Equation \ref{eq:dft} can be re-written as:
\begin{equation}
\label{eq:cooley_tukey}
\tilde{X}_{k_2,k_1} = \sum_{n_2=0}^{N_2-1} \omega_{N_2}^{n_2k_1} \omega_N^{n_2k_2} \sum_{n_1=0}^{N_1-1} \omega_{N_1}^{n_1k_2} \tilde{x}_{n_1,n_2}
\end{equation}

To explicitly show the mapping onto analog hardware, this equation can be cast in matrix form as:
\begin{equation}
\label{eq:cooley_tukey_matrix}
\mathbf{\tilde{X}} = (\mathbf{W}_{N_2} \left[\mathbf{T} \odot \left(\mathbf{W}_{N_1} \tilde{\mathbf{x}}\right)\right]^\text{T})^\text{T}
\end{equation}
where $\odot$ is the element-wise product and $\mathbf{T}$ is an $N_1 \times N_2$ matrix of twiddle factors defined by $\mathbf{T}_{mn} = \left(\omega_N\right)^{mn}$. Fig. \ref{fig:cooley_tukey}b shows a sequence of steps that implement the analog FFT in Equation \ref{eq:cooley_tukey_matrix}. In the first stage, an $N_1$-point DFT is computed for every row of the matrix $\mathbf{\tilde{x}}$ using a sequence of analog MVMs on a resistive memory array, which execute the matrix-matrix multiplication $\mathbf{W}_{N_1} \tilde{\mathbf{x}}$. The results are digitized and multiplied element-wise with the twiddle matrix $\mathbf{T}$ using digital multipliers. Next, a second stage of $N_2$-point DFTs is executed to perform the multiplication with $\mathbf{W}_{N_2}$, again via a sequence of analog MVMs. Finally, the result $\mathbf{\tilde{X}}$ is reshaped back into a vector to obtain $\mathbf{X}$. This Cooley-Tukey decomposition enables a large $N$-point DFT, which otherwise needs to be split across many arrays, to be computed with just two arrays that implement an $N_1$-point analog DFT and an $N_2$-point analog DFT, respectively. 

Alternatively, the analog MVMs that make up the two stages of the FFT can be executed in parallel on multiple arrays, as shown in Fig. \ref{fig:parallel_fft}. This enables the analog FFT to match the throughput of the direct MVM, while still requiring much less total area for the analog cores. This parallel scheme further enables the twiddle multiplications to be folded into the analog MVMs, eliminating the need for any digital multiplications. In Supplementary Section \ref{sec:parallel_fft}, we experimentally demonstrate this twiddle-folded analog FFT and show that it has no effect on the accuracy of the FFT.

To scale to a DFT size $N$ that is much larger than the array size, the Cooley-Tukey factorization can be applied recursively to further reduce the elementary DFT sizes, or radices, as shown in Fig. \ref{fig:cooley_tukey}c. The physical size of the memory array sets the upper bound on the size $K$ of an elementary analog DFT; to compute a $K$-point complex DFT using one MVM in a single array, the array needs to have at least $2K$ rows and $4K$ columns \cite{cai2016memristor}. The analog inverse FFT (IFFT) can be computed via the same Cooley-Tukey factorization as the analog FFT, since the inverse DFT (IDFT) matrix has the same symmetries as the DFT matrix. The analog IFFT uses the conjugate transpose of  $\mathbf{W}_N$ and the complex conjugate of $\mathbf{T}$.

Fig. \ref{fig:cooley_tukey}d compares how the number of required ADC conversions scales between the two analog DFT approaches, considering both a small array ($K=16$) and a large array ($K=256$). The latter corresponds to the largest analog MVM that can be executed by the SONOS array used in this work, and also matches the array size of other large analog IMC prototypes \cite{ambrogio2023analog,Fick2022}. For the direct MVM method, the array's dimensions set the critical DFT size where the scaling law transitions from $O(N)$ to $O(N^2)$. For the analog FFT, if the Cooley-Tukey decomposition is only applied once, i.e. $N \leq K^2$, the total number of ADC conversions in the analog FFT scales as $\mathcal{O}(N)$. When the DFT is recursively factorized, all of these operations increase proportionally with the number of Cooley-Tukey decompositions, which scales as $\mathcal{O}(\text{log}_K N)$. Combining these two trends, the energy of the analog FFT scales as $\mathcal{O}(N \text{log}_K N)$ in the limit of large $N$. The worst-case area overhead of the parallel analog FFT also scales as $\mathcal{O}(N \text{log}_K N)$, as explained in Supplementary Section \ref{sec:scaling_laws}. Therefore, both the energy and area scaling laws are superior to the $\mathcal{O}(N^2)$ scaling of the direct MVM approach. In general, the DFT size where the analog FFT and direct MVM have similar efficiency is $N = 2K$. Above this, the FFT approach is increasingly more efficient due to its more favorable scaling, leading to orders-of-magnitude lower energy consumption and area for large DFT sizes.

The overall $\mathcal{O}(N \text{log} N)$ scaling of the analog FFT is similar to that of the digital FFT, but there are essential differences in the algorithmic structure. In digital FFT implementations, recursively decomposing the DFT down to the smallest possible elementary DFTs (i.e. radix-2 or radix-4) optimally reduces the complexity from $\mathcal{O}(N^2)$ to $\mathcal{O}(N \text{log}_2 N)$ \cite{Duhamel1990}. Meanwhile, for analog FFTs, it is more optimal from an energy standpoint to minimize the depth of the Cooley-Tukey factorization tree by terminating the decomposition as soon as the factored DFT is small enough to fit onto one memory array. Further factorization would decrease the total number of arithmetic operations, which is helpful for digital systems, but would increase the number of intermediate DFT results and hence the number of ADC conversions, which is harmful for analog systems (see Fig. \ref{fig:factorizations}b).
This explains why having a larger physical array size is more energy-efficient for large analog FFTs, as shown in Fig. \ref{fig:cooley_tukey}d. 
However, because of the $\mathcal{O}(N \text{log}_K N)$ scaling of the analog FFT, small arrays (small $K$) can also scale to large FFTs with only a modest energy penalty compared to large arrays.

Analog FFT processors can be rapidly reconfigured to compute DFTs of various sizes without re-programming any memory devices.
This is possible due to the symmetries in the DFT matrix.
A memory array that has been programmed with a $K$-point DFT matrix can be re-used to compute a smaller DFT of size $K / (a \times b)$ simply by applying inputs to every $a^\text{th}$ row and measuring the outputs from every $b^\text{th}$ column.
This reconfigurability can be used to dynamically change the FFT radix or to switch to direct MVM mode (for small DFTs), and this usage is experimentally demonstrated on an audio FFT example in Supplementary Section \ref{sec:reconfigurability}.
The flexibility to support multiple FFT sizes is valuable for applications such as wireless communications, where it is often desirable to switch on the fly between different channel bandwidths or standards \cite{Yang2012,Liu2019}.
The analog FFT can also be used to efficiently process other linear transforms that can be computed using a DFT, such as the DCT \cite{Makhoul1980}.
More broadly, the reconfigurability of analog arrays enables a highly flexible accelerator architecture, shown in Fig. \ref{fig:cooley_tukey}e, where each analog IMC core within the fabric can be programmed to implement variable-length FFTs, DNN layers, finite impulse response filters \cite{Huang2025}, linear equation solvers \cite{Song2024}, and other linear algebra kernels. The versatility of the analog IMC core allows the architecture to efficiently process workloads that otherwise would have required the combination of multiple specialized digital or analog processors.

%% file: sonos.tex
\section{Analog DFTs using SONOS charge-trapping memory}
\label{sec:sonos}

\begin{figure}[t]
\centering
\includegraphics[width=\textwidth]{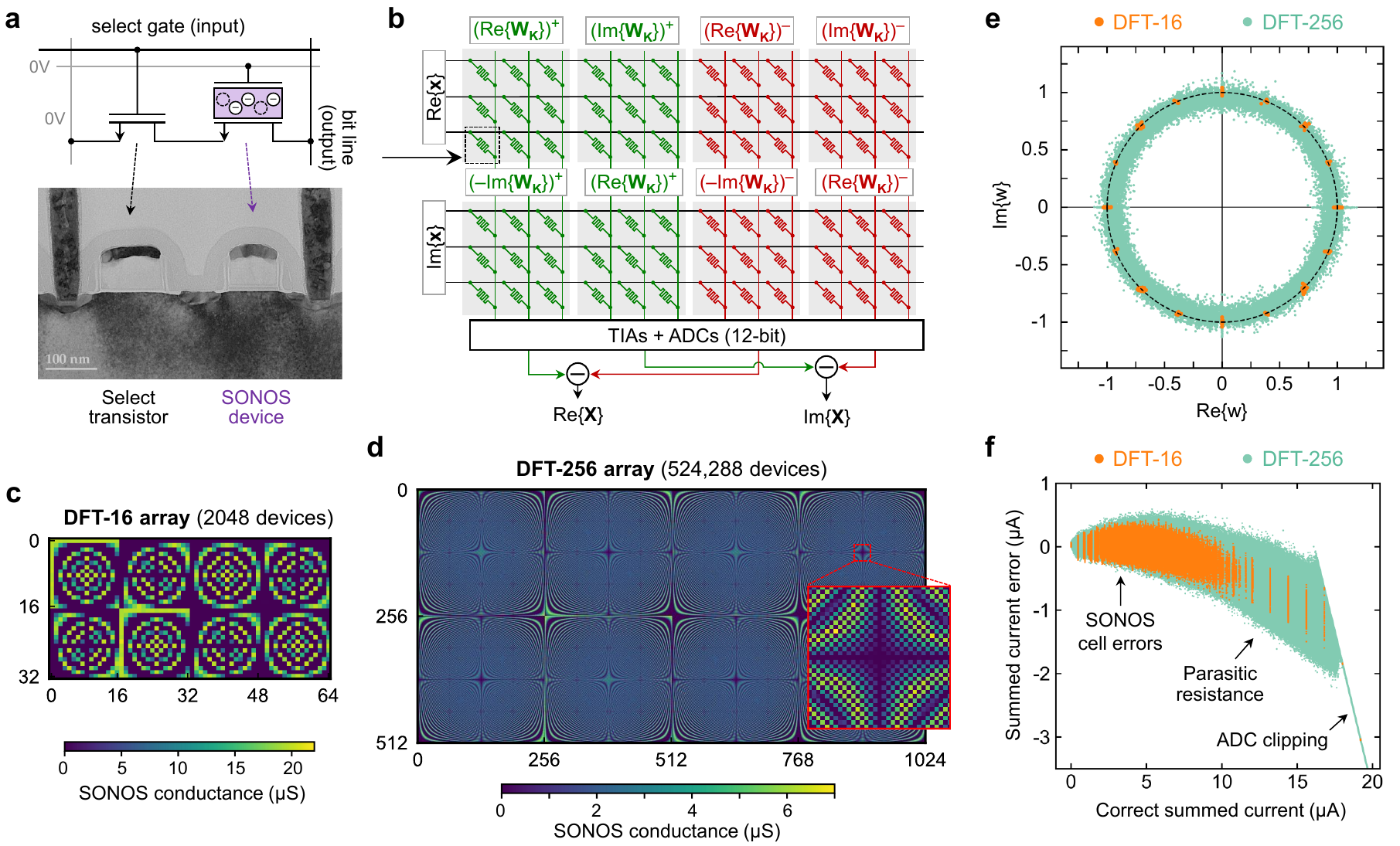}
\caption{\textbf{DFT mapping onto a SONOS charge-trapping memory array.} (a) Electrical schematic (top) and transmission electron microscope image (bottom) of the two-transistor SONOS memory cell. (b) Mapping a DFT operation with complex-valued weights and inputs to a resistive memory crossbar. The SONOS cell with the input-output connections in (a) is simplified in this schematic to a resistor. (c) Measured SONOS conductance profile for a DFT-16 matrix. (d) Measured SONOS conductance profile for a DFT-256 matrix. Inset shows a $32\times 32$ region of the programmed array. (e) Constellation of the complex-valued DFT weight values stored in the programmed SONOS devices, for DFT-16 and DFT-256. The ideal DFT weights lie along the unit circle (black dashed circle). (f) Accuracy of individual current-mode analog MVMs for 16-point and 256-point DFTs, showing the dominant sources of error in different current regimes. Data on more than $3.7\times10^7$ analog current sums are collected from processing the audio signal in Fig. \ref{fig:audio_spectrograms}.}
\label{fig:sonos_dft}
\end{figure}

To execute analog FFTs, we used an array of SONOS charge-trapping memory devices that have stable and precisely programmable analog conductance levels. The state variable of SONOS memory is the amount of charge that is confined in a silicon nitride charge-trapping layer, which modulates the electronic conductance of the underlying silicon channel through the field effect. A 1024$\times$1024 crossbar array of SONOS devices was fabricated in a 40-nm CMOS process that uses the compact two-transistor (2T) memory cell shown in Fig. \ref{fig:sonos_dft}a. The array was integrated with peripheral CMOS circuits to support analog MVMs and write-verify programming of each SONOS device to a target conductance. To minimize conductance change over time, the programming procedure selectively places charge in mid-gap electronic traps in the nitride layer that have large energy barriers for escape \cite{Agrawal2020,xiao2022accurate,Agrawal2022}. We operate the SONOS transistors primarily in the subthreshold region, where the device has a large conductance On/Off ratio ($>$$10^6$) and errors that approach zero at low conductance \cite{xiao2022accurate}. The statistically characterized, state-dependent device-to-device variation and drift in the programmed SONOS conductances are quantified in Supplementary Section \ref{sec:gradient}.

The analog DFT mapping in Fig. \ref{fig:sonos_dft}b is used to multiply a complex-valued DFT weight matrix with a complex-valued vector \cite{cai2016memristor}. The difference in conductance of two SONOS devices encodes the signed value of each real or imaginary weight component. To execute a DFT, the input vector $\mathbf{x}$ is applied bit-serially to the select gate lines, and each selected SONOS cell draws a current from its bit line (BL) that is proportional to its conductance. On this chip, the analog sum of currents collected on the BL is converted to a voltage by a transimpedance amplifier (TIA), then to digital outputs using an ADC \cite{Agrawal2022}. Digital post-processing accumulates the analog MVM results for different input bits and produces the complex-valued DFT output $\mathbf{X}$ (see Methods). The on-chip ADC has a resolution of 12 bits over the range from 0 to 17{~\textmu}A. 

Fig. \ref{fig:sonos_dft}c and Fig. \ref{fig:sonos_dft}d show the measured conductance profiles of a portion of the SONOS array after programming a DFT-16 matrix and a DFT-256 matrix, respectively. A maximum target SONOS conductance of 20{~\textmu}S was used for the DFT-16 subarray, while the DFT-256 subarray used a reduced maximum conductance of 6.2{~\textmu}S to ensure that most of the summed currents in our exemplar applications (described below) do not exceed the ADC's limits. Fig. \ref{fig:sonos_dft}e shows the locations in the complex plane of the DFT weights that were programmed onto the SONOS array, whose ideal values ($e^{i2\pi nk/K}$) would lie perfectly along the unit circle. In general, the SONOS-encoded DFT weights are tightly distributed around the circle; the DFT-16 weights have a mean absolute error of $\epsilon_{|\omega|} = 0.0118$ in magnitude and $\epsilon_{\angle \omega} = 0.338^\circ$ in phase. The DFT-256 weights have somewhat larger errors: $\epsilon_{|\omega|} = 0.0458$ and $\epsilon_{\angle \omega} = 1.037^\circ$. This is because the smaller conductances used for the DFT-256 matrix led to a smaller ratio between the signal and error components of the conductance, and a larger amount of conductance drift caused a systematic reduction in the weight magnitudes. Conductance statistics for DFT arrays of various sizes can be found in Supplementary Section \ref{sec:dft_weights}. Notably, large DFT matrices have values that are distributed approximately uniformly over the conductance range; this differs dramatically from DNN weight matrices, which generally have an abundance of near-zero weights and are exponentially skewed toward low conductance \cite{xiao2022accurate,OnTheAccuracy}. This implies that for the same array size, analog DFT operations will accumulate larger currents, and potentially larger errors.
Therefore, compared to DNN inference, the analog FFT benefits more strongly from memory technologies that have high precision at low conductances.
The desired memory device properties for accurate and efficient analog FFTs are described in more detail in Supplementary Section \ref{sec:device_requirements}.

To illustrate the sources of error in the SONOS-based analog DFTs, Fig. \ref{fig:sonos_dft}f shows the error statistics of the analog current sums (representing dot products) across a large number of 16-point and 256-point elementary DFT computations, performed as part of the audio processing FFT experiments in the next section. There are three main sources of error, which dominate at different regimes of the summed current. For small summed currents, the error is random and zero-centered, and originates from the accumulated random variability and noise in the conductances of the SONOS devices, as described above. At intermediate to large summed currents, the error is dominated by parasitic $IR$ drops across the resistances of the array's rows and columns, which cause a systematic reduction in current \cite{xiao2021analysis}. This effect is more pronounced for the 256-point analog DFTs due to the much larger MVM size (512$\times$1024), while the 16-point DFTs have much lower summed currents on average, and hence has negligible $IR$ drops. Finally, for the largest outlier currents in Fig. \ref{fig:sonos_dft}f, the ADC clips the current measurement to the limit of 17{~\textmu}A, leading to a negative error that increases linearly with the correct current value.

%% file: audio.tex
\section{Audio processing using analog FFTs}
\label{sec:audio}

\begin{figure}[t]
\centering
\includegraphics[width=\textwidth]{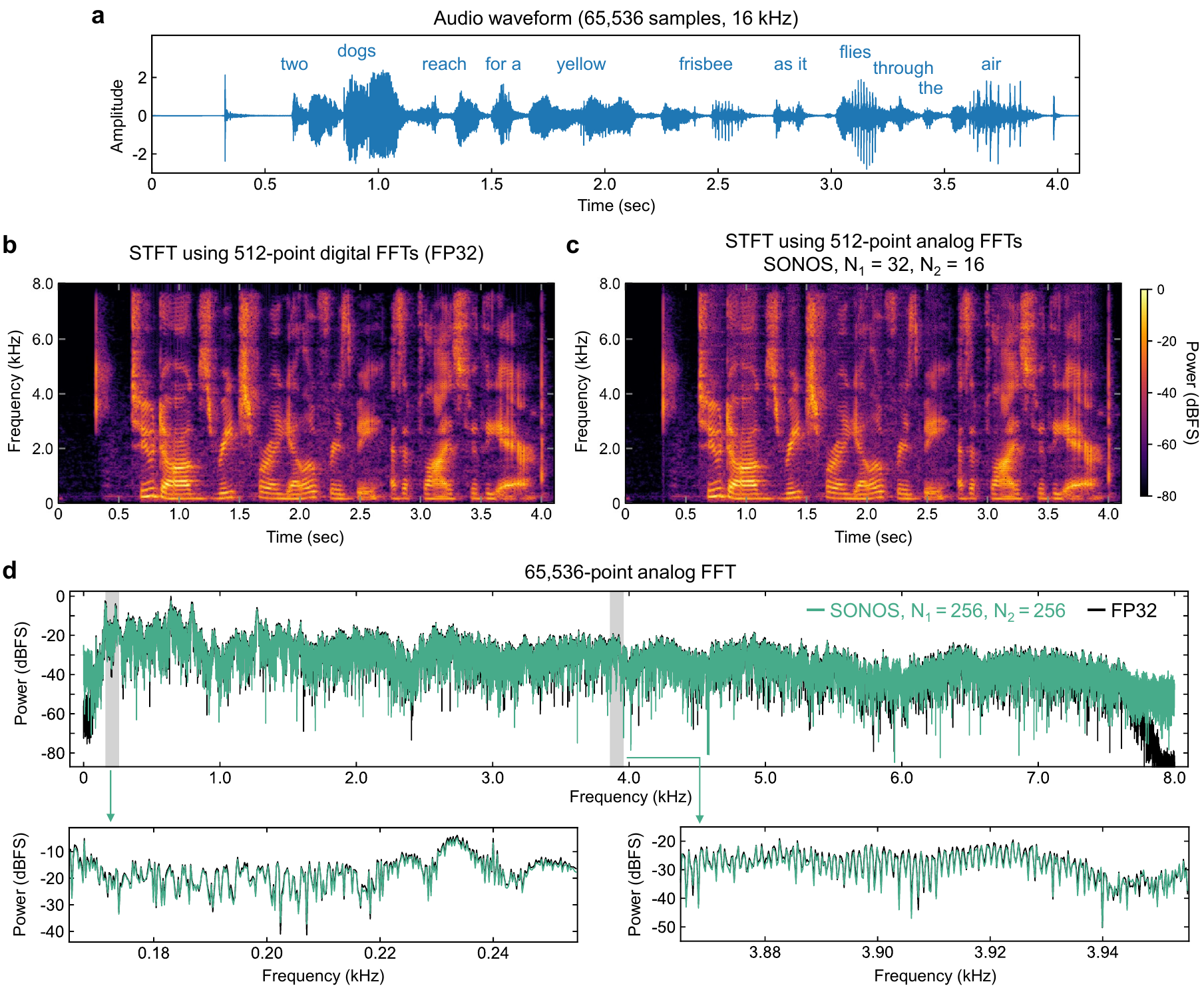}
\caption{\textbf{Audio processing with analog FFTs.} (a) Speech audio waveform with 65,536 samples. (b) Spectrogram of the audio waveform generated by FP32 STFTs, using a window size of 512 samples, a hop length of 128 samples, and a Hamming window function. The frequency resolution is 31.25 Hz. (c) Spectrogram generated experimentally using STFTs that are implemented with 512-point analog FFTs. The FFTs are factored into 32-point and 16-point analog DFTs that are executed on the SONOS array. (d) Power spectrum of the full audio waveform, computed by a 65,536-point analog FFT using the SONOS array (teal), compared to an FP32 digital FFT (black). The analog FFT was factored into 256-point analog DFTs that are executed on the SONOS array. Insets zoom in on two parts of the spectrum. The frequency resolution is 0.244 Hz. dBFS: decibels relative to full-scale.
}
\label{fig:audio_spectrograms}
\end{figure}

To experimentally demonstrate the analog Cooley-Tukey FFT and evaluate its accuracy, we used the SONOS array as a spectrum analyzer for audio waveforms. A useful representation of the spectrum of a signal is the spectrogram, which shows how the signal's frequency spectrum changes with time \cite{Koenig1946}. This is generated using a short-time Fourier transform (STFT), which computes the DFT over temporal sliding windows of the signal. The magnitude squared of the complex-valued STFT output yields the power spectrum for each time window, and these spectra are stacked in time to produce the final spectrogram. For this experiment, we used a four-second audio waveform from the Flickr8k Audio Caption dataset \cite{Flickr8k}, with slight zero-padding to 65,536 samples (4.096 seconds at 16 kHz sampling rate). The waveform is shown in Fig. \ref{fig:audio_spectrograms}a and contains the spoken caption, ``Two dogs reach for a yellow frisbee as it flies through the air." Fig. \ref{fig:audio_spectrograms}b shows the spectrogram of this signal on a logarithmic scale, computed using a digital STFT at single-precision floating-point (FP32) on a CPU. We used Hamming windows with 512 samples (i.e. uses 512-point FFTs), with 75\% overlap between windows, which are typical parameters for processing speech waveforms. Fig. \ref{fig:audio_spectrograms}c shows the spectrogram generated experimentally using an analog STFT, where each 512-point FFT was decomposed into 32-point and 16-point DFTs ($N_1 = 32$, $N_2 = 16$) that were computed using the SONOS array.

The SONOS-computed spectrogram reproduces all of the key features of the FP32 spectrogram and resolves the frequency signature of each spoken word. 
In this experiment, we are able to compute frequency components that span several orders of magnitude in dynamic range, because each component was accumulated over 12 bits of the input signal using digital shift-and-add operations.
The peak signal-to-noise ratio (PSNR), which measures the element-wise error between the two spectrograms, is 55.97 dB. There are somewhat subtle differences between the spectrograms, mainly visible as the presence of background noise in some parts of the SONOS spectrogram, which originates from device-to-device variability and cycle-to-cycle read noise in the SONOS conductances, accumulated over many devices in an analog MVM. As another assessment of the spectrogram's quality, we reconstructed the four-second audio waveform by digitally computing the IFFT on the SONOS-computed complex-valued spectrum of each sliding window. Both the original and reconstructed audio files are provided as Supplementary Information. Each spoken word is clearly reproduced in the reconstructed audio clip, with a nearly imperceptible difference compared to the original audio.

To demonstrate the scalability of the analog FFT, we use the SONOS array to compute a 65,536-point FFT, sufficient to generate the spectrum of the full audio waveform in Fig. \ref{fig:audio_spectrograms}a without breaking it up into sliding windows. The 65,536-point analog FFT is computed by factoring it into 256-point analog DFTs (i.e. $N_1 = N_2 = 256$) that were executed on the SONOS subarray in Fig. \ref{fig:sonos_dft}d. We note that a transform size of 65,536 points is far too large to be feasibly implemented in analog without the Cooley-Tukey FFT; if using a direct MVM approach, the partitioned collection of arrays would need to have $3.4 \times 10^{10}$ memory devices in total. The power spectrum of the audio waveform as computed experimentally by the SONOS array is shown in Fig. \ref{fig:audio_spectrograms}d, alongside the true spectrum computed at FP32 precision. The large size of the FFT yields a fine frequency resolution of 0.244 Hz. The left inset zooms into the frequency range of female human speech ($\sim$165 to 255 Hz \cite{baken2000clinical}), showing close point-by-point agreement between the SONOS and FP32 computations. Across the full range of frequencies, the SONOS-computed spectrum has a PSNR of 41.01 dB relative to FP32. The slightly lower PSNR compared to the STFT is due to the use of significantly larger elementary DFTs, which have larger analog MVM errors as shown in Fig. \ref{fig:sonos_dft}e. Additionally, we used both the magnitude and phase of the SONOS-computed spectrum to reconstruct the audio waveform via a single IFFT. The reconstructed audio clearly reproduces the spoken message (see Supplementary Information).

%% file: vector_radix.tex
\section{Image processing using the analog vector-radix FFT}
\label{sec:vector_radix}

For image processing applications, analog approaches to the DFT must be extended to two dimensions. The DFT of a two-dimensional $M\times N$ input $\mathbf{x}$ can be computed with two matrix-matrix multiplications:
\begin{equation}
\label{eq:2d_dft}
\mathbf{X} = [\mathbf{W}_N (\mathbf{W}_M \mathbf{x})^\text{T}]^\text{T}
\end{equation}
Using resistive memory arrays, a direct MVM approach to computing the above expression involves a sequence of $N$ analog MVMs ($M$-point DFTs) followed by $M$ analog MVMs ($N$-point DFTs). Considering the case of $N \times N$ square-shaped 2D DFTs for simplicity, the energy of this approach scales as $\mathcal{O}(N^3)$ due to the need to partition the MVMs across many memory arrays when $N$ is large.

\begin{figure}[t]
\centering
\includegraphics[width=\textwidth]{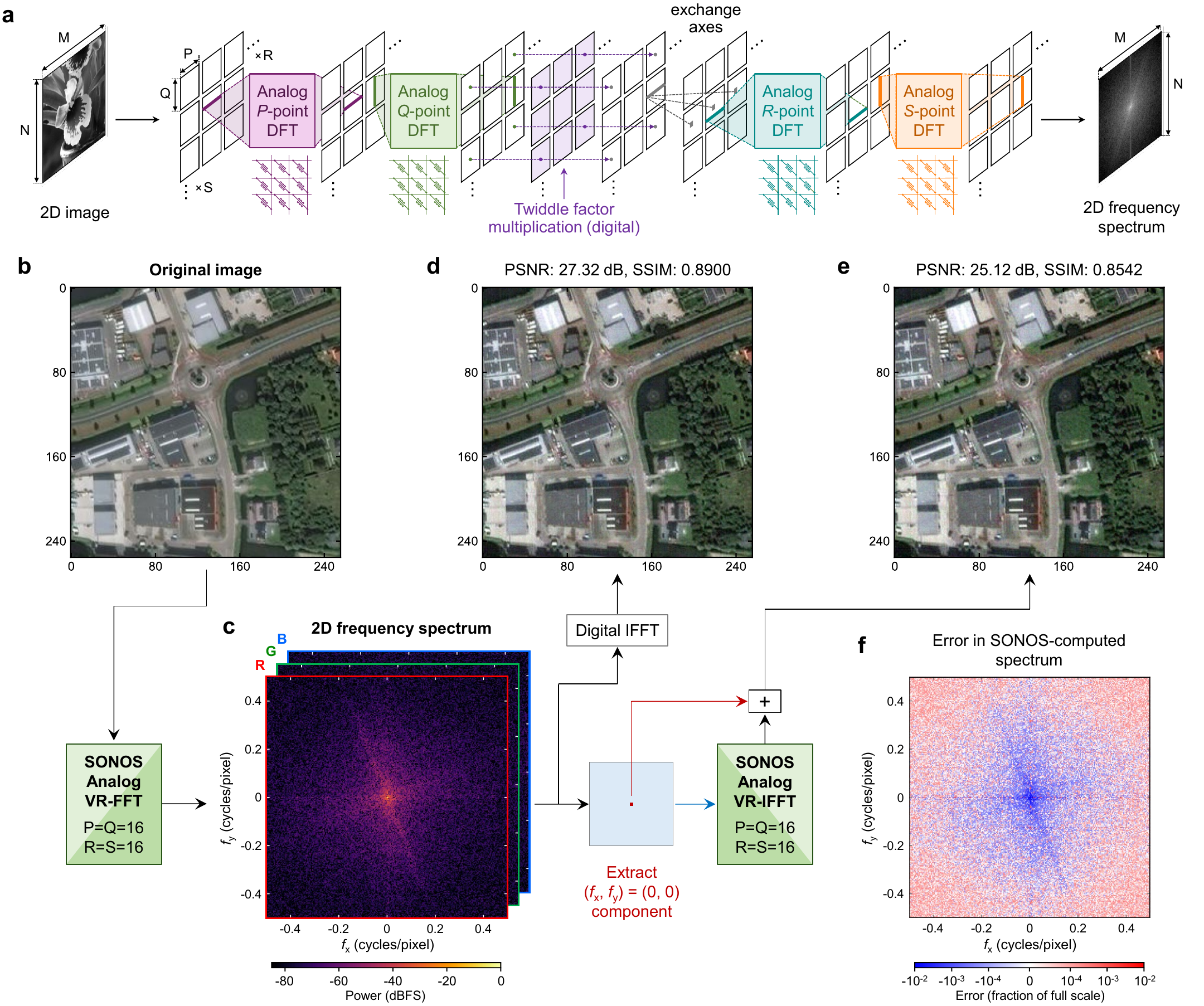}
\caption{\textbf{Analog vector-radix FFT for 2D image processing.} (a) Diagram of the analog 2D ($M \times N$) VR-FFT, which is composed of several analog DFT stages of smaller size. (b) $256 \times 256$ RGB satellite overhead image of Rotterdam, used for analog 2D FFT experiments. (c) Magnitude of the 2D frequency spectrum of the image in (b), computed using an analog VR-FFT by the SONOS array, showing one of three color channels. (d) Reconstruction of the image, obtained by digitally computing a 2D IFFT on the SONOS-computed spectrum. (e) Reconstruction of the image, obtained by computing an analog VR-FFT, then an analog VR-IFFT on the SONOS array. (f) Color-averaged error of the magnitude spectrum in (c) relative to that calculated by a 2D FFT at FP32 precision.
}
\label{fig:vector_radix}
\end{figure}

By applying the Cooley-Tukey decomposition to both dimensions together, the computational complexity of the 2D DFT can be reduced dramatically. This multi-dimensional generalization of the Cooley-Tukey FFT is the vector-radix FFT (VR-FFT) \cite{Rivard77,Harris77}, and its mapping onto analog in-memory computing is illustrated in Fig. \ref{fig:vector_radix}a for the 2D case. In this scheme, each dimension of the input is factored: $M = P \times R$ and $N = Q \times S$. The factors $P$, $Q$, $R$, and $S$ are the sizes of the elementary analog DFTs, which are performed sequentially on the entire image in four stages with one element-wise twiddle multiplication step after the second stage. For very large images, these smaller DFTs can be recursively decomposed. The mathematical details of the analog VR-FFT and its inverse, the analog VR-IFFT, are described in the Methods. If we again consider an $N \times N$ input, the energy of the analog 2D VR-FFT scales as $\mathcal{O}(N^2 \text{log}_K N)$, which is significantly more efficient than the analog direct MVM approach for large $N$, and is fundamentally similar to the $\mathcal{O}(N^2 \text{log}_2 N)$ energy scaling of digital 2D FFT implementations \cite{Duhamel1990}.

We experimentally demonstrate the analog VR-FFT on the SONOS array by computing the spectrum of the RGB image in Fig. \ref{fig:vector_radix}b -- a satellite aerial image of Rotterdam from the SpaceNet-6 dataset \cite{Shermeyer2020}. We factored the 256$\times$256 FFT using the parameters $P = Q = R = S = 16$, which allows all of the analog DFT steps in the VR-FFT to be computed by a single SONOS subarray that was programmed to the DFT-16 matrix. Each color channel was processed independently. Fig. \ref{fig:vector_radix}c shows the magnitude spectrum of the 2D spatial frequencies for the red channel of the image as computed by the SONOS array. The magnitude is largest for low spatial frequencies near zero, and for frequencies that lie along a diagonal through the origin. This diagonal corresponds to the angle of a straight road that spans the entire width of the original satellite image.

To evaluate the fidelity of analog image processing, we used the magnitude and phase of the SONOS-computed frequency spectrum to reconstruct the original image. Fig. \ref{fig:vector_radix}d shows the reconstruction that was obtained by applying an FP32 2D IFFT to the SONOS-computed spectrum using a digital processor. We note that unlike prior experiments that tiled smaller reconstructions into a larger image \cite{li2018analogue,zhao2023energy}, our reconstruction is based on computing the global spectrum of the full, unsegmented 256$\times$256 image in analog. To roughly compensate for signal loss caused by parasitic $IR$ drops, the reconstructed image was uniformly brightened using a scaling factor based on Parseval's theorem for Fourier transforms, as described in Supplementary Section \ref{sec:parseval}. In general, the SONOS VR-FFT preserves the spatial features in the images without introducing significant artifacts, though there is slight graininess caused by random analog conductance errors and circuit noise. Close inspection of the analog reconstructions also reveals that the edges are sharper compared to the original image. This effect is caused by parasitic $IR$ drops, which contribute negative MVM errors that grow with the summed current as shown in Fig. \ref{fig:sonos_dft}f, and therefore reduce the spectral components that have the highest magnitude. For these images, the $IR$ drops attenuate the low spatial frequencies, as shown in Fig. \ref{fig:vector_radix}f, resulting in an image sharpening effect in the reconstruction. 

In Fig. \ref{fig:vector_radix}, we report two similarity metrics between the original and the analog reconstructed images: the reconstruction PSNR, which measures pixel-wise errors, and the structural similarity index measure (SSIM), which evaluates statistical and structural differences between images \cite{Wang2004}.
A similar reconstruction experiment was performed on two other 256$\times$256 RGB images with similar results, as shown in Fig. \ref{fig:sandia_orchid_dmvm}.
We also experimentally computed 2D FFTs at several scales using both the analog VR-FFT and the analog direct MVM method.
Compared to analog FFTs, the reconstructed images using direct MVMs had lower quality metrics due to more significant error accumulation in large arrays, as shown in Supplementary Section \ref{sec:fft_dft_comparison}.

We additionally demonstrate an end-to-end analog reconstruction of the ``Rotterdam'' image, whose result is shown in Fig. \ref{fig:vector_radix}e. Here, the output of the analog VR-FFT was processed with an analog VR-IFFT, which was implemented by a SONOS subarray that was programmed to the IDFT-16 matrix. We note that for this image, the zero-frequency (DC) component of the spectrum is far larger than any other component. To avoid compressing all the non-DC components into a small signal range, the DC component was extracted prior to the VR-IFFT then added to the final IFFT result. This step adds minimal overhead, and would be unnecessary for many types of image sensors (such as synthetic aperture radar) where the inputs naturally have a nearly zero DC component. The additional analog MVM stages in the VR-IFFT only marginally reduce the quality of the reconstructed image in Fig. \ref{fig:vector_radix}e, compared to the result in Fig. \ref{fig:vector_radix}d. This shows that large, multi-dimensional analog FFT/IFFT operations can be accurately cascaded to efficiently implement more complex Fourier-domain signal processing pipelines, such as long-range convolutions, cross-correlations, and Fourier neural operators \cite{Li2020FNO}.

%% file: scaling.tex
\section{Scalability, performance, and efficiency}
\label{sec:scaling}

\begin{figure}[t]
\centering
\includegraphics[width=\textwidth]{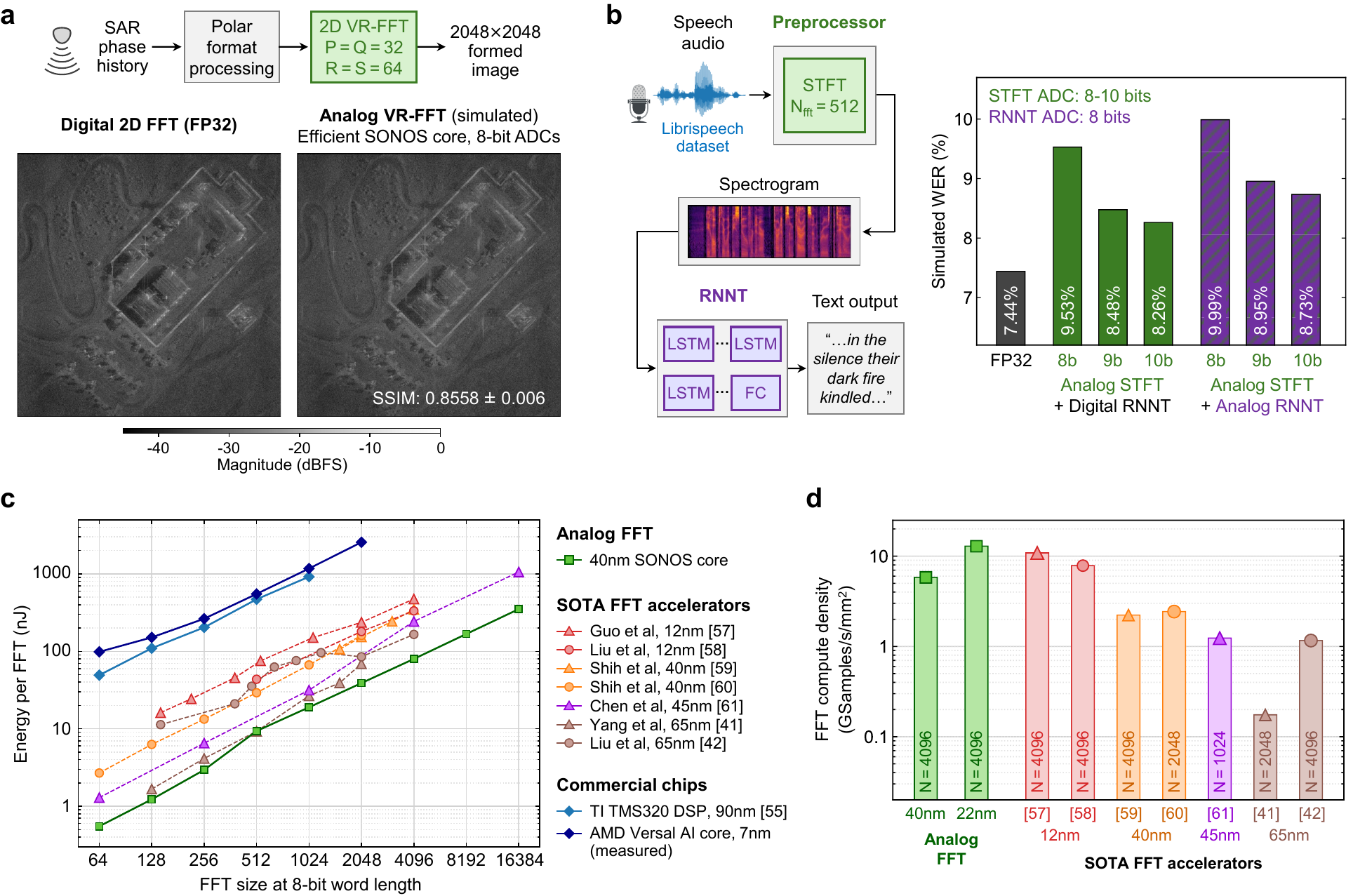}
\caption{\textbf{Simulated accuracy, energy, and performance scaling of analog Fourier transforms.} (a) Formation of a 2048$\times$2048 range-azimuth image from raw SAR phase history data, using an analog VR-FFT as part of the polar format algorithm. The SSIM of the simulated image using analog FFTs (right) is computed relative to the formed image using a 2D FFT at FP32 precision (left). The mean and standard deviation of the SSIM is reported, from ten Monte Carlo accuracy simulations of an optimized 40-nm SONOS core with 8-bit ADCs. (b) Automatic speech recognition using 512-point analog FFTs for spectrogram generation and analog MVMs to accelerate the RNNT speech-to-text neural network. The simulated WER on the Librispeech ``test-clean'' dataset (2620 audio samples) is reported, where SONOS-based analog MVMs are used for STFT only, and for both STFT and RNNT. The ADC resolution for the analog STFT is varied from 8 to 10 bits, while for all RNNT layers it is fixed at 8 bits. (c) Comparison of the FFT energy vs 1D FFT size for a SONOS-based analog FFT, commercial chips that can process DSP workloads \cite{mckeown2010fft,versal-fft}, and various state-of-the-art (SOTA) digital FFT processors from the literature that support a flexible FFT size \cite{Guo2023,Liu2025,Shih2018,Shih2018_2,Chen2018,Yang2012,Liu2019}. For the analog FFT, the ADC resolution is 8 bits and the maximum array size is 1024$\times$1024 (one array can process up to a 256-point DFT). (d) Comparison of the FFT compute density, quantified as the throughput normalized by chip area, between the SONOS-based analog FFT core and specialized digital FFT processors. The analog FFT was evaluated for a 4096-point FFT. The other processors were evaluated at their individual maximum supported FFT size that can fit onto one chip.
}
\label{fig:scaling}
\end{figure}

We now use simulations to assess the scalability of the analog FFT on more practical, larger-scale FFT workloads.
We have validated that our accuracy model (described in Methods), when configured to simulate the properties of the fabricated SONOS test chip, produces results that closely match our experimental data, as shown in Fig. \ref{fig:image_fft_dmvm}.
For the results in this section, we modify this accuracy model to simulate a more energy-efficient analog IMC core than our fabricated prototype, but retain the same error models for the SONOS devices and parasitic $IR$ drops.
This more efficient core, whose operation is described more fully in Supplementary Section \ref{sec:energy_supp}, uses 8-bit ADCs and performs subtraction and input bit accumulation in the analog domain.

We first evaluate how the accuracy of the analog 2D VR-FFT scales to a real application with large transform sizes, by applying it to an image formation algorithm for synthetic aperture radar (SAR), shown in Fig. \ref{fig:scaling}a.
We use raw phase history data from a SAR sensor and an implementation of the widely used polar-format image formation algorithm, both of which are publicly available \cite{ritsar}.
SAR systems use chirped pulses to sample information about the imaged scene in the spatial frequency domain.
The polar-format algorithm first interpolates this nonuniformly sampled, complex-valued data onto a rectangular grid in frequency space; this is a linear pre-processing step that can be done digitally or using resistive arrays with analog MVMs, and we do not explicitly evaluate it here.
Then, a 2D DFT is computed on the interpolated data to form an image along the axes that are perpendicular (range) and parallel (azimuth) to the sensor's flight path \cite{Jakowatz2012}.
Fig. \ref{fig:scaling}a shows the formed aerial SAR images, where the 2048$\times$2048 DFT was computed both using an FP32 FFT (left) and using the analog VR-FFT on a simulated SONOS IMC core (right), where 8-bit inputs and 8-bit ADCs were used in all four stages.
Even at this large transform size, the analog VR-FFT is able to maintain high fidelity (SSIM$\,>0.85$) and preserve all the salient features in the SAR image.

Next, we assess the quality of the analog FFT when its results are used for complex downstream workloads.
For this purpose, we simulate the end-to-end automatic speech recognition (ASR) pipeline in Fig. \ref{fig:scaling}b.
Raw audio waveforms are first pre-processed with STFTs using 512-point FFTs as in our audio processing experiment, then the resulting spectrograms are passed into a large recurrent neural network transducer (RNNT) model that outputs the predicted text transcription \cite{Graves2012}.
We select the pre-trained RNNT model from the MLPerf Inference Benchmark \cite{Reddi2020} and evaluate the word error rate (WER) of the transcription on the Librispeech test dataset \cite{Panayotov2015}.
Using the model for the efficient SONOS-based analog MVM core described above, we simulate two scenarios: analog processing of the STFTs only, and analog processing of both the STFTs and all of the long-short term memory (LSTM) and fully-connected (FC) layers in RNNT, which collectively have 45.3 million weights.
These layers are mapped to analog in-memory computing arrays in a similar manner as demonstrated in Ref.~\citenum{ambrogio2023analog}.
We find that when spectrograms are generated by analog STFTs with 8-bit ADCs, the WER increases by $\sim$2\% relative to a fully FP32 baseline, but this difference decreases to $\sim$1\% when using 9-bit ADCs, which provide slightly higher precision and dynamic range in the spectral coefficients.
The WER saturates for an ADC resolution higher than 9 bits, as the SONOS conductance errors become the bottleneck for MVM precision.
Meanwhile, we find that 8-bit ADC precision provides sufficient accuracy for the RNNT layers, and that combining analog STFTs with analog RNNT processing only marginally increases the WER.
This demonstrates that a fabric like that of Fig. \ref{fig:cooley_tukey}e, where highly efficient analog MVM cores can be used flexibly as ML or DSP engines, can yield accurate results for commercially relevant edge applications.

In Fig. \ref{fig:scaling}c, we compare the energy efficiency of the 40-nm-SONOS based analog FFT to state-of-the-art designs from the literature for specialized FFT processors, as well as two commercial chips with DSP cores.
The digital FFT accelerators, which are not matrix multiplication engines, implement a variety of architectures and algorithms to optimally compute the FFT.
Since these FFT implementations use varying numbers of operations, their efficiency is not well captured by the TOPS/W (TeraOperations/s/Watt) metric typically used for ML processors.
We also compare the area-normalized performance of these FFT processors in Fig. \ref{fig:scaling}d.
For the analog FFT, we focus on the 8-bit ADC design, which is close to the analog error dominated regime as described above.
To ensure an iso-precision comparison, we linearly scaled both the energy efficiency and compute density of all the digital FFT processors from their original FFT word length to 8 bits \cite{Guo2023}.
For the analog implementation, Cooley-Tukey decomposition is used only for DFT sizes larger than 256.
See Supplementary Section \ref{sec:energy_supp} for details on the analog IMC core's parallelized and pipelined operation, and the projections for its energy, area, and performance.

Across a wide range of practical FFT sizes, the analog FFT processor can achieve a higher energy efficiency than the most efficient digital ASICs for the FFT.
It also provides higher FFT throughput per area than the digital FFT ASICs at a comparable process node.
When implemented using a more scaled 22-nm SONOS memory technology \cite{Agrawal2022}, the analog FFT can deliver higher performance per area than a digital FFT ASIC design in a 12-nm node, as shown in Fig. \ref{fig:scaling}d.
While significant, we note that these are not orders-of-magnitude improvements because, as described earlier, the low-radix digital FFTs and the high-radix analog FFTs offer distinct, mutually exclusive paths to high efficiency and performance: the first leads to a greater reduction in operation count, while the second drastically reduces the energy per operation.

The key unique advantage of the analog FFT is that it delivers best-in-class efficiency and performance while being far more flexible than digital ASICs that are optimized exclusively for the FFT: the same analog arrays can be re-programmed as matrix multiplication engines for ML algorithms that can deliver up to 249 TOPS/W of energy efficiency (depending on the matrix size, see Fig. \ref{fig:TOPSW}). 
The analog core can also be re-programmed to implement other linear transforms that are useful in edge signal processing, such as the DCT, Chirp-Z transform, discrete wavelet transform, and interpolation -- all of which can be computed using MVMs.
When compared to the AMD Versal, which is also a flexible processor for AI and DSP workloads, the analog IMC core can execute FFTs at $>$59$\times$ higher efficiency, and matrix multiplications at $>$63$\times$ higher efficiency, for typical DNN matrices that have more than 256 rows.
The analog FFT can be realized using any technology for analog IMC, such as memristors \cite{Wang2025} or capacitive matrix multiplication engines \cite{Lee2024}.
Future implementations that can achieve higher TOPS/W or higher compute density for matrix multiplication would also improve the efficiency and throughput-per-area of analog FFTs that are carried out at the same precision.

%% file: conclusion.tex
\section{Conclusion}
\label{sec:conclusion}

We have shown that the fast Fourier transform can be implemented on analog in-memory computing accelerators, enabling analog systems to scale efficiently and accurately to practically useful DFT sizes that were otherwise inaccessible. 
By decoupling the full DFT size from the physical array size, we have experimentally demonstrated a 65,536-point analog DFT computed using SONOS memory, which is more than 500$\times$ larger than the largest previous demonstration \cite{Wang2025}.
The analog FFT technique allows the same resistive crossbar array to serve both as an efficient, scalable, and flexible FFT engine, and as an efficient matrix multiplication engine for ML and linear algebra workloads.
This versatility can transform the capabilities of edge systems that often need both DSP and ML processing within a low power envelope.
A homogeneous fabric of reprogrammable, multi-purpose accelerators would also simplify the hardware design of such systems, while providing superior efficiency and performance to a heterogeneous combination of state-of-the-art digital ASICs.
Applying similar principles to other divide-and-conquer algorithms may further expand the versatility of analog in-memory computing hardware beyond its traditional domain of matrix multiplication.

%% file: methods.tex
\newpage

\section*{Methods}
\label{sec:methods}

\subsection*{SONOS analog IMC demonstration system}

The SONOS IMC chip used in this work was fabricated in a commercial 40-nm foundry process. The formation of the embedded SONOS memory cell in Fig. \ref{fig:sonos_dft}a was integrated into the front-end-of-line CMOS logic process. The IMC chip contains a 1024$\times$1024 array of SONOS memory cells along with analog and digital circuitry to support write, read, and analog MVM operations. Fig. \ref{fig:demo} shows the analog IMC demonstration system used in this work. The fabricated die was wire-bonded to a 100-pin thin quad flat package (TQFP), and the packaged SONOS IMC chip was mounted on a socket on a custom-designed circuit board. Commands and input/output data were sent to and from the IMC chip through an Infineon Technologies ARM microcontroller on a second custom-designed board. Four Keysight N6705C DC power supplies and a 9V power adapter were used to power the IMC chip, microcontroller, and boards. The microcontroller communicates with the host PC through a USB interface. All experiments were conducted through sequences of read, write, and analog MVM commands that were sent to the microcontroller using a Python application programming interface.

\subsection*{Programming and characterizing the SONOS array}

The conductances of the SONOS devices were programmed with a write-verify algorithm that was designed to minimize variability and drift. To obtain a precise and stable conductance, the algorithm selectively places charge in deep, mid-gap electronic traps within the nitride that have a large energy barrier for charge confinement, while vacating charge from shallow traps near the band edges that de-trap over short timescales \cite{Agrawal2020,xiao2022accurate}. All conductances and currents reported in this work were measured with the same voltage bias at the four terminals of the SONOS cell in Fig. \ref{fig:sonos_dft}a: 0.06V on the bit line ($V_{BL}$), 2.5V on the select transistor gate ($V_{SG}$), 0V on the source line ($V_{SL}$, the select transistor source), and 0V on the control gate ($V_{CG}$, the SONOS transistor gate). The resolution of the SONOS cell current measurements is 0.88 nA, set by the fine-resolution setting of the 12-bit ADC on the IMC chip. Our systematic characterization of the state-dependent programming variability and conductance drift in the SONOS cells is described in Supplementary Section \ref{sec:gradient}.

When programming a DFT matrix onto a section of the SONOS array, the complex-valued matrix $\mathbf{W}_K$ is first decomposed into four real matrices as shown in Fig. \ref{fig:sonos_dft}b to support multiplication with a complex-valued input vector $\mathbf{x}$. Each real value is encoded by the difference in conductance of two SONOS cells. We use the convention where for a positive-valued weight, one cell encodes the absolute value while the other cell is programmed to a target of 0{~\textmu}S, and vice versa for negative weights \cite{OnTheAccuracy}. The maximum SONOS conductance $G_\text{max}$ used to represent DFT weights was set primarily by the ADC input current limit of 17{~\textmu}A. A large $G_\text{max}$ would cause excessive clipping of analog current sums by the ADC, while a small $G_\text{max}$ reduces the effective DFT weight precision because the conductance errors (shown in Fig. \ref{fig:gradient}e) become a larger fraction of the utilized conductance range. To balance these errors, we first simulated all of the analog DFT operations for each of our FFT experiments prior to programming. We then set $G_\text{max}$ to the largest value which ensured that 99.99\% of the simulated analog sums were not clipped by the ADC, then programmed the SONOS array using this value. In general, the selected $G_\text{max}$ decreases with increasing DFT size because more SONOS cell currents are summed. For the results in Figs. \ref{fig:audio_spectrograms}, \ref{fig:vector_radix}, and \ref{fig:scaling}, we used $G_\text{max} = 20${~\textmu S} for the DFT-8 and DFT-16 arrays, $G_\text{max} = 16.7${~\textmu S} for DFT-32, and $G_\text{max} = 6.2${~\textmu S} for DFT-256. Supplementary Section \ref{sec:dft_weights} has additional experimental results on the conductance statistics for DFT arrays of various sizes.

\subsection*{SONOS chip analog DFT computation}

To compute analog MVMs, input values are applied bit-serially to the SONOS array to the select gate (SG) lines. If a bit is `1', the SG is driven to 2.5V to turn on the select transistor, and the SONOS transistor conducts current from the BL to the source line (SL). If a bit is `0', the SG is biased to 0V, and the cell does not conduct any current from the BL. Each BL is held at 0.06V by a voltage regulator. The control gate (CG) is held at 0V. The SONOS cell conducts current only in one direction (BL to SL), which keeps its drain current more robust to variations in the BL voltage. For each input bit, the summed current on each BL is converted to a voltage using a TIA, then converted to a 12-bit digital value by the ADC. In analog MVM mode, the on-chip ADC has an input current dynamic range of 17{~\textmu}A and a current spacing between ADC levels of 4.88 nA. Any summed current that exceeds this limit is clipped to the maximum value upon digital read-out.

To increase the throughput of analog MVMs on the test chip for smaller DFT or IDFT sizes ($K < 64$), we programmed multiple copies of the DFT/IDFT matrix onto the SONOS array and tiled them block-diagonally. This allowed multiple bit-wise MVMs to be processed concurrently in a single analog MVM. All SONOS cells not lying along the block diagonal were programmed to 0{~\textmu S}. The large On/Off ratio of the SONOS memory eliminates sneak currents and ensures that the bit-wise MVM results are independent. MVMs with positive and negative input values were computed in separate cycles then subtracted digitally. On this chip, the results of each analog bit-wise MVM were digitized. To obtain the final real and imaginary DFT outputs, the following digital post-processing steps were conducted: (1) for each analog MVM, subtraction of the summed results for positive and negative weights, (2) for each input magnitude bit, subtraction of the results for positive and negative inputs, and (3) for each input magnitude bit, power-of-two weighted accumulation of the result. Reshapes and transposes of intermediate data in the FFT dataflow, as well as uniform scaling steps (e.g. FFT/IFFT normalization), were also handled digitally. 

For all experiments, we used 13-bit signed integers (12 magnitude bits and a sign bit) for the inputs to each analog DFT or IDFT, with the exception of the first VR-FFT stage which used 8-bit unsigned integers since this was the original format of the JPEG images. Intermediate results between DFT stages were re-quantized to 13 bits before the next analog DFT stage, over a range that is set by their maximum value. The digital baseline FFTs used the same resolution for the input audio waveform or image as the analog FFTs, but used FP32 precision for the computation. In the experiments, twiddle factor multiplications between the DFT stages of the FFT were computed digitally at FP32 precision. As described in the main text and in Supplementary Section \ref{sec:parallel_fft}, it is possible to fold the twiddle factor multiplications into the analog MVMs for higher efficiency. Since the audio and image signals used for the experiments were purely real, their frequency spectra should be zero-symmetric with frequency. To produce the audio magnitude spectra in Fig. \ref{fig:audio_spectrograms}, we calculated the mean of the complex-valued positive- and negative-frequency spectral components that were computed by the SONOS array as: $|\frac{1}{2}\left(X_k + X^*_{-k}\right)|$.

\subsection*{Vector-radix FFT details}

We describe here the mathematical formulation of the 2D VR-FFT and its implementation using analog MVMs, which is summarized in Fig. \ref{fig:vector_radix}a. Analogous to the 1D case, the first step is to reshape the 2D input matrix $\mathbf{x}$ to a 4D matrix $\mathbf{\tilde{x}}$, with dimensions $R \times S \times P \times Q$, where $M = P \times R$ and $N = Q \times S$. The values of $R$, $S$, $P$ and $Q$ are the sizes of the elementary DFT operations in the factorization and are directly related to the dimensions of the constituent analog MVMs. As drawn in Fig. \ref{fig:vector_radix}a, this reshape can be visualized as partitioning $\mathbf{x}$ into an $R \times S$ grid of sub-matrices, each of which has dimensions $P \times Q$. The matrix $\mathbf{\tilde{x}}$ has four indices: the pair $(r, s)$ indexes a sub-matrix and the pair $(p, q)$ indexes an element of the sub-matrix. This is specified by: $\mathbf{\tilde{x}}_{r,s}(p,q) = \mathbf{x}_{r+Rp,s+Sq}$. In the VR-FFT, a $P \times Q$ 2D DFT is first performed on every sub-matrix $\mathbf{\tilde{x}}_{r,s}$, then the resulting matrix is element-wise multiplied by a matrix of twiddle factors $\mathbf{T}_{r,s}$ whose values also depend on the sub-matrix. The result is a 4D intermediate matrix $\mathbf{y}$ with the same dimensions as $\mathbf{\tilde{x}}$, where each sub-matrix is specified by:

\begin{equation}
\label{eq:vrfft_1}
\mathbf{y}_{r,s} = \mathbf{T}_{r,s} \odot [\mathbf{W}_Q ( \mathbf{W}_P \,\mathbf{\tilde{x}}_{r,s})^\text{T}]^\text{T}
\end{equation}
Next, the positions of the $(p, q)$ and $(r, s)$ axis pairs of $\mathbf{y}$ are exchanged to form a matrix $\mathbf{\tilde{y}}$ with dimensions $P \times Q \times R \times S$. Afterwards, an $R \times S$ 2D DFT is performed on every sub-matrix of $\mathbf{\tilde{y}}$:

\begin{equation}
\label{eq:vrfft_2}
\mathbf{\tilde{X}}_{p,q} = [\mathbf{W}_{S}(\mathbf{W}_{R} \,\mathbf{\tilde{y}}_{p,q})^\text{T}]^\text{T}
\end{equation}
Finally, the 4D matrix $\mathbf{\tilde{X}}$ is re-shaped to obtain the 2D frequency spectrum $\mathbf{X}$: $\mathbf{X}_{q+Qs,p+Pr} = \mathbf{\tilde{X}}_{p,q}(r,s)$. More details on the VR-FFT dataflow are provided in Supplementary Section \ref{sec:vector_radix_overhead}.

Each of the smaller 2D DFTs in Equation \ref{eq:vrfft_1} and \ref{eq:vrfft_2} can be computed using a sequence of analog MVMs within a resistive memory array. The four matrix-matrix multiplications in the two equations comprise the four stages of the VR-FFT that must be computed sequentially. The number of required ADC conversions, without any further application of Cooley-Tukey decomposition, is proportional to the product $PQRS$, or $\mathcal{O}(MN)$. In the limit of a very large 2D DFT, these constituent 1D DFTs can be computed using 1D analog FFTs, and their energy would each scale as $\mathcal{O}(N \text{log}_K N)$, where $K$ is the size of the elementary analog DFT. The total number of ADC conversions needed for the VR-FFT would scale in this regime as $PQRS \times \left( \text{log}_K P + \text{log}_K Q + \text{log}_K S + \text{log}_K R \right)$, which is equivalent to $\mathcal{O}(MN \, \text{log}_K (MN))$. For square DFTs where $M = N$, this asymptotic scaling law simplifies to $\mathcal{O}(N^2 \, \text{log}_K N)$.

The analog VR-IFFT can be computed using exactly the same steps as the VR-FFT, with two changes: (1) replace the DFT matrices with their conjugate transposes, which are the IDFT matrices, and (2) replace the twiddle matrix with its element-wise complex conjugate.

\subsection*{Accuracy simulations of the SONOS FFT/DFT}

Accuracy simulations of the analog FFT in Fig. \ref{fig:scaling} and the Supplementary Information were conducted using the CrossSim modeling tool \cite{CrossSim}. These simulations were conducted in two modes: (1) behavioral replication of the fabricated 40-nm SONOS test chip to validate the realism of our simulations, and (2) modeling of a more efficient 40-nm SONOS IMC core that is described in Supplementary Section \ref{sec:energy_supp}.

In mode 1, we chose the analog hardware parameters in CrossSim to model: the bit-serial operation of analog MVMs; the SONOS cell conductance variability, drift, and read noise; 12-bit ADC quantization and clipping for each bit-wise MVM; and parasitic resistances in the array. 
Modeling of the variability and drift in the SONOS cells is based on analytical fits to the conductance characterization data in Fig. \ref{fig:gradient}e and \ref{fig:gradient}f; for replicating an experiment, the drift time was set to one or three days, depending on the actual time since programming for a specific experiment.
Random errors due to SONOS variability and drift were re-sampled by running multiple Monte Carlo simulations of the full workload.
By contrast, random cycle-to-cyle read noise was re-sampled on every bit-wise analog MVM, and was modeled based on the 40-nm SONOS noise properties reported in Ref.~\citenum{Agrawal2020}. 
The simulated accuracy values in Fig. \ref{fig:scaling}a-b report the mean of ten Monte Carlo runs; the variance on all the accuracies were small.
To model the effects of parasitic $IR$ drops on accuracy, we added an error to the summed currents that increased quadratically with the correct value of the summed current, similar to the method used in Ref.~\citenum{rasch2023hardware}.
This simple model does not capture the complicated input data dependence and spatial non-uniformity of the $IR$ drops \cite{xiao2021analysis}, but fits well to the average trend of the errors in Fig. \ref{fig:sonos_dft}f.
Fig. \ref{fig:image_fft_dmvm} shows that simulations using this mode match well to our experimental results.

In mode 2, we retained the same models for the SONOS device variability, read noise, and array parasitic resistance. We modeled one day of SONOS drift, as we expect a more optimized chip to have much higher throughput, and a weight refresh interval of one day would have a very small power and endurance overhead in real edge deployments. Additionally, we modified the MVM accuracy model to subtract the currents from positive and negative weight columns, and from positive and negative inputs, in the analog domain prior to the ADC. We also assumed charge-domain analog accumulation of partial results from different input bits. These changes reduce the number of ADC conversions as well as the dynamic range of inputs to the ADC; the required circuit modifications are discussed in Supplementary Section \ref{sec:energy_supp}. The ADC resolution was changed from the test chip's resolution of 12 bits to 8, 9, or 10 bits for the different results in Fig. \ref{fig:scaling}a-b. To minimize both quantization and clipping errors at these lower resolutions, the input dynamic range for each ADC was optimized based on profiling of the input data for the individual analog DFT stages, as well as the neural network layers in RNNT.

\subsection*{DSP workload accuracy simulations}

The analog FFT operations that were simulated for the workload results in Fig. \ref{fig:scaling}a and \ref{fig:scaling}b used mode 2 of the analog FFT accuracy simluation, described above.

For the SAR results in Fig. \ref{fig:scaling}a, we used raw SAR sensor data that was collected on a test flight by Sandia National Laboratories, available at Ref.~\cite{ritsar}. We also used the Python implementation of the polar-format algorithm in this repository, without modification until the 2D DFT step. The complex-valued SAR phase history contains data that are sampled at discrete points on a polar annulus in the spatial frequency domain. The polar-format algorithm interpolates this data onto a rectangular grid, first along the range then along the azimuth direction. Then, a 2D DFT is performed to obtain a real-space image. We simulate analog processing only for the 2D DFT step while the other parts of the algorithm are done in digital. We quantized the floating-point inputs to the DFT to 8 bits (sign-magnitude), for both the FP32 FFT and the analog VR-FFT; we found that this made a negligible difference in the final image. No Parseval correction was applied for the analog VR-FFT. The magnitude of the FFT result is plotted on a logarithmic scale in Fig. \ref{fig:scaling}a, normalizing by the maximum pixel value of the FP32 formed image. The SSIM was computed using the logarithmic-scale images in Fig. \ref{fig:scaling}a. We note that the raw data was collected using 1999 chirp pulses with 1800 samples per pulse, where each pulse had the same center frequency of 2.67 GHz. This was slightly upsampled to form a 2048$\times$2048 image.

For the ASR results in Fig. \ref{fig:scaling}b, STFTs were simulated on raw Librispeech audio waveforms using 512-point analog FFTs, with a Cooley-Tukey decomposition of $N_1 = 32$ and $N_2 = 16$. These were applied to zero-padded windows with a window size of 320 samples, a hop length of 160 samples, and a Hanning window function. This was followed by a few digital steps prior to passing the inputs to RNNT: conversion to a power spectrogram, conversion to a Mel-frequency spectrogram with 80 logarithmically spaced frequency bins (features), element-wise logarithm, and per-feature normalization to a mean of zero and standard deviation of one. Within RNNT, all LSTM and FC layers were simulated using CrossSim, using the same hardware parameterizations as the analog FFT but with a fixed ADC resolution of 8 bits. Transcendental functions in the LSTMs were assumed to be computed digitally, and weight matrices larger than 1024$\times$1024 were partitioned across multiple SONOS arrays with separate ADCs.

\subsection*{SONOS IMC core energy, performance, and area projections}

The fabricated 40-nm test chip used for our experiments was designed to demonstrate high-precision SONOS-based analog MVMs, but was not optimized for performance, power efficiency, or area. Therefore, the energy, performance, and area metrics for the analog FFT in Fig. \ref{fig:scaling}c-d are modeled based on a more optimal SONOS-based IMC core design with faster and more efficient peripheral circuits. The details of these energy, performance, and area models are described in Supplementary Section \ref{sec:energy_supp}.

\subsection*{Versal AI Engine benchmarking}

The FFT energy values in Fig. \ref{fig:scaling}c for the AMD Versal were obtained by executing 2D-FFT designs on a Versal AI core (VC1902) chip, following the documentation in Ref.~\citenum{versal-fft}.
The designs implemented 2D row-column FFTs of various sizes ($N \times N$, from $N=64$ to $N=2048$) at CINT16 precision on the AI Engines, and were compiled using the AMD Vitis AI tool (version 2023.1) for the VCK190 evaluation board.
The compiled programs were serially written to bootable program images, which were written to SD cards using the BalenaEtcher tool, and then booted up one at a time on the VCK190's PetaLinux environment, then finally executed on the AI Engines.
The AI Engines were run at 1.00 GHz and communicated with a data mover kernel via a 128-bit AXI4-Stream interface at 312.5 MHz.
We report results from runs that contained five parallel instances of the same design on the chip, which reduced the per-instance power consumption.
The latencies of individual 2D FFTs were obtained from the runtime trace data, and these were scaled to derive the latencies of 1D $N$-point FFT in the batch.
Power estimates were obtained from the Xilinx Power Design Manager tool in a vector analysis mode, which used switching data from detailed timing simulations to produce accurate estimates of signal toggle rates and dynamic power consumption.
We include only the power consumption of the AI Engines to ensure a fair comparison with the other FFT accelerators.
The energy per FFT was obtained by multiplying the 1D FFT latency by the per-instance power consumption.
For Fig. \ref{fig:scaling}c, the energy was further reduced by $2\times$ to estimate the power for an 8-bit (rather than 16-bit) word length.

%% file: suppmat.tex
\section*{\centering{\Large Supplementary Information}}

\begin{appendices}

\renewcommand{\thesection}{\arabic{section}}
\renewcommand{\thefigure}{S\arabic{figure}}
\renewcommand{\thetable}{S\arabic{table}}

\input{_SI_parallel_FFT}
\newpage
\input{_SI_platform}
\newpage
\input{_SI_sonos}
\newpage
\input{_SI_dft_weights}
\newpage
\input{_SI_device_requirements}
\newpage
\input{_SI_reconfigurability}
\newpage
\input{_SI_factorizations}
\newpage
\input{_SI_dot_products}
\newpage
\input{_SI_parseval}
\newpage
\input{_SI_2dfft}
\newpage
\input{_SI_sar}
\newpage
\input{_SI_energy_performance_area}
\newpage
\input{_SI_vector_radix_overhead}
\newpage
\input{_SI_scaling_laws}
\newpage
\end{appendices}

%% file: _SI_parallel_FFT.tex
\section{High-throughput analog FFTs with analog twiddle multiplications}
\label{sec:parallel_fft}

\subsection{Parallelized, twiddle-folded FFT mapping}

\begin{figure}[h]
\centering
\includegraphics[width=\textwidth]{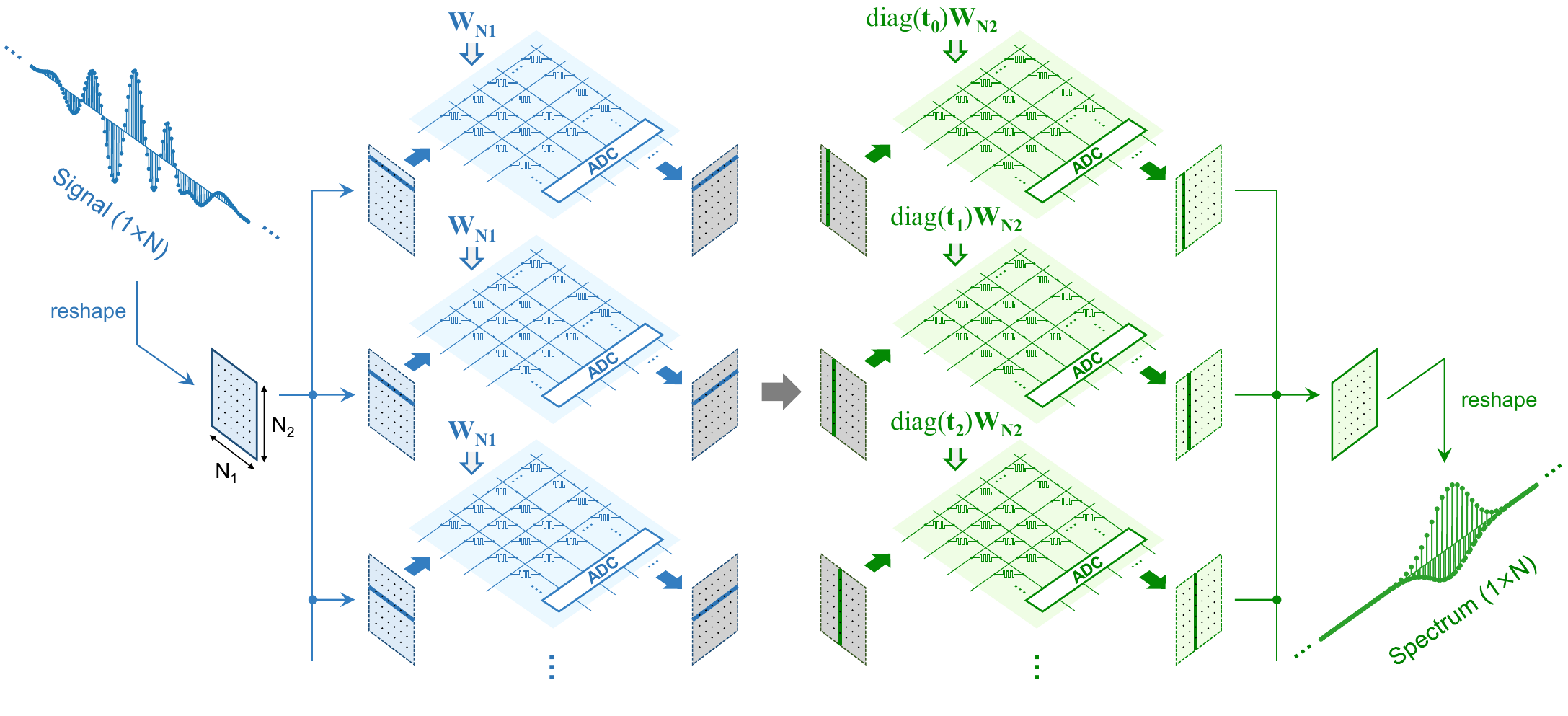}
\caption{High-throughput parallelized implementation of an analog FFT with a single Cooley-Tukey decomposition.
}
\label{fig:parallel_fft}
\end{figure}

In the most compact implementation of the analog FFT, all of the elementary DFTs within an FFT stage are computed sequentially using a single resistive memory array. However, this implementation comes at a penalty to latency and throughput, due to the sequential execution of matrix-vector multiplications (MVMs) that make up each matrix-matrix multiplication. To attain higher speed at the cost of area, the elementary DFTs within an FFT stage can be executed in parallel across multiple arrays, rather than sequentially using a single array. This solution is shown in Fig. \ref{fig:parallel_fft}. In this scheme, the $N_2$ MVMs of the first stage are computed in parallel across $N_2$ arrays that each store the matrix $\mathbf{W}_{N_1}$. Likewise, the $N_1$ MVMs of the second stage are computed in parallel across $N_1$ arrays.

In addition to high throughput, another benefit of this parallelized scheme is that \textbf{it eliminates the need for explicit digital twiddle factor multiplications}. Instead, the twiddle factor multiplications can be folded into the analog MVMs of the second stage so that they introduce no additional energy, latency, or storage overhead during runtime. This can be done by first decomposing the $N_1 \times N_2$ twiddle matrix into its column vectors: $\mathbf{T} = \left[ \mathbf{t}_0 \,\,\, \mathbf{t}_1 \,\,\, \mathbf{t}_2 \,\, ... \,\, \mathbf{t}_{N_1-1}\right]$. Each of the $N_1$ memory arrays that are allocated for the second stage is then programmed with a unique matrix; the $i^\text{th}$ array is programmed to store $\text{diag}(\mathbf{t}_i) \, \mathbf{W}_{N_2}$, where $\text{diag}(\mathbf{t}_i)$ is a diagonal matrix whose main diagonal consists of the elements of $\mathbf{t}_i$.
This multiplication simply scales the rows of $\mathbf{W}_{N_2}$ by the vector $\mathbf{t}_i$.
In this hardware implementation, the twiddle multiplications come at zero additional cost in time, energy, or area. We also note that the cost of an element-wise multiplication pre-processing step (e.g. a window function) can also be eliminated using the same method, by folding it into the programmed matrices of the first stage. The parallel MVMs in the first stage, the storage of intermediate results, and the parallel MVMs in the second stage can all be pipelined. This allows the parallel analog FFT to match the throughput of an analog direct MVM approach.

In mathematical terms, this hardware implementation can be understood to implement the equation below, which yields the same result as Equation \eqref{eq:cooley_tukey_matrix}:
\begin{equation}
\label{eq:parallel_fft_matrix}
\mathbf{X} = \left[ \mathbf{I}_{N_1} \otimes_i \left( \text{diag}(\mathbf{t}_i) \, \mathbf{W}_{N_2} \right) \right] \mathbf{P} \left(\mathbf{I}_{N_2} \otimes \mathbf{W}_{N_1} \right) \mathbf{x}
\end{equation}
where $\mathbf{I}_k$ is the $k \times k$ identity matrix and $\otimes$ is the Kronecker product.
The product $\mathbf{I}_{N_2} \otimes \mathbf{W}_{N_1}$ is equivalent to tiling the matrix $\mathbf{W}_{N_1}$ block-diagonally $N_2$ times.
The notation $\mathbf{I}_{N_1} \otimes_i \mathbf{F}_i$ indicates a direct sum of matrices, i.e. tiling the matrices $\mathbf{F}_0, \mathbf{F}_1, ..., \mathbf{F}_{N_1-1}$ block diagonally, following Ref.~\citenum{Puschel2008}. 
In hardware, the block diagonals are implemented in separate arrays while the zero-valued off-diagonal elements do not need to be physically implemented. In this mathematical formulation, $\mathbf{x}$ is transformed into $\mathbf{X}$ through two large MVMs rather matrix-matrix multiplications.
$\mathbf{P}$ is a matrix that represents a permutation operation on the intermediate $N\times 1$ vector, and performs the equivalent function to the inner transpose in Equation \eqref{eq:cooley_tukey_matrix}.
The 1D vectors $\mathbf{x}$ and $\mathbf{X}$ differ only by a possible permute operation from the original input signal and the final DFT output, respectively.

\begin{figure}[t]
\centering
\includegraphics[width=0.9\textwidth]{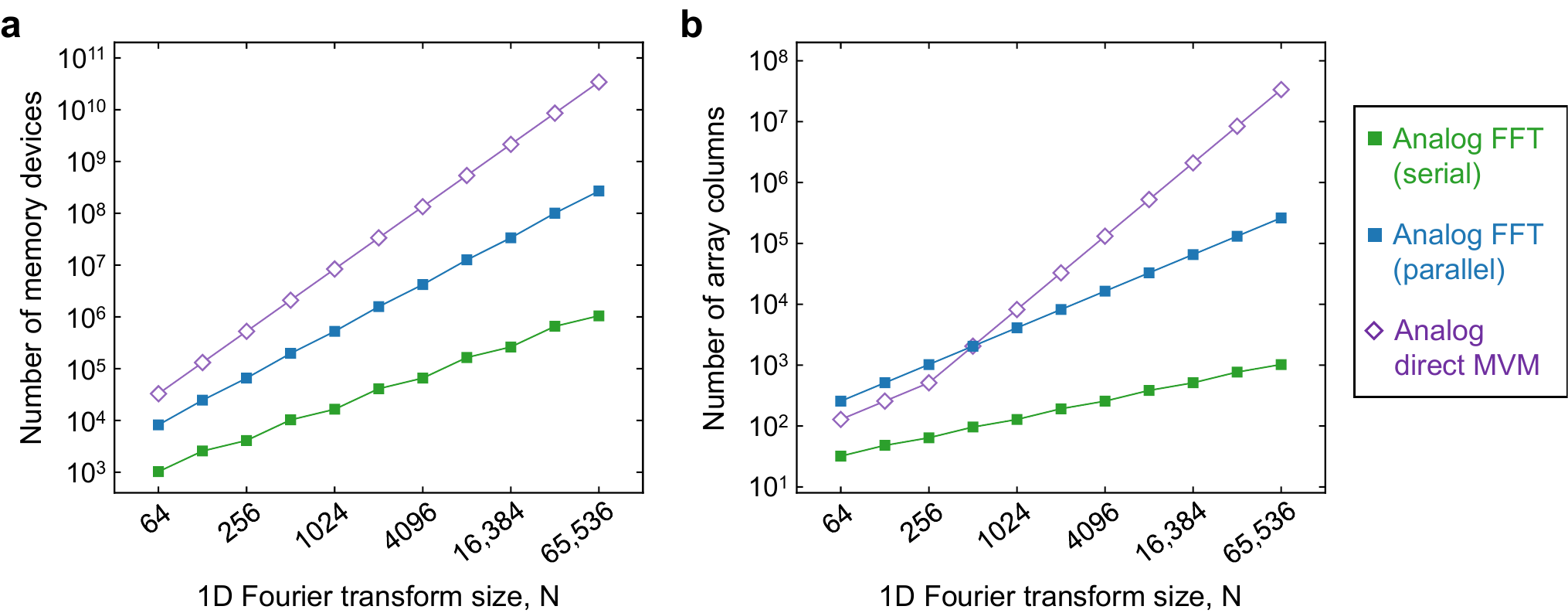}
\caption{Comparison of the area scaling of the serial implementation of the analog FFT, parallel implementation of the analog FFT, and the analog direct MVM implementation of the DFT, in terms of (a) total number of memory devices, and (b) total number of array columns. The maximum array size is assumed to be $512\times1024$, which implements an elementary DFT size of 256.
}
\label{fig:area_scaling}
\end{figure}

Even after provisioning for multiple arrays, the parallel analog FFT has a more favorable area scaling than the analog direct MVM while achieving the same throughput.
This is shown in Fig. \ref{fig:area_scaling}, which compares the area scaling of the analog direct MVM, and two implementations of the analog FFT: the serial implementation in Fig. \ref{fig:cooley_tukey}b, and the parallel implementation in Fig. \ref{fig:parallel_fft}.
For this scaling analysis, we consider two proxies for the total area: the total number of resistive memory devices (if the memory array area dominates), and the total number of memory array columns (if the column peripheral circuitry area dominates).
By both measures, the area overhead of the parallel FFT scales much more slowly with the transform size compared to the direct MVM; for large $N$, the area overhead of the parallel FFT is several orders of magnitude smaller.

We note that the dynamic reconfigurability property of the analog FFT that was described in Section \ref{sec:cooleytukey} still holds even when the twiddle matrices are folded into the DFT matrices to implement this parallel FFT scheme. This is shown mathematically in Supplementary Section \ref{sec:reconfigurability}.

\subsection{Experimental demonstration of folded-twiddle analog FFT}

\begin{figure}[t]
\centering
\includegraphics[width=\textwidth]{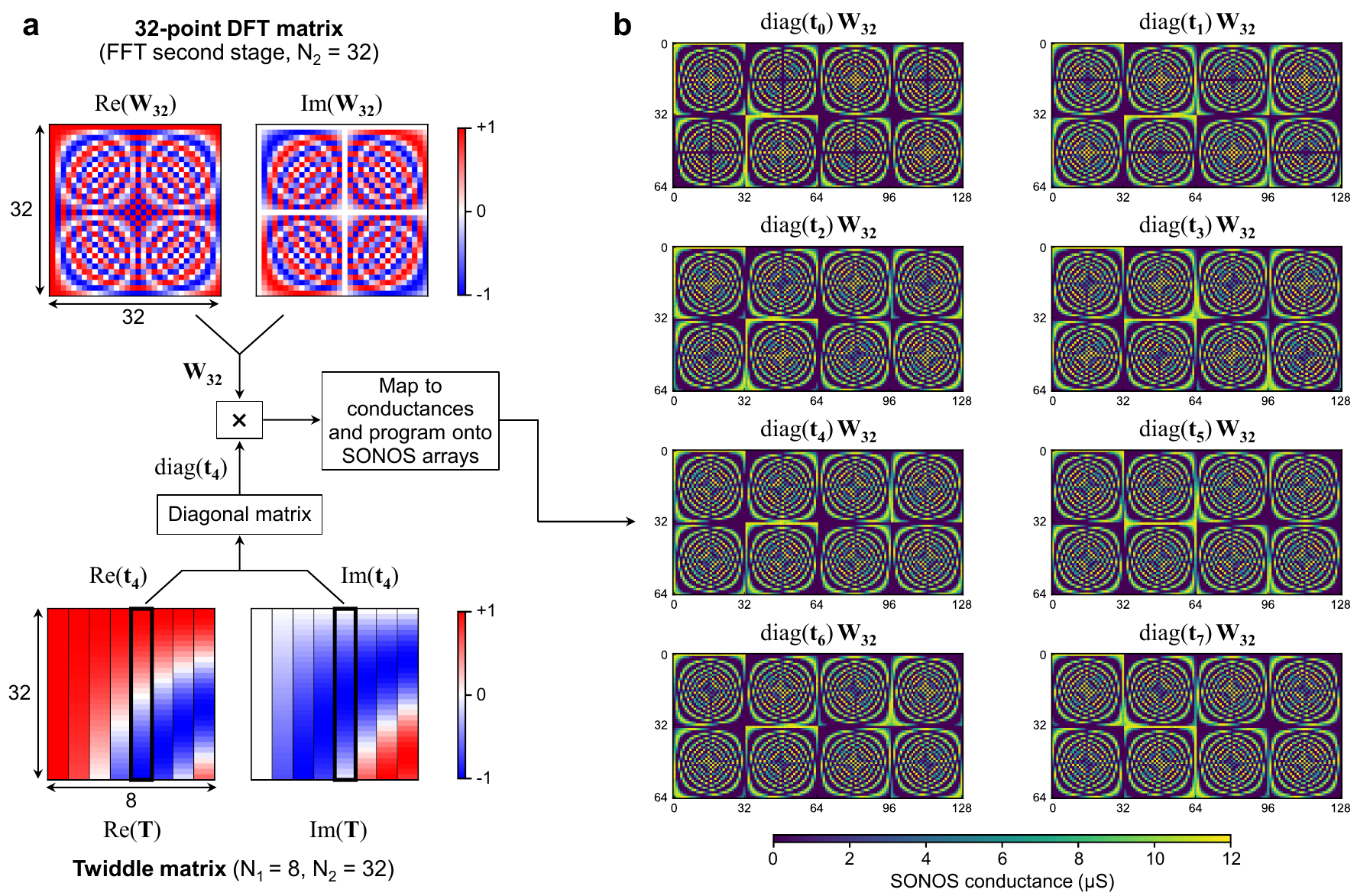}
\caption{Experimental demonstration of a folded-twiddle, 256-point analog FFT with $N_1 = 8$ and $N_2 = 32$. (a) The twiddle factors are folded into the second stage by multiplying the DFT-32 matrix by diagonal matrices formed from each column of the twiddle matrix $\mathbf{T}$. The eight resulting matrices are converted to conductances and programmed onto different 64$\times$128 sectors of the SONOS array. (b) Measured SONOS conductances that store each of the eight twiddle-folded DFT-32 matrices used for the 256-point FFT.
}
\label{fig:twiddle_folding}
\end{figure}

We experimentally demonstrate that the twiddle factor multiplications in the FFT can be folded into the analog MVMs without any adverse affect on accuracy.
For this experiment, we compute a 256-point 1D FFT that is factored into $N_1 = 8$ and $N_2 = 32$.
The first FFT stage is computed using a SONOS subarray that was programmed with the DFT-8 matrix.
We do not perform any digital twiddle multiplications prior to the second stage.
Instead, we fold the twiddle factors into the SONOS conductances that are used to process the second stage, using the method shown on the right side of Fig. \ref{fig:parallel_fft}.

Fig. \ref{fig:twiddle_folding} shows the details of the twiddle folding process for this example, where the complex-valued twiddle matrix $\mathbf{T}$ has dimensions 32$\times$8.
Each 32$\times$1 column vector $\mathbf{t}_i$ is expanded into a 32$\times$32 diagonal matrix $\text{diag}(\mathbf{t}_i)$, then multiplied by the complex-valued DFT-32 matrix to obtain $\text{diag}(\mathbf{t}_i)\mathbf{W_{32}}$, where $i \in [0,7]$.
Then, each of these complex-valued product matrices is mapped to conductances in the manner shown in Fig. \ref{fig:sonos_dft}b.
These eight conductance target matrices are programmed onto eight 64$\times$128 sectors (65,536 total devices) of the SONOS array.
Fig. \ref{fig:twiddle_folding}b shows the measured conductances of the SONOS devices for each of these sectors after programming.
Since the first column of the twiddle matrix is unity, $\text{diag}(\mathbf{t}_0)\mathbf{W_{32}} = \mathbf{W_{32}}$, so the corresponding sector of conductances simply maps the DFT-32 matrix.
The remaining sectors are structurally similar to the DFT-32 matrix, but have some subtle variations due to multiplication by the twiddle factors prior to programming.

\begin{figure}[t]
\centering
\includegraphics[width=\textwidth]{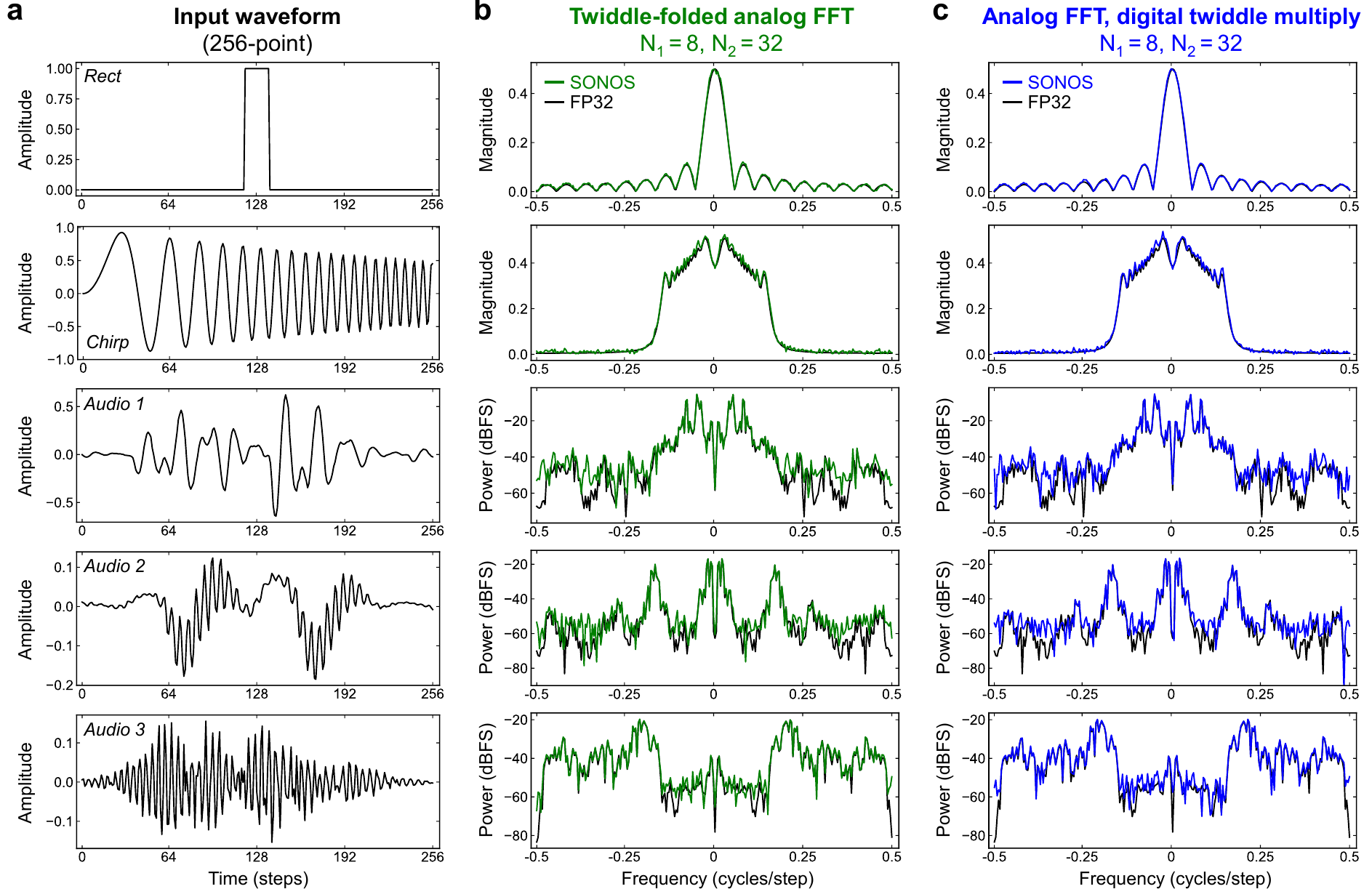}
\caption{Experimental results of twiddle-folded, 256-point analog FFT. (a) Five of forty exemplar 256-point 1D input waveforms used for the experiment. (b) SONOS-computed frequency spectra of the waveforms in (a) using twiddle-folded analog FFTs. (c) SONOS-computed frequency spectra from a control experiment using analog FFTs with unfolded, digital twiddle factor multiplication. For the first two waveforms, we plotted the magnitude of the spectra using linear scale, while for the three audio waveforms we plotted the signal power in logarithmic scale due to the large fluctuations in magnitude.
}
\label{fig:twiddle_folding_results}
\end{figure}

We used these programmed sectors of the SONOS array to compute the folded-twiddle analog FFT of a collection of forty 256-point, one-dimensional waveforms.
These waveforms included: (1) Four rectangular pulses with different pulse widths (9, 18, 27, and 36 samples); (2) Four linear chirp signals with zero initial frequency and different chirp rates ($3.125\times10^{-4}$, $6.25\times10^{-4}$, $9.375\times10^{-4}$, and $1.25\times10^{-3}$ cycles/step$^2$). Each chirp also has an exponentially decaying amplitude envelope which smoothly reduces the amplitude to a final value of 0.5; (3) Thirty-two 256-sample audio snippets from the full 65,536-sample speech audio waveform in Fig. \ref{fig:audio_spectrograms}a.
Five representative input waveforms are shown in Fig. \ref{fig:twiddle_folding_results}a, including one rectangular pulse, one chirp signal, and three audio waveforms.
We note that the inputs that were applied to the SONOS array were globally normalized, rather than separately normalized for each waveform.
This means some audio waveforms (covering fainter sound snippets) have a lower peak amplitude than others.

For the experiment, the 256-point analog FFT of these waveforms was computed by first running the first FFT stage with a DFT-8 matrix using analog MVMs on the SONOS array.
The remainder of the FFT was computed using two methods for comparison purposes, using the same intermediate results from the first stage: (1) using analog MVMs on the SONOS array with twiddle factors folded into the conductances, and (2) with digital twiddle factor multiplication, followed by analog MVMs with the DFT-32 matrix on the SONOS array.
The second experiment, which is similar in methodology to all the other experiments in this paper, is a control case.
The experimentally computed spectra for the five representative waveforms are shown in Fig. \ref{fig:twiddle_folding_results}b and \ref{fig:twiddle_folding_results}c for the two methods, respectively, alongside the ideal FP32 result.
In both cases, the SONOS-computed spectra agree closely with the FP32 results wherever the signal power is above a noise floor.
Similarly to the spectrum of the full audio waveform that was shown in Fig. \ref{fig:audio_spectrograms}d, this noise floor is between about $-50$ dB and $-40$ dB.

To more precisely assess and compare the quality of the 256-point analog FFTs, we quantify several metrics for the accuracy of the individual complex-valued spectral components:
\begin{gather}
\text{Complex normalized RMS error (NRMSE)} = \frac{\text{RMS}(|\mathbf{Y}_\text{SONOS} - \mathbf{Y}_\text{FP32}|)}{\text{max}(|\mathbf{Y_\text{FP32}}|)} \label{eq:metric1}\\
\text{Magnitude NRMSE} = \frac{\text{RMS}(|\mathbf{Y}_\text{SONOS}| - |\mathbf{Y}_\text{FP32}|)}{\text{max}(|\mathbf{Y_\text{FP32}}|)} \label{eq:metric2}\\
\text{Weighted phase mean absolute error (MAE)} = \frac{\sum_i |Y_{\text{FP32},i}|^2 \left| \angle Y_{\text{FP32},i} - \angle Y_{\text{SONOS},i} \right|}{\sum_i |Y_{\text{FP32},i}|^2} \label{eq:metric3}
\end{gather}
where $\mathbf{Y}$ are a set of complex dot products, $|\cdot|$ is the (element-wise) magnitude, $\angle(\cdot)$ is the phase, and RMS($\cdot$) is the root-mean-square operation. We use a power weighting (i.e. magnitude squared) in the phase MAE since phase error in components that have nearly zero power is inconsequential. We do not average the positive and negative frequency components in these metrics.

\def\arraystretch{1.3}
\begin{table}[t]
\centering
    \begin{tabularx}{1\textwidth}{|>{\hsize=1.8\hsize}X|>{\hsize=0.8\hsize}X|>{\hsize=0.8\hsize}X|>{\hsize=0.8\hsize}X|>{\hsize=0.8\hsize}X|}
        \hline
      & \textbf{PSNR} & \textbf{Complex NRMSE} & \textbf{Magnitude NRMSE} & \textbf{Weighted phase MAE} \\
        \hline
     \textbf{Twiddle-folded analog FFT} & 45.183 dB & $4.76\times10^{-3}$ & $5.51\times10^{-3}$ & 1.009$^\circ$    \\
        \hline
     \textbf{Analog FFT with digital twiddle multiplications} & 45.815 dB & $4.27\times10^{-3}$ & $5.12\times10^{-3}$ & 0.913$^\circ$ \\
        \hline
    \end{tabularx}
\caption{Comparison of accuracy metrics between the twiddle-folded analog FFT and the analog FFT with digital twiddle multiplications, evaluated on forty 256-point waveforms.
}
\label{tab:twiddle_folding_metrics}
\end{table}

Table \ref{tab:twiddle_folding_metrics} compares the results of the two analog FFT experiments along these accuracy metrics.
The two methods have nearly identical accuracy; the difference is small enough to be a result of random variations in SONOS programming error and noise.
This result experimentally demonstrates that the twiddle factors can be folded into the conductances to eliminate digital operations without affecting the accuracy of the analog FFT.

The mathematical reason that the twiddle folding does not affect the FFT's accuracy is as follows.
The twiddle factors -- like the DFT weights -- are by definition values that lie along the unit circle in the complex plane.
When the twiddle factors are multiplied by the DFT weights in the folding process, the weights are simply rotated to a different position along the unit circle without changing the magnitude.
The folding process therefore does not significantly change the distribution of programmed conductances, and thus the effective signal-to-noise ratio of the weights that are stored in SONOS devices remains the same.
This is shown in Fig. \ref{fig:twiddle_conductances} and discussed in Supplementary Section \ref{sec:dft_weights}.

%% file: _SI_platform.tex
\section{SONOS analog IMC chip and demonstration system}
\label{sec:sonos_demo}

\begin{figure}[h!]
\centering
\includegraphics[width=\textwidth]{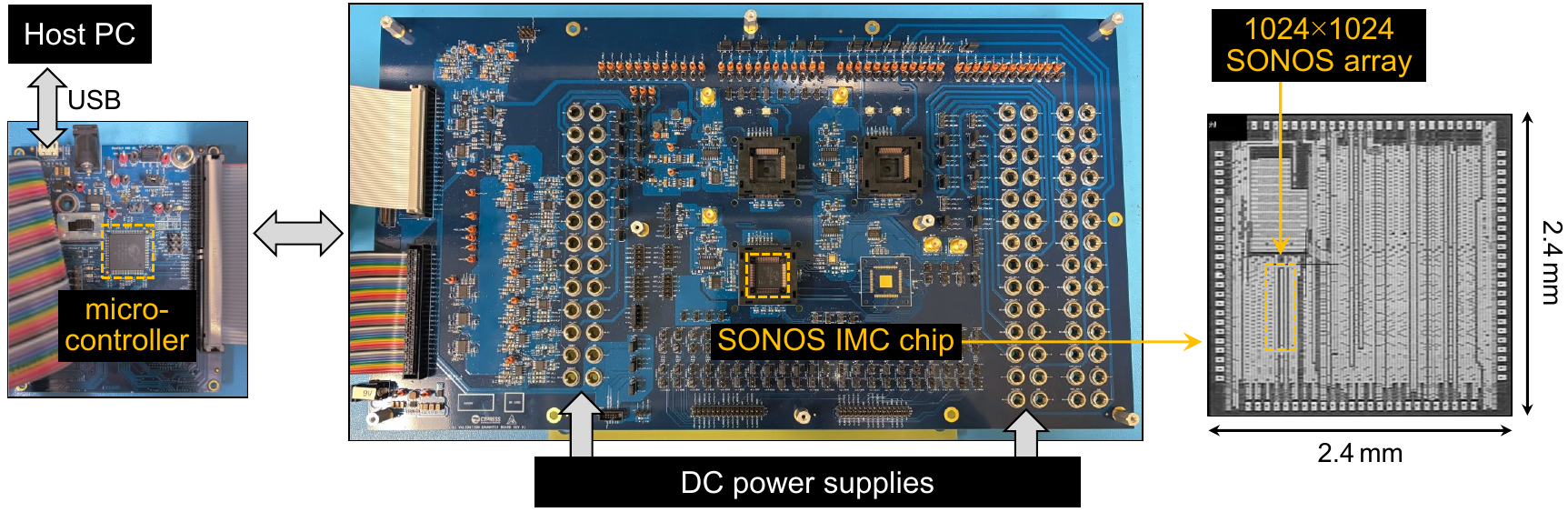}
\caption{Diagram of the hardware demonstration system used for the analog DFT and FFT experiments in this work. The packaged SONOS in-memory computing (IMC) chip was mounted on a 100-pin TQFP socket on a custom-designed board (center) that included components necessary to power the IMC chip. Only one of multiple sockets on the board was used for this experiment. Data and commands were sent to and from the IMC chip to an Infineon Technologies ARM microcontroller on a second custom board (left), connected to the center board through ribbon cables. The microcontroller communicated with the host PC through a USB interface. The boards were powered by four Keysight N6705C power supplies and a 9V power adapter, which were left disconnected in the photo. The right side shows a die photo of the IMC chip. The yellow rectangle shows the location of the $1024\times1024$ SONOS array, though the structure of the array is not visible due to metal over-layers.
}
\label{fig:demo}
\end{figure}

Fig. \ref{fig:demo} shows the hardware demonstration system used for the experiments in this work. Additional details are described in Methods. We note that this IMC chip was designed to demonstrate accurate analog MVMs using a large, 1024$\times$1024 SONOS array, but the chip was not optimized for power efficiency, speed, or area.

%% file: _SI_sonos.tex
\section{SONOS charge-trapping memory and analog state characterization}
\label{sec:gradient}

\begin{figure}[h]
\centering
\includegraphics[width=0.95\textwidth]{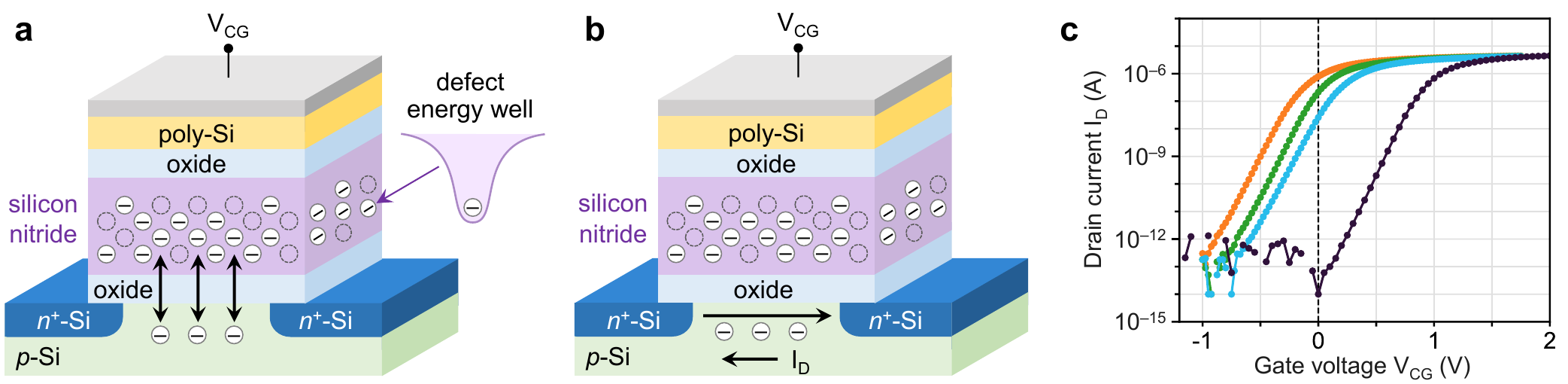}
\caption{Simplified diagram of a SONOS charge-trapping memory device. (a) During programming, charge is injected into electronic traps in the silicon nitride layer by tunneling. (b) During read or analog MVM, current flows between the source and drain (left and right) terminals, through a channel whose conductance is electrostatically modulated by the trapped charge through the field effect. (c) $I_{D}$-$V_{CG}$ transfer characteristics of a SONOS memory cell, which has four terminals as shown in Fig. \ref{fig:sonos_dft}a. The curves correspond to four non-volatile states which draw a current of 800 nA (orange), 200 nA (green), 25 nA (blue), and $<1$ pA (black) at $V_\text{CG} = 0$V. Here, the select gate is fixed at 2.5V, the bit line voltage is 0.1V, and the select transistor source is 0V. Note that this bias differs from the bit line voltage of 0.06V used for reads and analog MVMs in the rest of the paper.
}
\label{fig:sonos_device}
\end{figure}

This work used a non-volatile charge-trapping memory technology based on the SONOS (silicon-oxide-nitride-oxide-silicon) material stack, shown in Fig. \ref{fig:sonos_device}a. The state variable in this memory is the quantity of charge that is stored in a silicon nitride layer, which is intentionally rich with defects that act as electronic trapping sites. During programming, charge is added or removed from the trapping layer by Fowler-Nordheim tunneling of electrons or holes through the lower oxide, as shown in Fig. \ref{fig:sonos_device}a. The amount of trapped charge changes the threshold voltage of the underlying transistor channel, which is shown in Fig. \ref{fig:sonos_device}b. While the stored charge is not part of the conductive channel, it electrostatically modulates its conductance. The quantity of charge within the structure remains stable over time due to the double confinement provided by the energy wells of the electronic traps and the potential barriers at the nitride-oxide interfaces. The traps span a range of energies inside the nitride bandgap; ``deep" traps close to the midgap have larger energy barriers than ``shallow" traps close to the band edges.

Fig. \ref{fig:sonos_device}c shows the gate transfer characteristics of an individual SONOS cell, which also contains a select transistor as shown in Fig. \ref{fig:sonos_dft}a. The curves correspond to different programmed non-volatile states, showing multiple values of the threshold voltage. For this measurement, we used a semiconductor parameter analyzer to bypass the resolution limits of the on-chip ADC. As described in Methods, during read and MVM operations, a fixed voltage bias ($V_{SG} = 2.5$V, $V_{CG} = 0$V, and $V_{BL} = 0.06$V, relative to the select transistor source) is applied to the terminals, and the conductance of the SONOS device is defined under this operating bias. This condition is approximately represented by the dashed line in Fig. \ref{fig:sonos_device}c, with the difference that $V_{BL} = 0.1$V in this measurement. Nonetheless, this shows that a very large conductance dynamic range spanning about six orders of magnitude is accessible.

A SONOS device that is programmed to operate in the subthreshold regime can have extremely low conductance, down to $G < 1$ pS; this state is used to store zero-valued DFT weights. For the non-zero DFT weights, we still primarily operate the SONOS device in the subthreshold or weak inversion regime, where the silicon channel has not fully inverted but its conductance can nonetheless be tuned with high precision. For the higher conductance values used in this paper ($G > \, \sim$10 {\textmu}S), the devices operate in the strong inversion regime, where the channel has a high density of free electrons.

\begin{figure}[t]
\centering
\includegraphics[width=\textwidth]{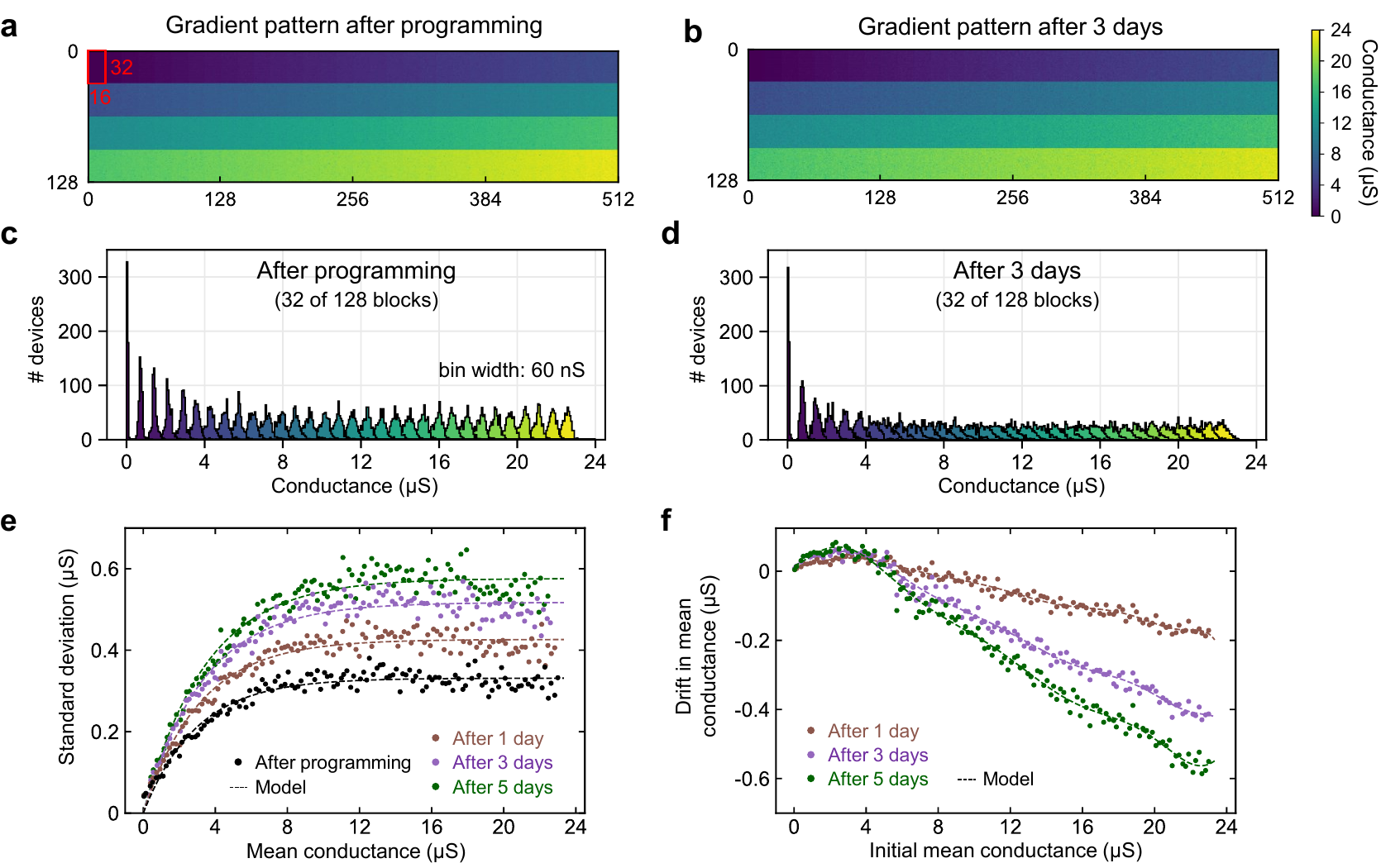}
\caption{(a) Spatial profile of measured SONOS conductances in a 128$\times$512 sub-array that was programmed with a gradient pattern to characterize the state dependence of conductance variability and drift. Measurement was taken just after programming. A single block (out of 128) with a uniform target conductance is labeled. (b) Conductances in the same sub-array measured three days after programming. (c)-(d) Histograms of measured SONOS conductances in 32 of 128 blocks (every fourth block) in (a) and (b), respectively. Each block contains 512 devices. (e) Standard deviation of the conductance in all 128 blocks after programming and after one, three, and five days. (f) Drift in the mean conductance of all 128 blocks after one, three, and five days, as a function of the initial mean conductance of each block. In (e) and (f), dashed curves are fits to the data.
}
\label{fig:gradient}
\end{figure}

Separately from the FFT experiments, we characterized the state-dependent conductance variability and drift characteristics of the SONOS analog memory cells. This was done by programming a 128$\times$512 portion of the array to a gradient pattern of conductance targets. The gradient pattern consisted of a 4$\times$32 tiling of 128 blocks, where each block is a 32$\times$16 rectangular group of SONOS cells. All 512 cells within a block were programmed to the same current target. Across the blocks, 128 linearly spaced current targets were selected from 0 nA to 1400 nA (0{~\textmu}S to 23.3 {~\textmu}S). Fig. \ref{fig:gradient}a shows the measured conductances of all 65,536 SONOS cells in the gradient pattern just after programming. The conductances are read out using the on-chip ADC, with a conductance resolution of 14.7 nS. Fig. \ref{fig:gradient}b shows a second measurement of the same cells taken three days after programming.

Fig. \ref{fig:gradient}c shows the distribution of SONOS conductances for 32 of the 128 blocks, measured just after programming. Because the target conductance is uniform within a block, the ideal width of each distribution is zero. The actual non-zero width of the distributions is due to random conductance variability in the SONOS cells caused by device-to-device process variation, write noise, and read noise. This variability directly causes random errors in the values of DFT weights used for analog computation. Notably, the distributions asymptotically approach a width of zero in the limit of zero conductance. This is a unique property of the SONOS memory device when operated in the subthreshold conductance regime. In this regime, random variability in the quantity of stored charge translates linearly to a variability in $V_T$ but exponentially to a variability in the channel conductance. When the target conductance is very low, a large random variability in $V_T$ may induce a large \textit{relative} variability in the conductance, but a very small \textit{absolute} variability due to the small conductance. It can be shown that as a result, the subthreshold conductance variability is proportional to the conductance \cite{xiao2022accurate}. We quantify the width of each block's conductance distribution by its standard deviation, and this is plotted in Fig. \ref{fig:gradient}e by the black points for all 128 blocks. The proportionality between conductance and variability at low conductance can be clearly seen. At larger conductances, the width of the distribution eventually saturates to the conductance tolerance used in the write-verify algorithm.

Fig. \ref{fig:gradient}d shows the same distributions measured after three days. During this time, stored charge in the nitride layer can escape thermally from the electronic traps, and subsequently migrate within the nitride layer or leave the nitride via trap-assisted tunneling through the oxides. The drift dynamics can be complex due to the presence of both trapped electrons and trapped holes, and the fact that a defect's influence on the channel conductance depends on its physical location within the nitride layer. The programming algorithm is designed to minimize de-trapping by selectively storing charge in deep traps near the middle of the nitride bandgap, but a small number of carriers can nonetheless de-trap over this time period. The above effects induce a change in the channel's threshold voltage and conductance. Since these effects are thermal in origin and hence stochastic, the variability in conductance tends to increase over time. This causes the conductance distributions in Fig. \ref{fig:gradient}d to be wider in general than the ones in Fig. \ref{fig:gradient}c. This can also be observed directly by comparing the black and purple points in Fig. \ref{fig:gradient}e.

We also observed that the conductance drift over time had some systematic state dependence. Fig. \ref{fig:gradient}f shows how the mean conductance of each of the 128 blocks shifted over the first day, over the first three days, and over the first five days after programming. Overall, the conductances tend to decrease with time indicating a net loss of holes or net gain of electrons with time. The state with the minimum conductance is the most stable with time for the same reason that it has the highest conductance precision; drift in $V_T$ for deep subthreshold states induce a very small absolute change in conductance. The rate of conductance drift slows down with time as the loosely bound stored charge in the shallow traps of the silicon nitride are gradually depleted.

For accuracy simulations, the time-dependent conductance variability and the mean drift of conductance over time were modeled using analytical fits to the data in Fig. \ref{fig:gradient}e and \ref{fig:gradient}f, respectively. The variability was fit using a saturating exponential function of $G$: $\sigma(G,t) = A(t)\times\left(1 - e^{-G/B(t)}\right)$. The mean drift at each measured time point was fit using a tenth degree polynomial function of $G$. These analytical fits are shown by the dashed curves in Fig. \ref{fig:gradient}e and \ref{fig:gradient}f.

For the audio and image processing experiments in this paper, drift was an ongoing effect during the computation, i.e. later parts of the computation experienced more errors due to drift than the earlier parts. For simulation purposes, we modeled either one day or three days of drift, depending on which was closer to the actual time after programming of these experiments.

%% file: _SI_dft_weights.tex
\section{Precision and distribution of SONOS DFT weights}
\label{sec:dft_weights}

\begin{figure}[h]
\centering
\includegraphics[width=\textwidth]{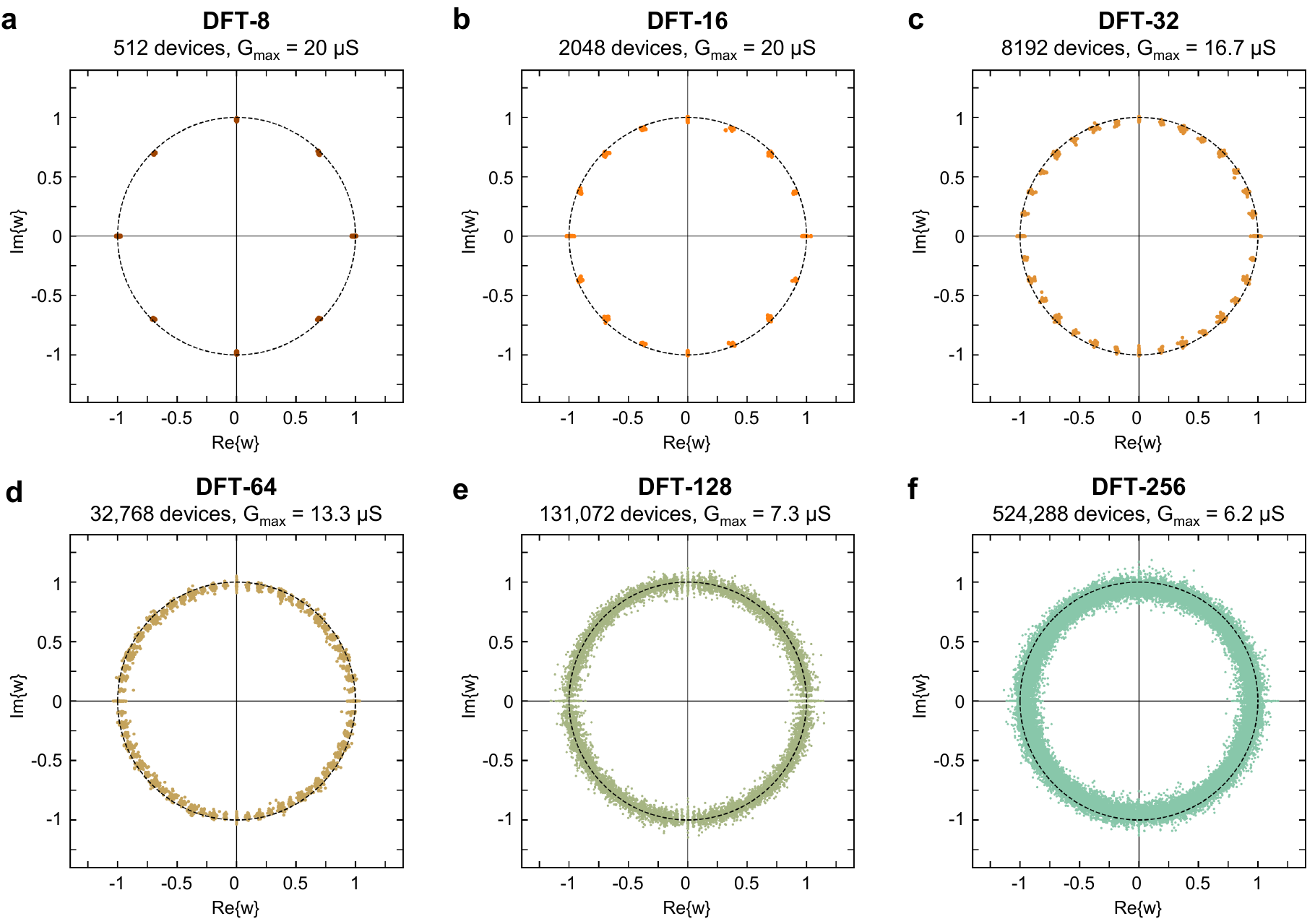}
\caption{Constellation of complex-valued DFT weight values stored in SONOS devices, for several sizes of the programmed DFT matrix. Weights are measured just after programming. The ideal DFT weights lie along the unit circle (dashed).
}
\label{fig:dft_circles}
\end{figure}

\begin{figure}[t]
\centering
\includegraphics[width=0.84\textwidth]{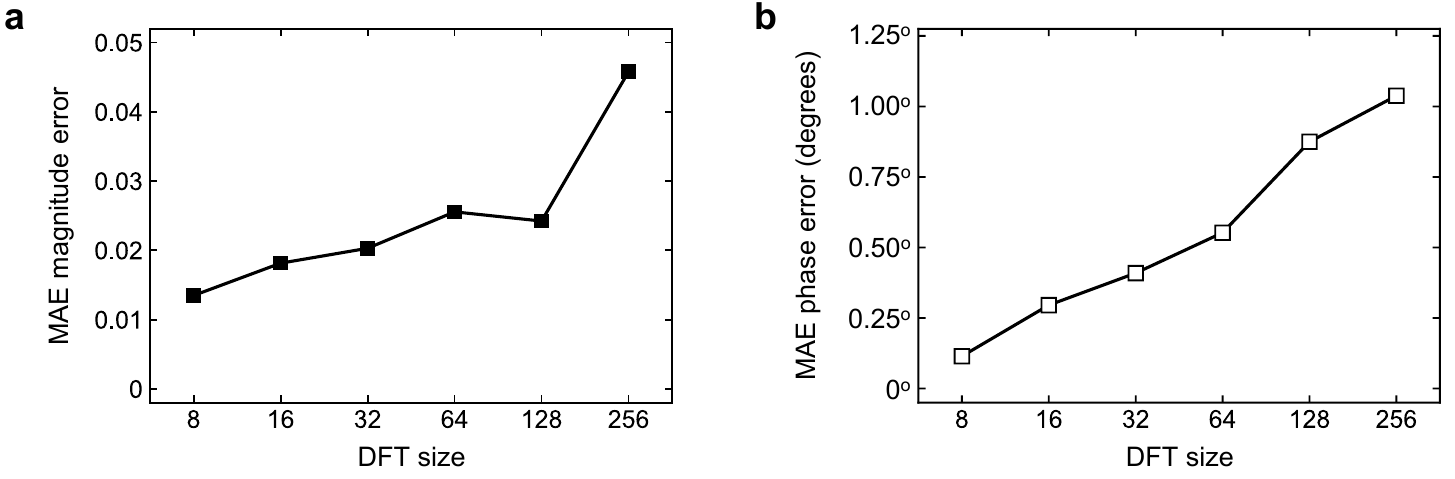}
\caption{Mean absolute error in the (a) magnitude and (b) phase of DFT weights stored in the SONOS array vs DFT size, based on the same SONOS weights in Fig. \ref{fig:dft_circles}.
}
\label{fig:dft_weight_err}
\end{figure}

Fig. \ref{fig:dft_circles} shows the constellation in the complex plane of the DFT weights that are stored physically in programmed SONOS devices for various DFT sizes. For all matrices, the ideal weights have the form $\exp(-i2\pi nk/N)$ where $n$ and $k$ are indices into the 2D matrix. These complex exponentials all have a magnitude of 1, so the ideal weights all lie along the unit circle in the complex plane. The weights stored in the SONOS devices are generally close to their target locations along the circle. Errors in the radial direction correspond to errors in magnitude, while errors along the circumferential direction correspond to errors in the phase of the complex-valued weights. 
Fig. \ref{fig:dft_weight_err} shows the mean absolute error (MAE) of both the magnitude (red) and phase (green) of the DFT weights for the different DFT sizes.

Since the DFT arrays were used for analog DFT computation in addition to weight storage, we selected the maximum SONOS conductance $G_\text{max}$ to ensure optimally accurate analog MVMs. Too large a value for $G_\text{max}$ induces errors due to ADC saturation and parasitic $IR$ drops. Meanwhile, a small $G_\text{max}$ reduces the precision of the DFT weights, because the conductance variability, noise, and drift become a larger proportion of the utilized conductance range. The chosen values of $G_\text{max}$ in Fig. \ref{fig:dft_circles} were optimized for the distribution of summed column currents for the speech processing experiments. For processing the 2D spatial images, which have a much larger zero-frequency component, we selected different values for $G_\text{max}$ which are shown in Table \ref{tab:2d_dft_sonos}. For larger DFT sizes, which require larger arrays, the maximum conductance was reduced to keep the summed currents low. At the level of individual DFT weights, because the range of SONOS conductances was reduced with increasing DFT size, both the magnitude and phase errors increase with DFT size, as shown in Fig. \ref{fig:dft_weight_err}.

\begin{figure}[t]
\centering
\includegraphics[width=\textwidth]{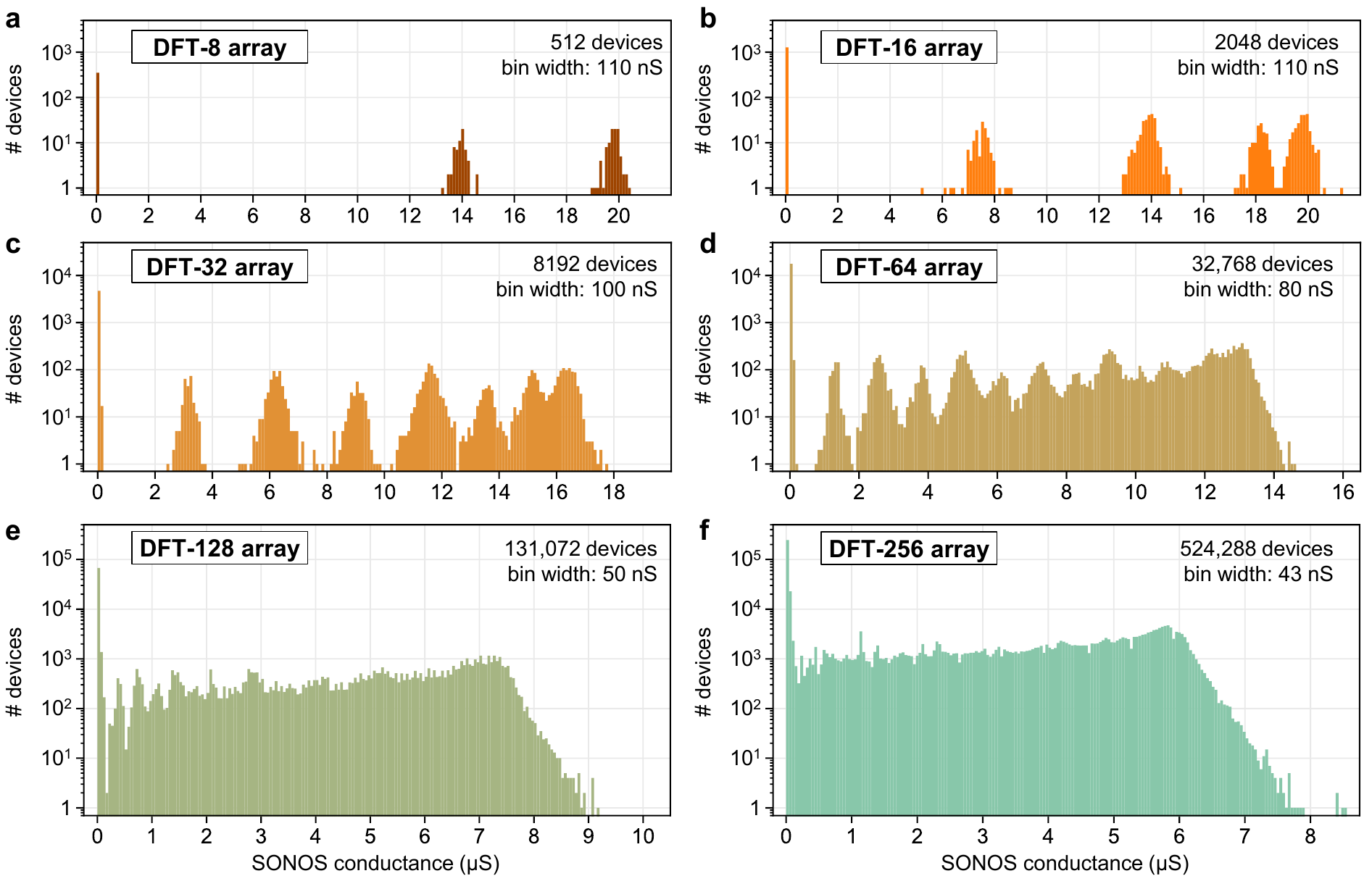}
\caption{Distribution of the measured SONOS conductances corresponding to the programmed DFT matrices in Fig. \ref{fig:dft_circles}.
}
\label{fig:dft_conductance_dist}
\end{figure}

Fig. \ref{fig:dft_conductance_dist} shows the distributions of the individual SONOS conductances that implement the DFT matrices in Fig. \ref{fig:dft_circles}. We note that the target conductances for an IDFT matrix would have exactly the same distribution as the target conductances for a DFT matrix of the same size. As the DFT size increases, more values are sampled along the unit circle in the complex plane, causing a gradual transition from a discrete to a nearly continuous conductance distribution. Each distribution contains a large peak at the minimum conductance, because our scheme for mapping signed weights leads to at least half the SONOS devices being programmed to nearly zero conductance.

There is a highly consequential difference between the conductance distributions of DFT matrices and the conductance distributions of deep neural network (DNN) weight matrices. Across many different DNN models, it has been observed that the weight value distributions tend to be heavily peaked around zero, with the number of weights decaying exponentially with increasing magnitude. When using our differential mapping scheme, this leads to conductance distributions that also decay exponentially from the minimum conductance level \cite{OnTheAccuracy, xiao2022accurate}. This is not the case for DFT weights. Because the complex-valued DFT weights are distributed uniformly around the unit circle, the conductance distribution is closer to uniform than exponential, with a much higher average conductance compared to DNNs. This observation leads to an important difference in the error tolerance of the accuracy of analog FFTs and analog DNN inference. Analog DNN inference benefits enormously from having minimal error at the low end of the conductance range, while the error at the high conductance end is less consequential \cite{OnTheAccuracy}. Meanwhile, for analog FFTs, it is important to keep the conductance errors low across the entire utilized dynamic range of the device.

\begin{figure}[t]
\centering
\includegraphics[width=0.85\textwidth]{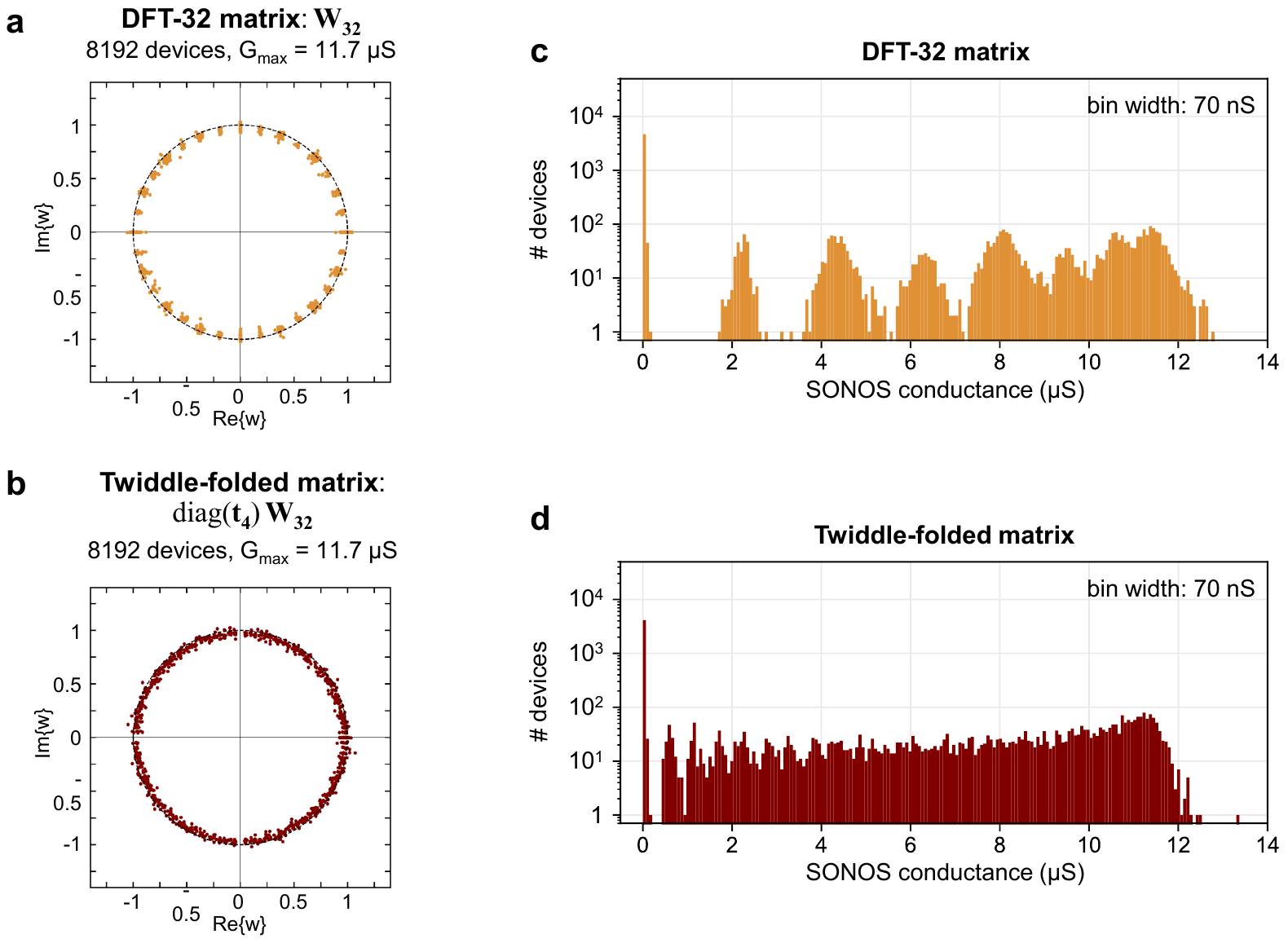}
\caption{(a-b) Constellation of the complex-valued weights for a DFT-32 matrix and one of the twiddle-folded DFT-32 matrices used for the experiment shown in Fig. \ref{fig:twiddle_folding}. (c-d) Corresponding distributions of the measured SONOS conductances for the two matrices. Note that a lower $G_\text{max}$ was used here compared to the DFT-32 matrix in Fig. \ref{fig:dft_circles}-\ref{fig:dft_conductance_dist}.
}
\label{fig:twiddle_conductances}
\end{figure}

We next discuss how the process of folding the FFT twiddle factors into the analog MVMs of the FFT second stage (described in Supplementary Section \ref{sec:parallel_fft}) changes the programmed conductance distributions.
Fig. \ref{fig:twiddle_conductances}a and \ref{fig:twiddle_conductances}b show the complex-plane constellations of the unfolded DFT-32 matrix and one of the eight twiddle-folded DFT-32 matrices, respectively, for the programmed SONOS devices that were used in the experiment in Fig. \ref{fig:twiddle_folding}b.
Since both the DFT weights and the twiddle factors are both complex exponentials that lie along the unit circle in the complex plane, their product also lies along the unit circle, with a phase that is equal to the sum of the phase of the two values.
Therefore, the twiddle folding process simply rotates the weights along the unit circle.
Since each row of the DFT-32 matrix is element-wise multiplied by a different twiddle factor, the rotation amount varies among the different rows.
This causes the constellation to smear out over the unit circle, as observed in the measurement results in Fig. \ref{fig:twiddle_conductances}b.
The effect of the folding on the programmed conductance distributions is similar.
Since the weights are now more uniformly distributed along the unit circle, the conductance distribution becomes more uniform: this can be seen by comparing Fig. \ref{fig:twiddle_conductances}c and \ref{fig:twiddle_conductances}d.
The same effect was observed in the other twiddle-folded matrices in Fig. \ref{fig:twiddle_folding}b.

%% file: _SI_device_requirements.tex
\section{Device requirements for the analog FFT}
\label{sec:device_requirements}

\begin{figure}[b]
\centering
\includegraphics[width=0.625\textwidth]{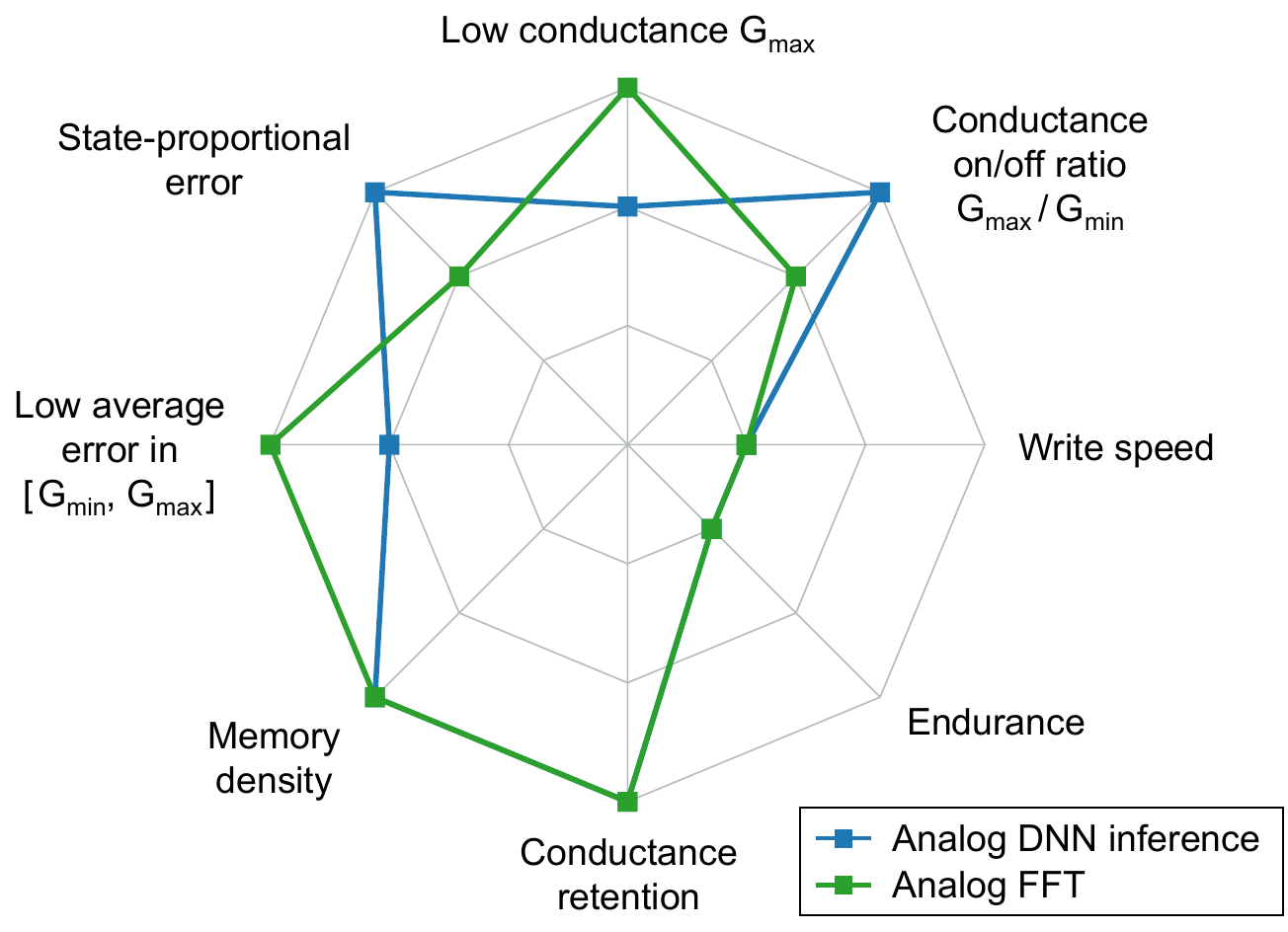}
\caption{Comparison of the relative importance of different memory device metrics for analog DNN inference vs. analog FFTs. Points that are further from the center indicate greater relative importance.
}
\label{fig:radar_plot}
\end{figure}

In Fig. \ref{fig:radar_plot}, we qualitatively compare the desired properties of the non-volatile memory device for analog FFTs to the desired properties for analog DNN inference. For both applications, the desired properties are those that enable higher accuracy, energy efficiency, and compute throughput per area. The two applications have similar requirements for several metrics, but differ in others.

The differences in the device requirements between the analog FFT and analog DNN inference stem from the statistical difference in the distribution of DFT matrix elements and that of DNN weights, as described in Supplementary Section \ref{sec:dft_weights}. Complex-valued DFT weights are uniformly distributed around the unit circle, and these are mapped to conductances that are approximately uniformly distributed over the utilized conductance range from $G_\text{min}$ to $G_\text{max}$. By contrast, DNN weights almost universally tend to heavily concentrate near zero, and map to a conductance distribution that peaks at $G_\text{min}$ and decays roughly exponentially with increasing conductance \cite{OnTheAccuracy, xiao2022accurate}.

For DNN inference, because low conductances near $G_\text{min}$ are far more abundant than high conductances, it is very important to ensure that the states near $G_\text{min}$ have low conductance errors. By comparison, the states near $G_\text{max}$ can tolerate higher errors. Inference therefore favors device technologies whose conductance error is state-proportional, i.e. the error $\sigma_G$ increases proportionally with the target conductance $G$ \cite{OnTheAccuracy}. SONOS memory has state-proportional errors below $\sim$4{~\textmu}S, as shown in Fig. \ref{fig:gradient}e. By contrast, for the DFT matrix, because all conductance values within the utilized range are used roughly equally, it is important for the conductance error to be low for the entire range from $G_\text{min}$ to $G_\text{max}$; conductances near $G_\text{min}$ are not more important than conductances near $G_\text{max}$. In other words, the analog FFT does not benefit strongly from a memory device with state-proportional error.

The energy efficiency of analog MVMs that is attainable with a given memory technology is largely determined by its ability to scale to large array sizes, since the energy consumption of the MVM tends to be dominated by peripheral circuits. A primary determinant of the maximum array size is the impact of parasitic $IR$ drops, which induce errors in the analog dot products that grow with the array dimensions, and eventually causes a total loss of accuracy for very large arrays. For a given metal interconnect technology, the effect of parasitic $IR$ drops can be reduced by decreasing the average memory device conductance within the array, which reduces the currents. How the average conductance can best be reduced depends on the distribution of conductances in the array, and therefore depends on the application. For DNN inference, since conductances near $G_\text{min}$ are abundant, the average conductance depends much more on the value of $G_\text{min}$ than on $G_\text{max}$. Therefore, memory devices with a high conductance on/off ratio ($G_\text{max}/G_\text{min}$) are extremely desirable for DNN inference. This includes SONOS flash, and other transistor-based memory such as floating-gate flash and ferroelectric FETs.

By contrast, for the analog FFT, the uniform conductance distribution of DFT matrices means that reducing the average conductance can only be achieved by reducing $G_\text{max}$, which reduces the overall utilized conductance range. Since the average utilized conductance (as a fraction of $G_\text{max}$) is much higher for the DFT matrix than for a DNN layer, parasitic $IR$ drops pose a more stringent limitation on array size scalability for the DFT compared to DNN inference. Therefore, the FFT application benefits more strongly from devices that can operate accurately at low conductance, but high conductance on/off ratio is not critical. We note that while the analog FFT can be recursively applied to factor any large transform into small MVMs, a factorization into large analog MVMs is more energy-efficient (see Supplementary Section \ref{sec:factorizations}).

Both analog FFTs and DNN inference benefit from high memory density to enable high compute density (performance per area) with analog in-memory MVMs. The memory density metric not only includes the physical size of the device, but also the layout of the full memory cell, and the amount of information that can be stored in a single cell. A memory that has truly analog states with fine conductance programmability in a small area is denser than a memory device that has a small footprint but intrinsically discrete states (such as standard magnetic tunnel junctions). A small memory cell also improves array size scalability, since it reduces the cell's incremental contribution to the parasitic interconnect resistances.

For both analog FFTs and DNN inference, the desired use case is that the matrix is programmed onto the array once, then used repeatedly for many analog MVMs before needing to be programmed again. This use case is possible since the FFT is a fixed transform, and most DNN layers have fixed, pre-trained weights during inference. The array only needs to be re-programmed when the desired matrix changes, which should be very infrequent by design, or when the device conductances have drifted significantly from their initially programmed values and need to be periodically refreshed. To minimize the accuracy loss due to drift and the overhead associated with frequent refresh, it is essential for the memory device to retain its programmed conductance (with multiple bits of precision) for as long as possible. Also, since programming operations on the array are infrequent in this use model, cycling endurance and write speed are not particularly important metrics.

%% file: _SI_reconfigurability.tex
\section{Dynamic reconfigurability of the analog FFT}
\label{sec:reconfigurability}

As described in Section \ref{sec:cooleytukey}, an analog FFT system can be dynamically re-configured to compute FFTs of smaller sizes without re-programming any memory devices. This cheap reconfigurability is possible by exploiting a property of the DFT matrix. If $N_1$ is a multiple of $N_2$, i.e. $N_2 = N_1 / (a \times b)$ where $a$ and $b$ are positive integers, then it can be shown that:
\begin{equation}
\left(\omega_{N_2}\right)^{nk} = e^{-i2\pi nk/N_2} = e^{-i2\pi (na\times kb)/N_1} = \left(\omega_{N_1}\right)^{na\times kb}
\end{equation}
Therefore, an $N_2$-point DFT can be computed using a memory array that implements $\mathbf{W}_{N_1}$ simply by applying the inputs to every $a^\text{th}$ row and measuring the outputs from every $b^\text{th}$ column.

\begin{figure}[h]
\centering
\includegraphics[width=0.85\textwidth]{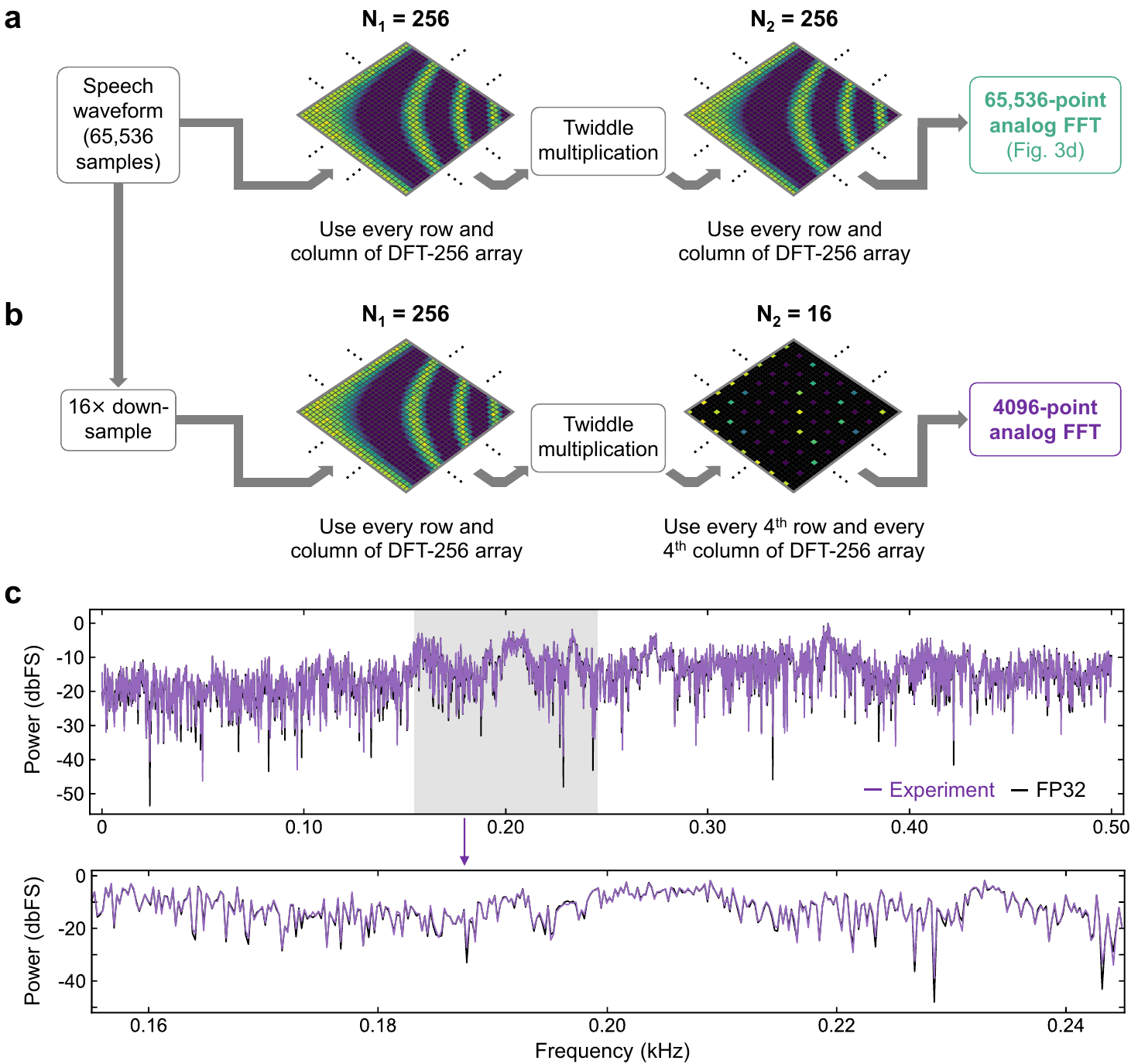}
\caption{(a) Fully utilizing a single programmed DFT-256 SONOS array to compute a 65,536-point FFT. The input waveform and frequency spectrum are shown in Fig. \ref{fig:audio_spectrograms}a and \ref{fig:audio_spectrograms}d, respectively. (b) A 16-point DFT can be computed by selectively activating the rows and columns of the same programmed DFT-256 array. Images show a $32\times 32$ portion of the DFT-256 array; unused devices are colored black. (c) SONOS-computed power spectrum of the 4096-point down-sampled audio signal (top), and a zoom into one portion of the spectrum (bottom).
}
\label{fig:reconfigurability}
\end{figure}

We demonstrate this capability by re-using the SONOS sub-array that was programmed with the DFT-256 matrix. This sub-array was initially used to compute the 65,536 analog FFT in Fig. \ref{fig:audio_spectrograms}d, using a Cooley-Tukey decomposition with $N_1 = N_2 = 256$. This configuration, which fully utilizes the programmed array, is shown in Fig. \ref{fig:reconfigurability}a. Next, without reprogramming any SONOS devices, we computed a 4096-point analog FFT on a 16$\times$ down-sampled version of the same audio signal, as shown in Fig. \ref{fig:reconfigurability}b. The first DFT stage used $N_1 = 256$, as before. The second DFT stage used $N_2 = 16$, where the 16-point analog DFTs were computed using the already programmed array by activating every fourth row and every fourth column. The power spectrum that is computed using the 4096-point analog FFT is shown in Fig. \ref{fig:reconfigurability}c. The SONOS-computed spectrum is in close agreement with the result of an FP32 digital FFT that was computed on the downsampled signal, with a PSNR of 36.19 dB.

This reconfigurability can be extended to the case where the technique in Supplementary Section \ref{sec:parallel_fft} is used to fold the twiddle factors into the DFT matrices before programming the resistive elements. In this case, the arrays in the first DFT stage would contain the normal DFT matrices that can be reconfigured as described above, but the second DFT stage would use multiple arrays that are each programmed with an element-wise product of a DFT matrix and a different column of a twiddle matrix.

Let us consider the case where we would like to compute two different FFT sizes: $F_1 = N_1 \times N_2$ and $F_2 = N_3 \times N_4$. 
We assume that $N_1$ is a multiple of $N_3$ and $N_2$ is a multiple of $N_4$: $N_1 = AN_3$ and $N_2 = BN_4$, where $A$ and $B$ are integers that are greater than or equal to 1. 
For the first stage, since $N_1 \geq N_3$, the arrays would be programmed with the $N_1$-point DFT matrix. Since $N_1$ is an integer multiple of $N_3$, the rows and columns of these arrays can be selectively activated as described above to compute the $N_3$-point DFTs.
Therefore, the same arrays can be re-used for the first stage of both the $F_1$-point and $F_2$-point FFTs.

Since $N_2 \geq N_4$, the arrays in the second stage would be programmed with the matrices $\left[ \text{diag}(\mathbf{t_j}) \, \mathbf{W}_{N_2} \right]$, where $j$ ranges from 0 to $N_1-1$. Here, $\mathbf{t_j}$ is one column of the twiddle matrix for an FFT of size $F_1$, so the elements of this column are: $\mathbf{t}_{\mathbf{j},k} = (\omega_{N_1 N_2})^{jk}$. Then, a single element at the index $(n,k)$ of this ``folded'' matrix represents the value:

\begin{align}
\left[ \text{diag}(\mathbf{t_j}) \, \mathbf{W_{N_2}} \right]_{n,k} &= (\omega_{N_1 N_2})^{jk} (\omega_{N_2})^{nk} \nonumber \\
&= \exp \left[  -i2\pi \left( \frac{jk}{N_1 N_2} + \frac{nk}{N_2} \right) \right] \label{eq:fold1}
\end{align}

Now, suppose we had instead programmed the arrays to execute the second stage of the $F_2$-point DFT. In this case, these arrays would have been programmed with a different set of folded matrices, $\left[\text{diag}(\mathbf{t'_{j'}}) \, \mathbf{W}_{N_4} \right]$. Here, $\mathbf{t'_{j'}}$ is one column of the twiddle matrix for an $F_2$-point FFT. A single element at the index $(n', k')$ of this different folded matrix represents the value:
\begin{align}
\left[ \text{diag}(\mathbf{t'_{j'}}) \, \mathbf{W_{N_4}} \right]_{n',k'} &= (\omega_{N_3 N_4})^{j'k'} (\omega_{N_4})^{n'k'} \nonumber \\
&= \exp \left[  -i2\pi \left( \frac{j'k'}{N_3 N_4} + \frac{n'k'}{N_4} \right) \right] \label{eq:fold2}
\end{align}

The question now is whether the value in Equation \ref{eq:fold2} can be accessed using the proper choice of indices in Equation \ref{eq:fold1}.
To find this, we can equate the arguments of the two exponentials:

\begin{equation}
\frac{jk}{N_1 N_2} + \frac{nk}{N_2} = \frac{j'k'}{N_3 N_4} + \frac{n'k'}{N_4}
\end{equation}

We make the following substitutions: $k = ak'$, $n = bn'$, and $j = cj'$, where $a$, $b$, and $c$ are integers. Then, by equating the two terms in each sum separately, the equation above can be satisfied with the following conditions on $a$, $b$, and $c$:
\begin{align}
a \times c &= A \times B \\
a \times b &= B
\end{align}

This has at least one integer solution: $a = B, b = 1, c = A$. If $A$ or $B$ can be factored, then there may be other solutions. In either case, this shows that by selectively activating the rows and columns of the properly selected arrays that were programmed with the folded matrices in Equation \ref{eq:fold1}, we can compute an MVM with the folded matrices in Equation \ref{eq:fold2}. Therefore, the parallel analog FFT technique in Fig. \ref{fig:parallel_fft} can still be re-configured to compute FFTs of different sizes.

%% file: _SI_factorizations.tex
\section{The effect of varying analog FFT factorizations}
\label{sec:factorizations}

\subsection{Efficient analog FFT factorizations}

\begin{figure}[h]
\centering
\includegraphics[width=0.9\textwidth]{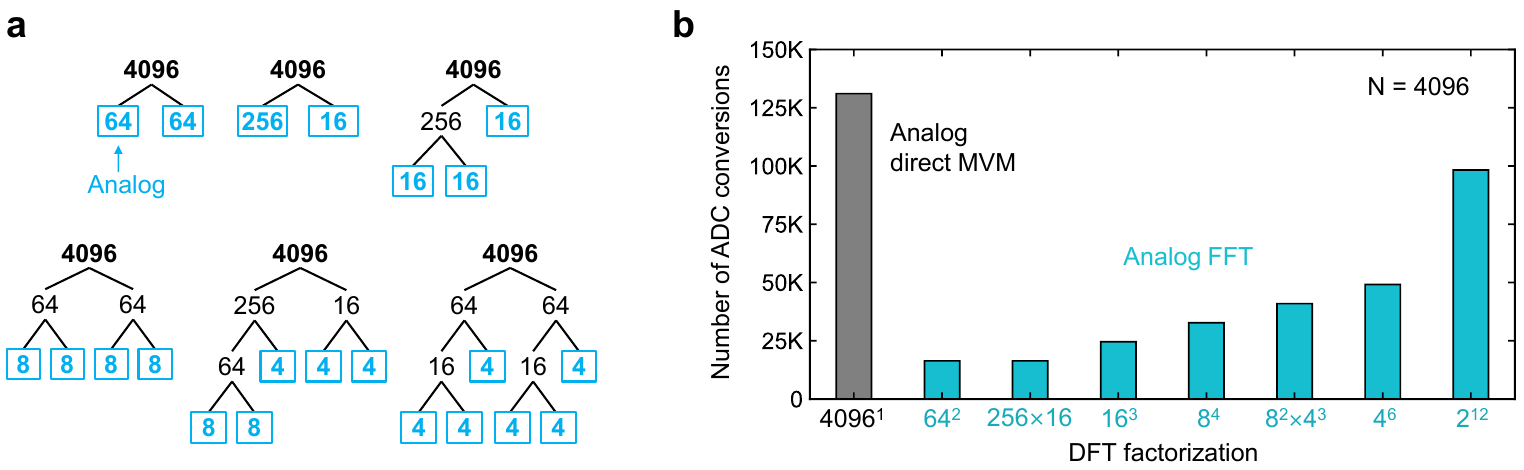}
\caption{(a) Several possible factorizations of a 4096-point DFT using two-factor Cooley-Tukey decompositions. Blue indicates the elementary analog DFT operations. (b) The effect of analog FFT factorization on the number of ADC conversions for a 4096-point DFT, assuming a maximum array size of 512$\times$1024.
}
\label{fig:factorizations}
\end{figure}

There are many possible ways to factor an $N$-point DFT using two-factor Cooley-Tukey decompositions. Several examples are shown in Fig. \ref{fig:factorizations}a for $N = 4096$, where the decomposition is applied once or multiple times to break up the computation into smaller elementary DFTs. Fig. \ref{fig:factorizations}b shows how the total number of ADC conversions depends on the factorization. While recursively applying Cooley-Tukey decompositions reduces the total number of arithmetic operations that are performed inside the analog MVMs, it also generates more intermediate results that must be converted to digital values. Therefore, as noted in the main text, for analog systems the optimal choice for energy efficiency is to terminate the decomposition as soon as the original DFT can be factored into DFTs that are small enough to fit into one memory array. For $N=4096$ and an array size of 512$\times$1024, it is optimal to apply the Cooley-Tukey decomposition exactly once, e.g. with $N_1 = N_2 = 64$.

By contrast, in digital systems, it is generally optimal to decompose the DFT into the smallest possible radices -- usually radix-2 or radix-4 -- since the energy expenditure generally scales with the total number of arithmetic operations.
The analogous case in Fig. \ref{fig:factorizations} is the $4096 = 2^{12}$ factorization, which decomposes the computation into 2-point analog DFTs, i.e. an analog MVM implementation of the basic 2$\times$2 butterfly operation.
This is an inefficient operating point for analog MVMs, since the matrix is too small to effectively amortize the energy cost of the peripheral circuits and ADCs.

We note that the processing of radix-2 decimation-in-time FFTs has previously been proposed inside NAND flash memory arrays \cite{Zhang2022}, but this method was based on analog element-wise vector multiplication rather than analog MVM operations, so it is fundamentally different from our approach.
Yet, like the case described above, this scheme involves at least one ADC operation for every intermediate result of a radix-2 FFT, which incurs a high energy cost.

\subsection{Audio processing accuracy with varying FFT factorizations}

The analog FFT factorization can also influence the accuracy of the FFT computation by affecting the size of the analog MVM operations.
To study this for FFTs with a single Cooley-Tukey decomposition, we experimentally computed the spectrograms of eight audio clips from Google’s Speech Commands Dataset \cite{warden2018speech}. Most of these audio clips have 16,384 samples, for a duration of 1.024 seconds (16 kHz sampling rate). 
For each clip, the spectrograms were computed using STFTs with a window size of 256 samples and no window function.
The STFTs were computed using the SONOS analog IMC array with different factorizations of the DFT, including the 256-point analog direct MVM.

\begin{figure}[h]
\centering
\includegraphics[width=\textwidth]{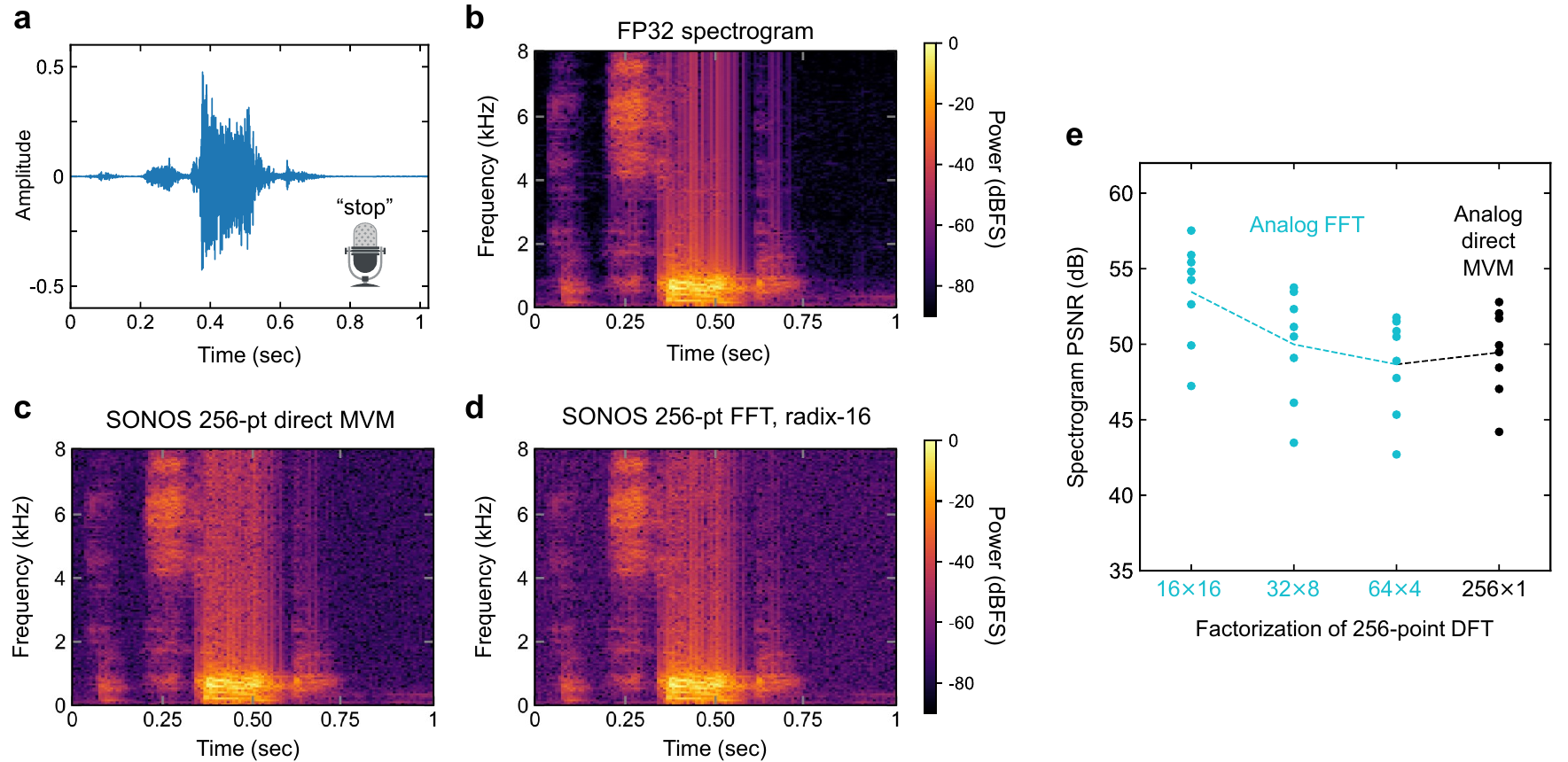}
\caption{(a) Audio waveform with 16,384 samples containing the spoken command ``stop.'' (b) Spectogram of the audio waveform generated by FP32 FFTs, using rectangular windows with a size of 256 samples and a hop length of 128 samples. (c) Same spectrogram computed experimentally using 256-point analog DFTs implemented using direct MVMs on the SONOS array. (d) Same spectrogram computed using 256-point analog FFTs, factored into 16-point analog DFTs that are executed on the SONOS array. (e) PSNR of the SONOS-computed spectrogram relative to the FP32 spectrogram, for several DFT factorizations. For each factorization, eight spectrograms were generated for eight distinct command words.
}
\label{fig:command_spectrograms}
\end{figure}

Fig. \ref{fig:command_spectrograms}a shows one of the waveforms, containing the word ``stop''. Fig. \ref{fig:command_spectrograms}b shows the ideal spectrogram as computed using digital 256-point FFTs at FP32 precision. Fig. \ref{fig:command_spectrograms}c shows the spectrogram computed experimentally using 256-point DFTs with analog direct MVMs on the SONOS array, and Fig. \ref{fig:command_spectrograms}d shows the spectrogram computed using 256-point analog FFTs with $N_1 = N_2 = 16$. Both of the SONOS-computed spectrograms reproduce all of the key features of the FP32 spectrogram for the spoken word ``stop'', including the initial high-frequency ``s'' sound and the louder, lower-frequency ``top'' sound.

We also computed spectrograms for seven other voice commands and two other factorizations of the 256-point DFT: ($N_1 = 32, N_2 = 8$) and ($N_1 = 64, N_2 = 4$). Fig. \ref{fig:command_spectrograms}e summarizes how the FFT factorization affects the PSNR metric of the SONOS-computed spectrograms relative to the FP32 spectrogram. In general, as the size of the elementary analog DFT (the larger of the two factored DFTs) increases, there is a gradual reduction in the accuracy of the computation. This is because the larger analog DFTs accumulate a greater amount of analog error from SONOS device conductance variations and noise, as well as accumulated parasitic $IR$ drops along the array interconnects.

The trend in Fig. \ref{fig:command_spectrograms}e shows that the choice of factorization or radix for the analog FFT brings a minor trade-off between accuracy and efficiency. At the small DFT size of 256, the most energy-efficient choice is the direct MVM because the full 256-point DFT can be computed by our SONOS array without having to split the DFT matrix. The radix-16 FFT (i.e. $N_1 = N_2 = 16)$ is somewhat more accurate due to the smaller MVM size, but not by a large margin. The optimal choice of DFT factorization is the one that maximizes energy efficiency while meeting the accuracy needs of the end application.

\subsection{Results of non-power-of-2 analog FFTs}

\begin{figure}[t]
\centering
\includegraphics[width=0.925\textwidth]{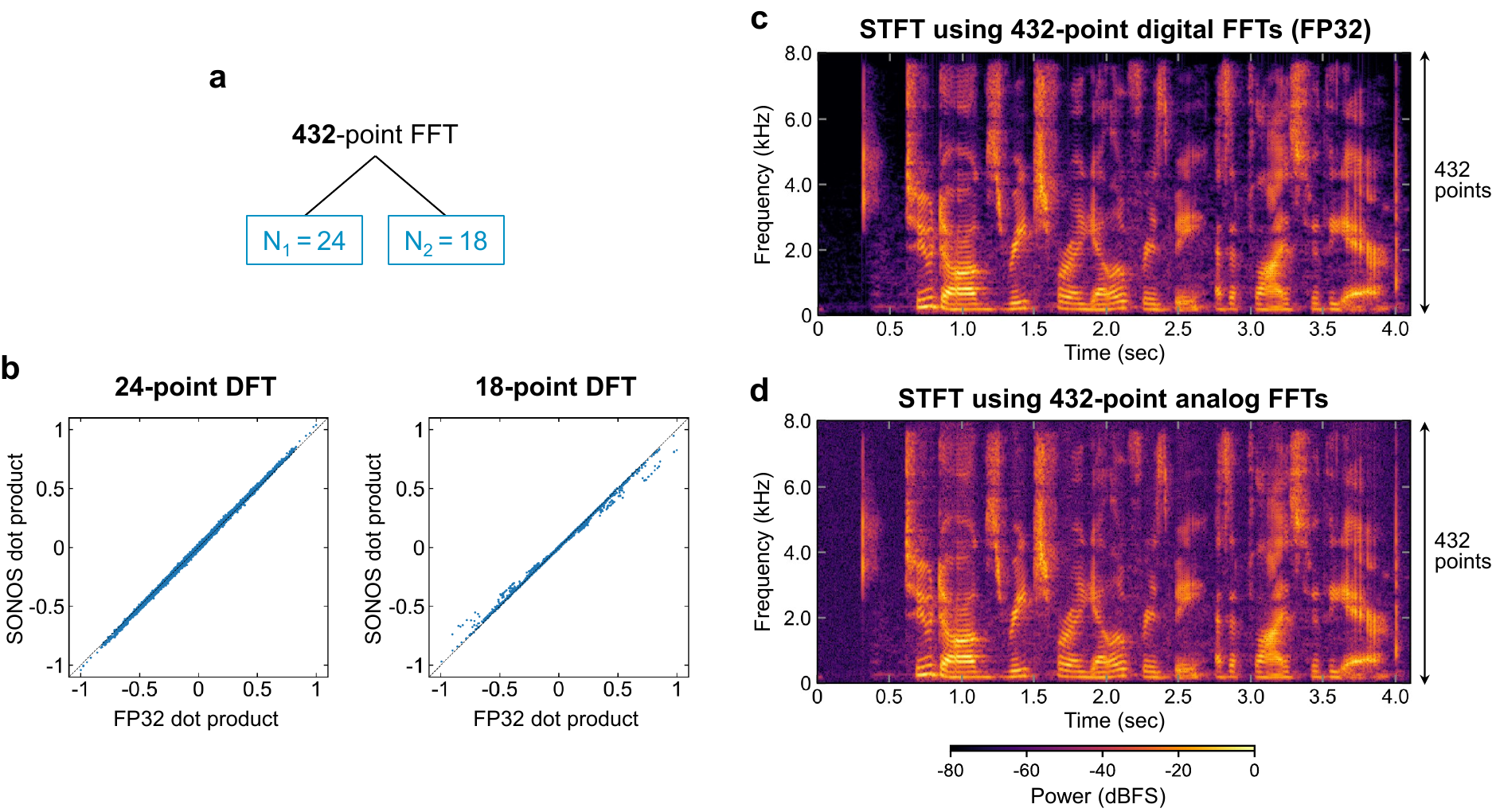}
\caption{(a) Factorization of a 432-point 1D FFT. (b) Correlation of FP32 and experimentally computed analog dot products using the SONOS array, for the two DFT stages of the FFT in (a). The real and imaginary dot product components are plotted separately. Each plot contains $5.21 \times 10^{5}$ dot products, computed in the process of generating the spectrogram in (d). (c) Spectrogram of the audio waveform in Fig. \ref{fig:audio_spectrograms}a, with a window size of 432 samples, a hop length of 108 samples, and a Hamming window function. The frequency resolution is 37.0 Hz. (d) Spectrogram generated experimentally using STFTs that are executed on the SONOS array with the FFT factorization in (a).
}
\label{fig:non_power_of_2}
\end{figure}

In Fig. \ref{fig:non_power_of_2}, we experimentally demonstrate a non-power-of-2 analog FFT that is factored into DFT stages with unequal radices.
We decompose a 432-point FFT into two DFT stages, with $N_1 = 24$ and $N_2 = 18$.
This FFT is used to compute the spectrogram of the same speech audio waveform in Fig. \ref{fig:audio_spectrograms}a, but this time using windows containing 432 audio samples.
The ideal FP32 spectrogram is shown in Fig. \ref{fig:non_power_of_2}c, while the spectrogram that is computed experimentally using the SONOS array is shown in Fig. \ref{fig:non_power_of_2}d.
This spectrogram has a PSNR of 53.78 dB, which is similar to the PSNR of the spectrogram in Fig. \ref{fig:audio_spectrograms}c that was computed using 512-point analog FFTs with $N_1 = 32$ and $N_2 = 16$.

The more fine-grained accuracy of the analog DFTs in each stage is shown in the dot product correlation plots in Fig. \ref{fig:non_power_of_2}b.
Table \ref{tab:non_power_of_2} compares the complex-valued dot product accuracy metrics for the different DFT stages that are used to process the same audio waveform.
These metrics are defined in Equations \ref{eq:metric1}-\ref{eq:metric3}.
There is no meaningful difference in error sensitivity between FFTs or DFT stages that have power-of-2 lengths and non-power-of-2 lengths.
In general, accuracy decreases with the size of the DFT stage, due to the greater amount of analog error accumulation in the MVM operation.
There are some small variations on this overall trend in the experimental results, due to random differences in SONOS device programming error or noise, and because the different DFT stages can have different data distributions in their MVM inputs.

\def\arraystretch{1.3}
\begin{table}[t]
\centering
    \begin{tabularx}{0.9\textwidth}{|>{\hsize=1.4\hsize}X|>{\hsize=0.9\hsize}X|>{\hsize=0.9\hsize}X|>{\hsize=0.9\hsize}X|>{\hsize=0.9\hsize}X|}
        \hline
     \textbf{FFT size} & \multicolumn{2}{c|}{$N = 512$}  & \multicolumn{2}{c|}{$N = 432$}    \\
        \hline
     \textbf{DFT stage size} & $N_1 = 32$ & $N_2 = 16$ & $N_1 = 24$ & $N_2 = 18$    \\
        \hline
     \textbf{Complex NRMSE} & $4.29\times10^{-3}$ & $1.91\times10^{-3}$ & $6.21\times10^{-3}$ & $2.48\times10^{-3}$ \\
        \hline
     \textbf{Magnitude NRMSE} & $3.21\times10^{-3}$ & $1.72\times10^{-3}$ & $5.14\times10^{-3}$ & $2.26\times10^{-3}$ \\
        \hline
     \textbf{Weighted phase MAE} & 0.904$^\circ$ & 0.581$^\circ$ & 1.003$^\circ$ & 0.667$^\circ$ \\
        \hline
    \end{tabularx}
\caption{Accuracy metrics for the experimentally computed SONOS dot products, comparing the stages of the 512-point FFT and the 432-point FFT. In both cases, the FFTs were used to compute the spectrum of the audio waveform in Fig. \ref{fig:audio_spectrograms}a.
}
\label{tab:non_power_of_2}
\end{table}

\subsection{Mapping prime-length FFTs}

Large prime-length Fourier transforms, which are much less commonly used than composite-length transforms, are more complicated to compute exactly since they cannot be factored.
Small to moderate-size prime DFTs can be computed efficiently using a direct analog MVM, if the array is large enough to support this operation; for our 1024$\times$1024 SONOS array, this mode can be used to compute prime-length DFTs up to $N = 251$, whose matrix maps to a 502$\times$1004 sub-array.
If an exact result is not required by the application, slightly zero-padding the prime-length signal to a nearby composite length may be acceptable.
Otherwise, computing long prime FFTs on analog hardware requires additional steps, involving Rader's algorithm or Bluestein's algorithm, described below.
These are also the most commonly used methods for computing prime-length FFTs using digital FFT implementations.

Rader's algorithm converts an $N$-point prime FFT into an $(N-1)$-point circular convolution, where $N-1$ is a composite number \cite{Rader68}.
To implement it, the zero-frequency component must be separately computed by summing the input signal; this is a small digital computation.
The rest of the spectrum is computed by taking the last $N-1$ points of the input signal, re-ordering this sequence, then computing a circular convolution between this sequence and an $(N-1)$-point kernel that depends only on the value of $N$.
Finally, the result is re-ordered and the first element of the input is digitally added to every element. 
Using the convolution theorem, the circular convolution can be computed using: an $(N-1)$-point analog FFT, an element-wise multiplication by the spectrum of the kernel, and an $(N-1)$-point analog IFFT.
As we showed experimentally in Fig. \ref{fig:vector_radix}, large analog FFTs and IFFTs can be accurately cascaded.
For a fixed value of $N$, the FFT of the convolution kernel can be pre-computed and stored.
Furthermore, since the kernel is fixed, the element-wise multiplications can be folded into the conductance matrices of a DFT (or IDFT) stage in either the FFT or IFFT, using the same scheme for folding the twiddle factor multiplications (see Supplementary Section \ref{sec:parallel_fft}).
In the best case, the energy, latency, and memory overhead of computing the $N$-point prime FFT is expected to be slightly more than 2$\times$ the overhead of a similarly sized composite-length FFT, due to the need for an $(N-1)$-point FFT, an $(N-1)$-point IFFT, and a small number of digital additions.

Some prime-length FFTs require more overhead than others.
The more expensive primes are ones where the only nontrivial factorization of the number $N-1$ involves large primes that cannot be directly computed using an analog MVM on a single resistive array.
For such cases, Bluestein's algorithm \cite{Bluestein70} is a more efficient alternative to recursively applying Rader's algorithm \cite{Bluestein70}. 
Bluestein's algorithm computes an $N$-point DFT using the following steps: (1) element-wise multiplication of the input by a fixed chirp signal, (2) linear convolution of the result with a fixed kernel, (3) extraction of the first $N$ points of the convolution result, and (4) element-wise multiplication of the result by a fixed chirp signal.
The linear convolution can be computed with a ($2N-1$)-point analog FFT, element-wise multiplication by the spectrum of the kernel, and a ($2N-1$)-point analog IFFT.
As with Rader's algorithm, the element-wise kernel multiplication within the convolution can be folded into the conductance matrices of either the last DFT stage or the first IDFT stage.
Additionally, because all of the above steps are linear operations and the chirp signals are also fixed for a given $N$, the element-wise multiplication before the convolution can be folded into the first DFT stage, and the element-wise multiplication after the convolution can be folded into the last IDFT stage.

Using a ($2N-1$)-point analog FFT and IFFT, the overhead of Bluestein's algorithm for computing the $N$-point prime FFT is slightly more than $4\times$ the overhead of a similarly sized composite-length FFT.
If the number $2N–1$ also has large prime factors, then the FFT length within the convolution can be increased to a number with more favorable factors by zero padding the input sequence to the convolution.
This slightly increases the overhead of the prime-length FFT but avoids an inefficient recursive application of the algorithm.

We note that the non-monotonic relationship between FFT size and its energy and latency overhead, induced by prime number considerations, applies to both digital and analog FFT implementations.
However, the analog implementation can compute the intermediate FFTs and element-wise multiplications within the convolutions much more efficiently.

\subsection{PSNR of audio spectra without positive/negative frequency averaging}

For a 1D signal $\mathbf{x}$, the computed FFT $\mathbf{X}$ contains the spectral coefficients for both positive and negative frequency components: $X_k$ and $X_{-k}$.
When the signal is purely real, as in our audio signal processing experiments, the negative frequency components are redundant because they obey Hermitian symmetry: $X_k = X^*_{-k}$.
Therefore, it is standard for audio spectra to show only the positive frequencies, as we have done in Fig. \ref{fig:audio_spectrograms}, even though the analog FFT computes both $X_k$ and $X_{-k}$.
Since analog computation can introduce errors, the SONOS-computed analog FFT result does not perfectly satisfy Hermitian symmetry.
As mentioned in the Methods, the magnitude spectra in Fig. \ref{fig:audio_spectrograms}c and \ref{fig:audio_spectrograms}d were computed from the full complex-valued analog FFT result by averaging the positive and negative frequency components: $|\frac{1}{2}\left(X_k + X^*_{-k}\right)|$.
The PSNR metrics are computed on these magnitude spectra.

Here, for completeness, we report the PSNR values without averaging the positive and negative frequencies.
For the audio spectrogram in Fig. \ref{fig:audio_spectrograms}c, lack of frequency averaging changes the PSNR from 55.97 dB to 54.98 dB.
For the audio spectrum in Fig. \ref{fig:audio_spectrograms}d, the PSNR changes from 41.01 dB to 40.36 dB.
For the audio spectrogram in Fig. \ref{fig:non_power_of_2}d, the PSNR changes from 53.78 dB to 52.49 dB.
For the audio spectrum in Fig. \ref{fig:reconfigurability}c, the PSNR changes from 36.19 dB to 33.94 dB.
Without averaging, the PSNR is reduced slightly because it additionally penalizes deviations from Hermitian symmetry between two terms that contribute to the same frequency bin in the magnitude spectrum.

%% file: _SI_dot_products.tex
\section{Analog dot product accuracy}
\label{sec:mvm_errors}

\begin{figure}[h]
\centering
\includegraphics[width=0.8\textwidth]{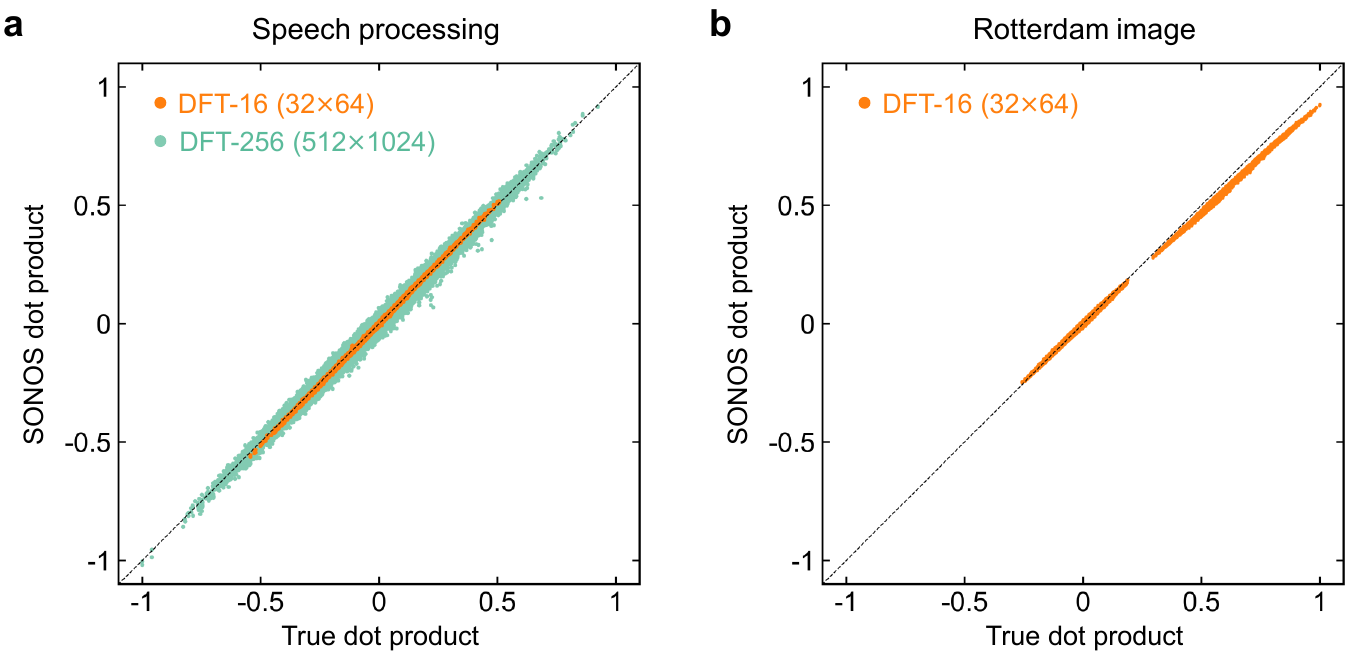}
\caption{Correlation between the true and SONOS-computed dot products using the analog IMC test chip, (a) for the first stage of the speech processing experiments, and (b) for the first stage of the 2D image processing experiment on the ``Rotterdam'' image. In (a), the dot product values have been normalized by $1.75\times 10^5$, the maximum true dot product value for the 256-point DFTs. In (b), the values have been normalized by $2.25\times 10^3$, the maximum true dot product value for the 16-point DFTs. The dimensions in parentheses are the size of the SONOS sub-array utilized for the computation.
}
\label{fig:mvm_corr}
\end{figure}

Fig. \ref{fig:mvm_corr} shows the correlation between the ideal and SONOS-computed dot products from the analog IMC test chip, for different input signals and elementary DFT sizes.

The dot products in Fig. \ref{fig:mvm_corr}a were obtained from processing the speech waveform in Fig. \ref{fig:audio_spectrograms}a; the DFT-16 operations were used for spectrogram generation and the DFT-256 operations were used to compute the one-shot 65,536-point FFT. These dot products were computed from the same summed currents shown in Fig. \ref{fig:sonos_dft}f, after subtracting the contributions from positive and negative inputs and weights, and accumulating the result over input bits. Interestingly, while parasitic IR drops introduced systematic errors into the current sums in Fig. \ref{fig:sonos_dft}f, the dot products do not show any significant systematic errors. This is because the errors induced by IR drops cancel to a large extent when the digitized current sums for the positive and negative components are subtracted. This cancellation was faciliated by the fact that the audio input signal had a mean value close to zero. We note that while this cancellation appears on average, the $IR$ drops are not fully canceled within the individual dot products so their effects are still non-zero. Also, this cancellation effect may be weaker for systems that have much larger parasitic metal resistance.

The dot products in Fig. \ref{fig:mvm_corr}b were obtained from the 2D image processing experiment on the ``Rotterdam'' image in Fig. \ref{fig:vector_radix}b. The dot products are from the first stage of the VR-FFT, which used 16-point DFTs. Compared to the dot products in Fig. \ref{fig:mvm_corr}a, the effects of parasitic IR drops are more visible: they manifest as a systematic negative error that grows with the value of the dot product. The greater effect of parasitic IR drops in this example comes from the fact that the pixel values in the input image are strictly positive with a significant positive mean value. This positive bias means that there is less cancellation of errors due to parasitic IR drops compared to the speech processing example. This is consistent with the error profile in the 2D frequency spectrum shown in Fig. \ref{fig:vector_radix}c, where the large low-frequency components have a systematic negative error. This positive bias is somewhat specific to this particular image processing example: in practical sensor DSP applications, we do not expect such a large bias to be present. This includes both the SAR image formation example in Fig. \ref{fig:scaling}a and the automatic speech recognition example in Fig. \ref{fig:scaling}b (which would have a similar error profile to Fig. \ref{fig:mvm_corr}a). 

The SONOS-computed dot products show an overall close agreement with the ideal values. For the 16-point DFTs in Fig. \ref{fig:mvm_corr}a, the RMS error in the dot product is 0.35\% of the maximum true dot product value in this experiment. For the 16-point DFTs in Fig. \ref{fig:mvm_corr}b, the corresponding normalized RMS dot product error is 0.81\%. The larger error in the latter case is due to the lack of cancellation of errors due to parasitic IR drops, described above. For the 256-point DFTs in Fig. \ref{fig:mvm_corr}a, the normalized RMS dot product error is 1.60\%. As discussed in the main text, the 256-point DFT has a larger error due to the greater amount of error accumulation in the analog sum and the smaller signal-to-noise ratio in the programmed conductances.

%% file: _SI_parseval.tex
\section{Image correction using Parseval's theorem}
\label{sec:parseval}

\begin{figure}[h]
\centering
\includegraphics[width=0.96\textwidth]{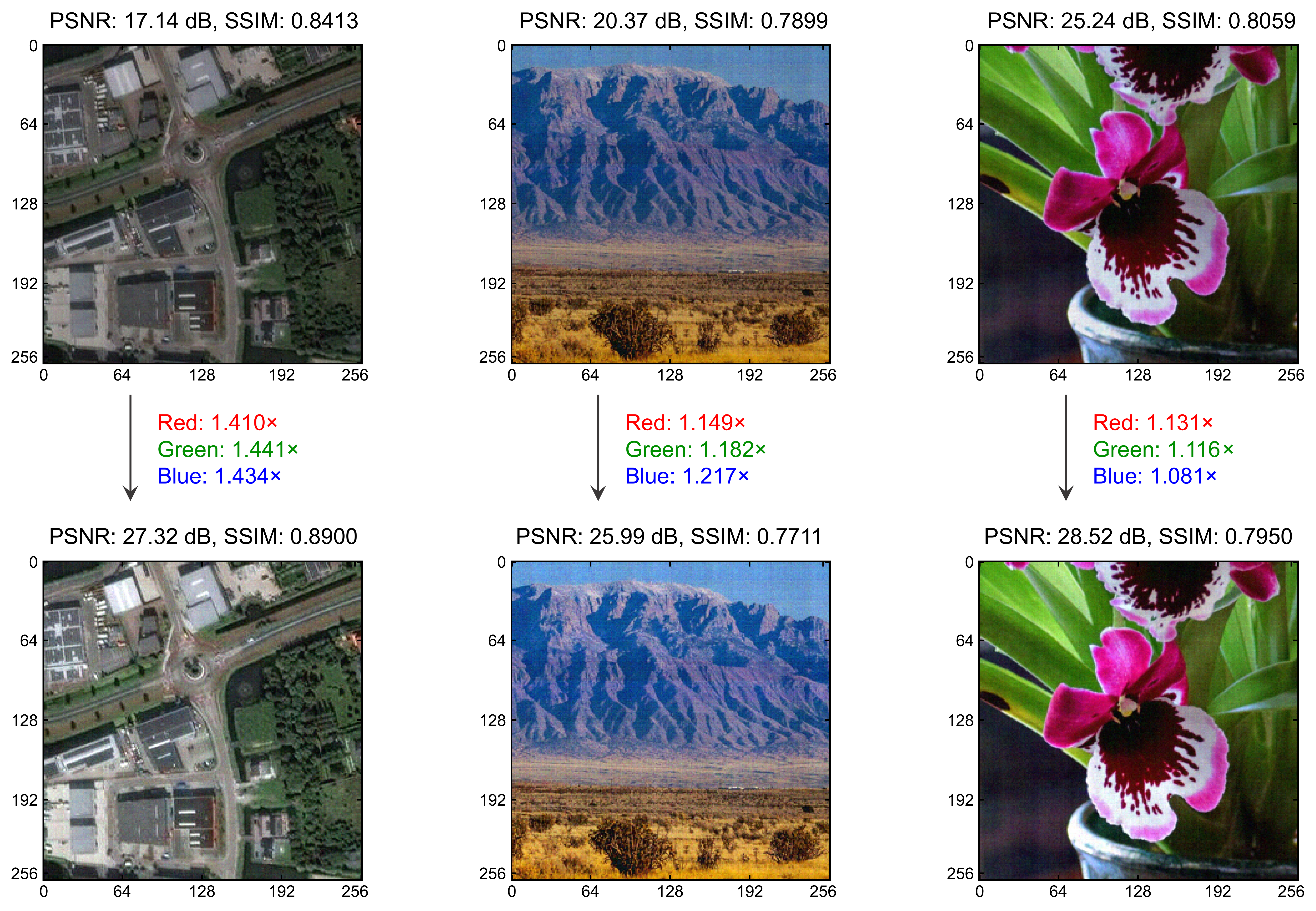}
\caption{(Top row) Uncorrected images reconstructed from 2D spectra computed by the SONOS array using analog VR-FFTs. The images were reconstructed from the spectra using digital 2D IFFTs. (Bottom row) Reconstructed images after the channel-wise scalar correction.
}
\label{fig:parseval}
\end{figure}

Parseval's theorem states that the total energy of the signal is preserved through an ideal DFT \cite{oppenheim1997signals}. For an $N \times N$ 2D DFT, this is expressed as:
\begin{equation}
\label{eq:parseval}
\frac{1}{N^2} \sum_j \sum_k |X_{jk}|^2 = \sum_j \sum_k |x_{jk}|^2
\end{equation}
For an analog FFT, this does not exactly hold due to errors induced by the physical hardware. The loss of stored charge from the SONOS device and power dissipation in the array parasitic resistances cause a systematic loss of signal energy during the computation. For the image reconstruction examples in Fig. \ref{fig:vector_radix}, this signal loss has the effect of reducing the brightness of the reconstructed image. We apply a first-order correction for this effect that takes advantage of Parseval's theorem. The sums on the right and left side of Equation \eqref{eq:parseval} can be evaluated digitally before and after the analog Fourier transform computation, respectively. Then, we brighten the analog reconstructed image $\mathbf{x}_\text{R}$ by uniformly multiplying every pixel by a scalar that depends only on the ratio of the two sums:
\begin{equation}
\label{eq:parseval_correction}
\mathbf{x}_\text{R, corrected} = \mathbf{x}_\text{R} \times \sqrt{\frac{\sum_j \sum_k |x_{jk}|^2}{\sum_j \sum_k |X_{jk}|^2 / N^2}}
\end{equation}
As shown by the results in Fig. \ref{fig:parseval} for three 256$\times$256 images, we find that this brightening improves the reconstructed image PSNR, though it does not always improve the SSIM. This is because a constant multiplication does not actually change any of the spatial features that were present in the uncorrected image. Therefore, the correction is unlikely to be necessary for applications that are not specifically optimizing for metrics like PSNR.

%% file: _SI_2dfft.tex
\section{Additional results and details on 2D image processing}
\label{sec:fft_dft_comparison}

\subsection{Comparison of the analog VR-FFT and direct MVM method}

\begin{figure}[b]
\centering
\includegraphics[width=\textwidth]{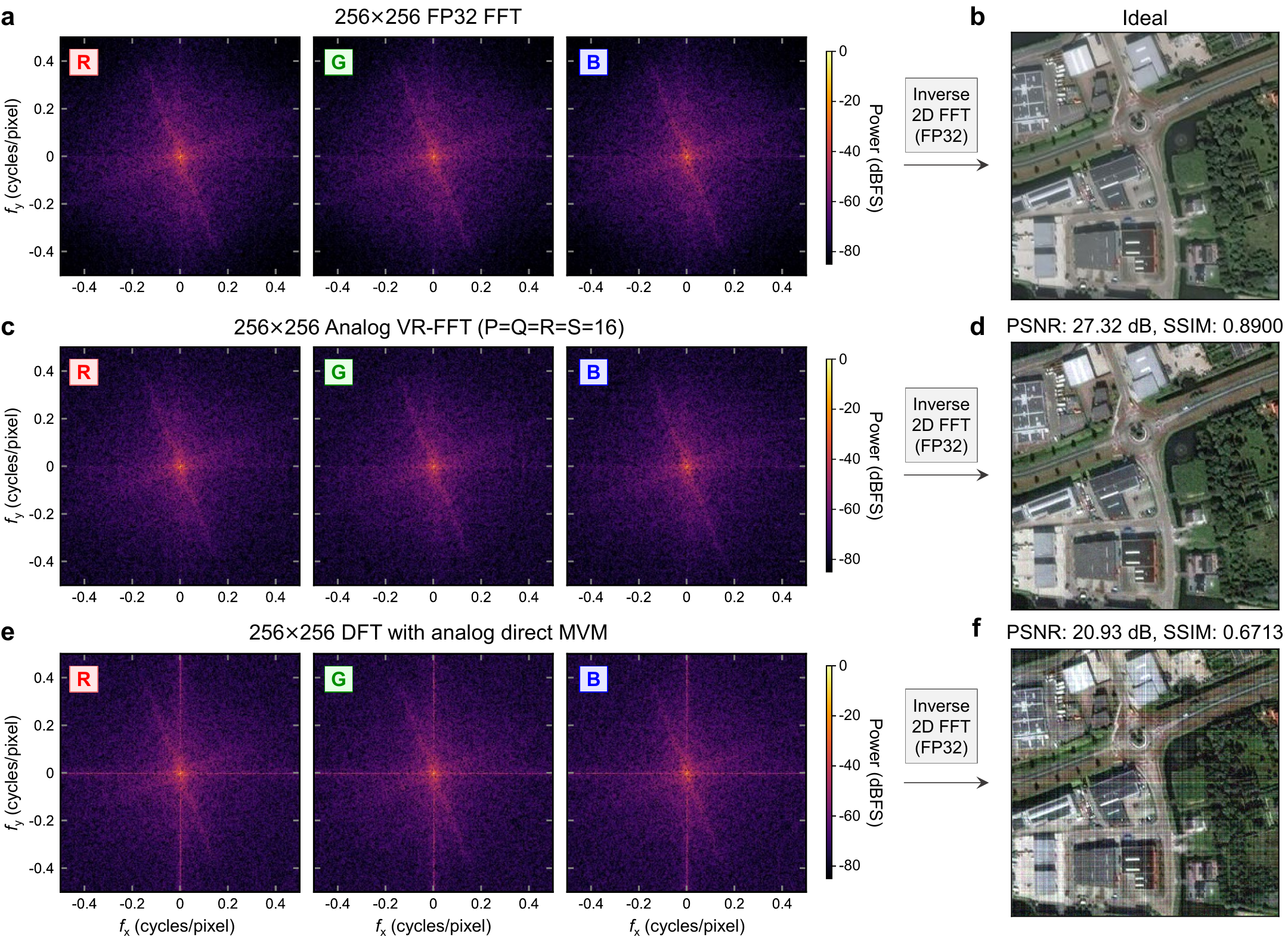}
\caption{(a) Ideal 2D power spectrum of the 256$\times$256 ``Rotterdam'' image computed using an FP32 FFT. (b) The ideal reconstructed image, which is identical to the original image. (c) 2D power spectrum computed by the SONOS chip using the analog VR-FFT. (d) Reconstructed image from the SONOS-computed complex-valued spectrum corresponding to (c). (e) 2D power spectrum computed by the SONOS chip, via a 2D analog DFT implemented directly using MVMs. (f) Reconstructed image from the SONOS-computed complex-valued spectrum corresponding to (e).
}
\label{fig:rotterdam}
\end{figure}

In this section, we compare the numerical precision of 2D DFTs computed by the analog VR-FFT and the direct MVM method for the 256$\times$256 RGB image (``Rotterdam'') in Fig. \ref{fig:vector_radix}. 
For the analog direct MVM, we used a SONOS sub-array programmed to the DFT-256 matrix (512$\times$1024 devices).
Fig. \ref{fig:rotterdam} compares the computed 2D frequency spectra and the reconstructed ``Rotterdam'' images for the two analog DFT methods.
Because of the much larger array size of the 256-point analog DFT, we used a more compressed conductance range (0 to 1.67{~\textmu}S vs. 0 to 20{~\textmu}S for 16-point DFTs) to avoid saturating the ADC for this application.
The direct MVM method produces significantly more degraded reconstructions with visible horizontal and vertical streaks, while the VR-FFT produced much cleaner reconstructions with higher PSNR and SSIM metrics.

The prominent horizontal and vertical streaks in Fig. \ref{fig:rotterdam}f trace their origin to the erroneously large values for the spectral components along $f_x = 0$ and $f_y = 0$ in Fig. \ref{fig:rotterdam}e.
When computing these specific components, the analog sum is conducted using a column of the matrix with strictly positive DFT weights that all map to $G_\text{max}$.
Since SONOS conductance errors increase roughly proportionally with conductance (see Fig. \ref{fig:gradient}e), the devices that are programmed to $G_\text{max}$ have larger absolute conductance errors than the other devices.
These facts together lead to a large analog current sum on this column, with a correspondingly larger accumulation of conductance errors and $IR$ drops compared to the other columns, which have both positive and negative weights and smaller average conductances.
This explains the relatively large errors for $f_x = 0$ and $f_y = 0$ in Fig. \ref{fig:rotterdam}e.
By contrast, this feature does not appear in the analog VR-FFT's frequency spectrum in Fig. \ref{fig:rotterdam}c for two reasons.
First, the elementary DFTs are much smaller (16-point rather than 256-point) so both the current sums and the accumulated errors in the analog MVMs will be smaller.
Second, the use of a larger $G_\text{max}$ for the 16-point DFTs improved the accuracy by increasing the signal-to-noise ratio in the DFT weights.
This is shown in Table \ref{tab:G_snr} below, and will be explained shortly.
These two factors lead to lower errors across all spectral components, but the difference is largest for the $f_x = 0$ and $f_y = 0$ components.

\begin{figure}[t]
\centering
\includegraphics[width=\textwidth]{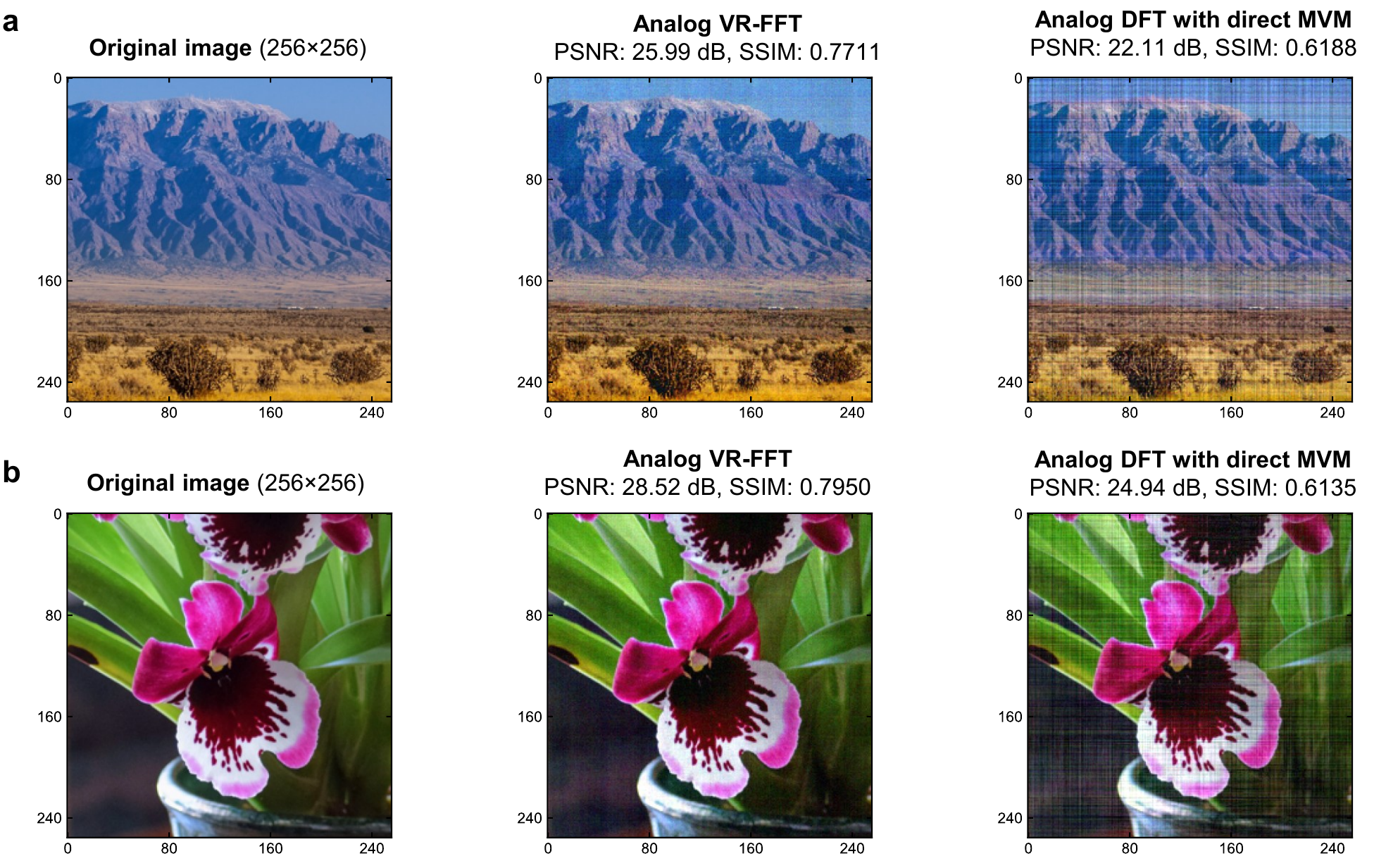}
\caption{Reconstructions of two additional 256$\times$256 images from 2D frequency spectra computed by the SONOS chip: (a) the ``Sandia'' image and (b) the ``orchid'' image. The analog VR-FFT or direct MVM method was used to compute the frequency spectra, then the image was reconstructed using a digital 2D IFFT. (``Orchid'' photograph was taken by the author. ``Sandia'' photograph from Dorothy Harris, Wikimedia Commons, CC-BY-2.0 license.)
}
\label{fig:sandia_orchid_dmvm}
\end{figure}

In addition to the ``Rotterdam'' image, we used the SONOS chip to compute the 2D frequency spectra of two other $256\times 256$ images, using both the analog VR-FFT and the direct MVM methods. The results are shown in Fig. \ref{fig:sandia_orchid_dmvm}. As with the ``Rotterdam'' image, we observe that the VR-FFT produces significantly higher-quality images with greater PSNR and SSIM.
Comparing the reconstructed images that are based on the analog VR-FFT (``Rotterdam'', ``Sandia'', and ``Orchid''), all three have similarly high values of PSNR $>25$ dB, but the SSIM has a greater variance. This is likely due to the metric's sensitivity to the presence of high-contrast edges \cite{Nilsson2020}. 

\subsection{Image reconstruction quality vs transform size}

\begin{figure}[t]
\centering
\includegraphics[width=0.85\textwidth]{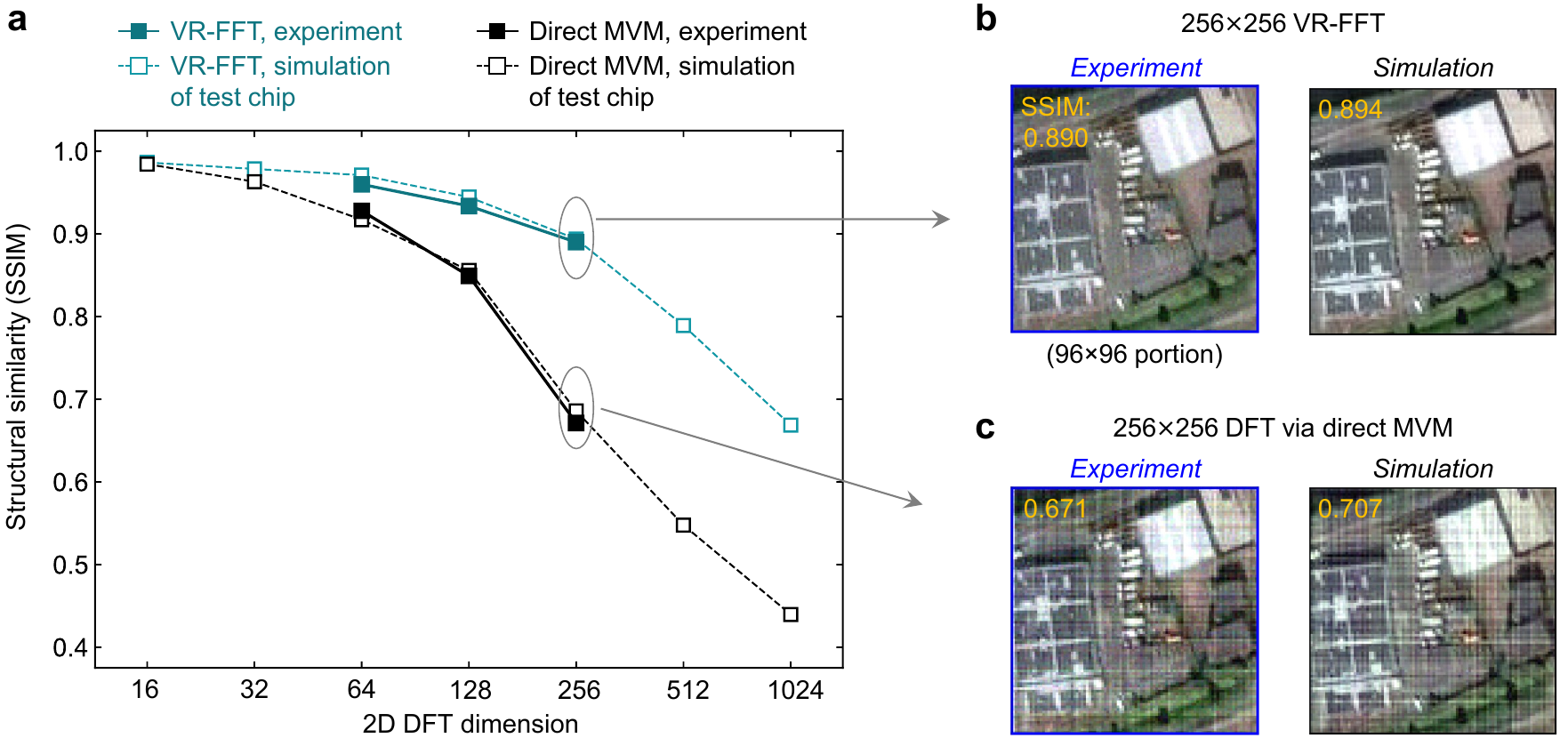}
\caption{(a) SSIM of the reconstructed Rotterdam image, generated from experimental and simulated analog VR-FFTs and direct MVMs at various sizes. The input was rescaled from a native resolution of 1024$\times$1024 pixels. The simulations were configured to match the fabricated SONOS test chip. Simulated SSIM values are the average of ten Monte Carlo runs with resampled random conductance errors. The VR-FFT points are the same as the points in Fig. \ref{fig:scaling}a that correspond to the test chip. (b) A 96$\times$96 cropped region of the experimental and simulated reconstructions using the analog VR-FFT. (c) Same region of the experimental and simulated reconstructions using analog 2D DFTs implemented as direct MVMs.
}
\label{fig:image_fft_dmvm}
\end{figure}

To project how the quality of the 2D DFT scales with image size for the two methods, we experimentally processed both the analog VR-FFT and the direct MVM on the ``Rotterdam'' image at three different sizes (64$\times$64, 128$\times$128, and 256$\times$256), then digitally computed the 2D IFFT of this result to reconstruct the images.
We also extended these experimental results by simulating the analog reconstruction over a larger range of sizes (from 16$\times$16 to 1024$\times$1024).
The utilized array sizes for these experiments and simulations are shown in Table \ref{tab:2d_dft_sonos}: for the direct MVM, only the 512-point and 1024-point DFTs are partitioned across multiple arrays.
The accuracy model, which is described in Methods, replicates the characterized properties of the SONOS devices, ADCs, and metal interconnects in our fabricated chip.
Fig. \ref{fig:image_fft_dmvm}a summarizes the quality of the reconstructions vs. image size of the two methods. 
Two simulated reconstructions are shown in Fig. \ref{fig:image_fft_dmvm}b and \ref{fig:image_fft_dmvm}c, showing that the simulation accurately captures the differences between the two methods both in terms of SSIM and the experimentally observed artifacts.
For both methods, the accuracy decreased for larger DFTs due to the use of larger programmed DFT matrices, which have greater accumulated conductance errors and parasitic $IR$ drops.

\def\arraystretch{1.2}
\begin{table}[t]
\centering
    \begin{tabularx}{\textwidth}{|>{\centering\arraybackslash\hsize=0.75\hsize}X|>{\centering\arraybackslash\hsize=1.05\hsize}X|>{\centering\arraybackslash\hsize=1.05\hsize}X|>{\centering\arraybackslash\hsize=1.05\hsize}X|>{\centering\arraybackslash\hsize=1.05\hsize}X|>{\centering\arraybackslash\hsize=1.05\hsize}X|}
        \hline
     \textbf{DFT size} & \textbf{Array size, VR-FFT} & \textbf{Max conductance, VR-FFT}  & \textbf{Array size, direct MVM} \newline (no partitioning for $N\leq 256$) & \textbf{Max conductance, direct MVM} \newline (no partitioning for $N\leq 256$) & \textbf{Max conductance, direct MVM} \newline (same array size as VR-FFT)  \\
        \hline
     16 	& 8 $\times$ 16 	& 20.00{~\textmu}S & 32 $\times$ 64  	& 20.00{~\textmu}S & 20.00{~\textmu}S \\
     32  	& 16 $\times$ 32 	& 20.00{~\textmu}S & 64 $\times$ 128 	& 10.00{~\textmu}S & 20.00{~\textmu}S \\
     64  	& 16 $\times$ 32 	& 20.00{~\textmu}S & 128 $\times$ 256 	& 5.00{~\textmu}S  & 20.00{~\textmu}S \\
     128  	& 32 $\times$ 64 	& 20.00{~\textmu}S & 256 $\times$ 512 	& 2.67{~\textmu}S  & 10.00{~\textmu}S \\
     256  	& 32 $\times$ 64 	& 20.00{~\textmu}S & 512 $\times$ 1024 & 1.67{~\textmu}S  & 10.00{~\textmu}S \\
     512  	& 64 $\times$ 128 	& 10.00{~\textmu}S & 512 $\times$ 1024 & 1.67{~\textmu}S  & 5.83{~\textmu}S \\
     1024  	& 64 $\times$ 128 	& 10.00{~\textmu}S & 512 $\times$ 1024 & 1.67{~\textmu}S  & 5.83{~\textmu}S \\
        \hline
    \end{tabularx}
\caption{Utilized SONOS array size and max conductance for the simulated and experimental results in Fig. \ref{fig:image_fft_dmvm} and \ref{fig:fft_dmvm_isoarea}. For VR-FFTs with unequal radices, the array size and $G_\text{max}$ are given for the larger of the two elementary DFT sizes.}
\label{tab:2d_dft_sonos}
\end{table}

\def\arraystretch{1.5}
\begin{table}[t]
\centering
    \begin{tabularx}{0.8\textwidth}{|>{\hsize=2.5\hsize}X|>{\hsize=0.75\hsize}X|>{\hsize=0.75\hsize}X|>{\hsize=0.75\hsize}X|>{\hsize=0.75\hsize}X|>{\hsize=0.75\hsize}X|>{\hsize=0.75\hsize}X|}
        \hline
     \textbf{Max conductance} $\mathbf{G_\text{max}}$ & 20.00{~\textmu}S & 10.00{~\textmu}S & 5.83{~\textmu}S & 5.00{~\textmu}S & 2.67{~\textmu}S & 1.67{~\textmu}S   \\
        \hline
     \textbf{Conductance SNR} & 141.4 & 84.3 & 62.2 & 58.1 & 47.0 & 42.4   \\
        \hline
    \end{tabularx}
\caption{Theoretical conductance SNR for the utilized values of $G_\text{max}$ in Table \ref{tab:2d_dft_sonos}.}
\label{tab:G_snr}
\end{table}

Table \ref{tab:2d_dft_sonos} (excluding the rightmost column) shows the SONOS array sizes and conductance ranges that were used for both the experimental and simulated 2D DFT results.
The utilized SONOS conductance range was optimally reduced with array size to balance the effects of ADC saturation, parasitic $IR$ drops, and loss of weight precision (see below).
The direct MVM method requires the largest array size possible to stay efficient, but as a result, it more quickly loses reconstruction fidelity at large image sizes due to the factors above.
By comparison, the analog VR-FFT remains significantly more accurate at large image sizes by enabling the array size to scale much more slowly with the image size.

We can quantify the overall loss of weight precision with decreasing $G_\text{max}$ by computing a theoretical value for the conductance signal-to-noise ratio (SNR). We define this as the ratio of $G_\text{max}$ to the average value of the condutance error $\sigma_G$, assuming for simplicity that the DFT weight magnitudes map to a uniform distribution of conductances:
\begin{equation}
\text{Conductance SNR} = \frac{G_\text{max}}{\sigma_\text{avg}} = \frac{2G_\text{max}^2}{\int_0^{G_\text{max}}\sigma_G(G) \, dG}
\end{equation}
For $\sigma_G(G)$, we use the analytical fit to the characterized SONOS programming errors shown in Fig. \ref{fig:gradient}e. The factor of 2 comes from the fact that when representing signed weights using a differential pair of SONOS cells, only one of the two cells contributes error: the other device can be programmed to a very low conductance state which has effectively zero error. This is a unique benefit of the SONOS device. 

Table \ref{tab:G_snr} shows that as $G_\text{max}$ is reduced, the conductance SNR first decreases quickly, then saturates.
This can be explained by the fact that in Fig. \ref{fig:gradient}e, the error is proportional to the conductance when $G$ is low, but saturates when $G$ is high.
This is because the highly conductive SONOS states are operating in the strong inversion regime, while the less conductive states are in subthreshold or weak inversion.
As a result, the SNR will be higher when more of the utilized SONOS states are in the strong inversion regime.
The decrease in conductance SNR at reduced $G_\text{max}$ leads to accuracy loss in the analog DFTs, but this can be partially compensated by the fact that a lower $G_\text{max}$ reduces the parasitic $IR$ drops.
The chosen values of $G_\text{max}$ in Table \ref{tab:2d_dft_sonos} approximately balance these errors to minimize the accuracy loss.

\subsection{VR-FFT and direct MVM comparison at the same array size}

\begin{figure}[h]
\centering
\includegraphics[width=0.45\textwidth]{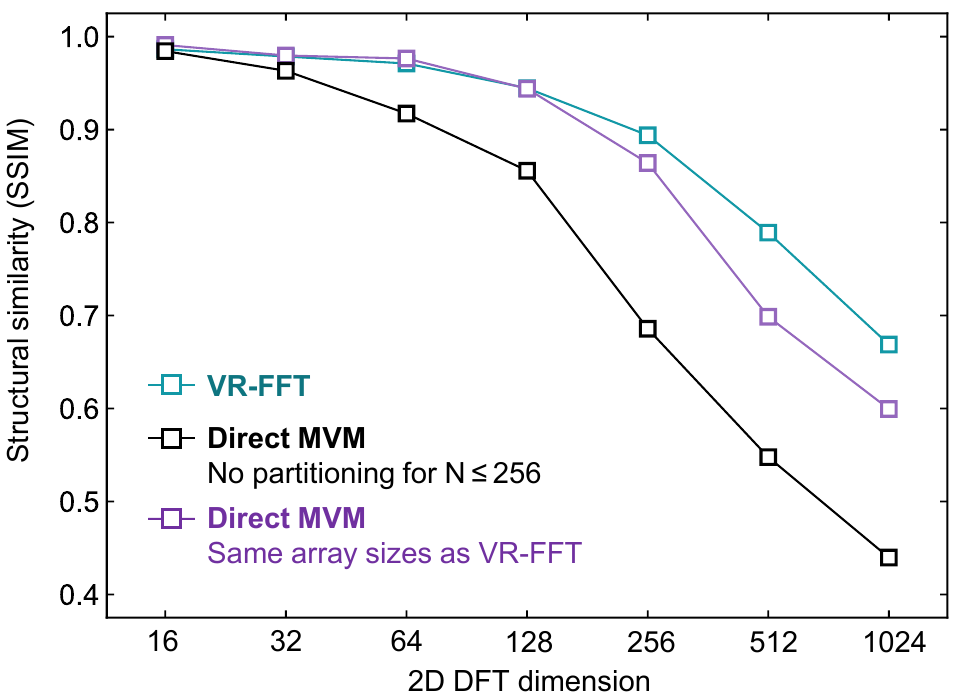}
\caption{Simulated SSIM of the reconstructed Rotterdam image, comparing the analog VR-FFT with two configurations of the direct MVM. For the black curve, the DFT matrix was not partitioned for DFT sizes $N \leq 256$ (same as Fig. \ref{fig:image_fft_dmvm}a). For the purple curve, the DFT matrix was partitioned at all DFT sizes to use the same array size as the VR-FFT. In all cases shown, the simulator was configured to model the fabricated SONOS test chip.
}
\label{fig:fft_dmvm_isoarea}
\end{figure}

For the analog VR-FFT and direct MVM comparison in Fig. \ref{fig:image_fft_dmvm}, we assumed for the direct MVM that the DFT matrix is partitioned only when the matrix size exceeds the maximum size of the SONOS array, which is 1024$\times$1024.
This assumption represents the most energy-efficient mode of the direct MVM, but leads to some accuracy loss as a result of the large summed currents and low $G_\text{max}$.
In Fig. \ref{fig:fft_dmvm_isoarea}, we compare to another mode of the direct MVM (purple curve), where the DFT matrix is partitioned to use the same array size as was used for the VR-FFT.
We note that this mode of operation has significantly larger area and energy consumption compared to both the VR-FFT and the unpartitioned direct MVM, due to the much larger number of ADC conversions needed for partial results.
This can be roughly seen in Fig. \ref{fig:cooley_tukey}d, where this case roughly corresponds to the curve for a direct MVM using a small array size.

Fig. \ref{fig:fft_dmvm_isoarea} shows that using a smaller array size increases the accuracy of the direct MVM method, because the smaller arrays can have less conductance error accumulation, smaller parasitic $IR$ drops, and better precision in the DFT weights because of a larger $G_\text{max}$.
However, the direct MVM is still less accurate than the VR-FFT at the same array size.
In the VR-FFT, each SONOS array is programmed with a full, small-radix DFT matrix, which has a collection of both high and low conductance values.
For the direct MVM case in the purple curve, each one of the many SONOS arrays is programmed with a small partition of a large DFT matrix.
The DFT weight values change only by a limited amount within these partitions, so some of these partitions will have a much higher average conductance than a small-radix DFT matrix of the same dimensions, leading to larger error accumulation and ADC clipping.
The ADC clipping has been mitigated by re-optimizing the $G_\text{max}$ values, leading to smaller conductances compared to the VR-FFT, as shown on the rightmost column of Table \ref{tab:2d_dft_sonos}.
However, this reduces the precision of the DFT weights as shown in Table \ref{tab:G_snr}, so the VR-FFT still achieves a moderately higher SSIM especially for large DFT sizes.

\subsection{Error decomposition of analog VR-FFT results}

\begin{figure}[h]
\centering
\includegraphics[width=0.45\textwidth]{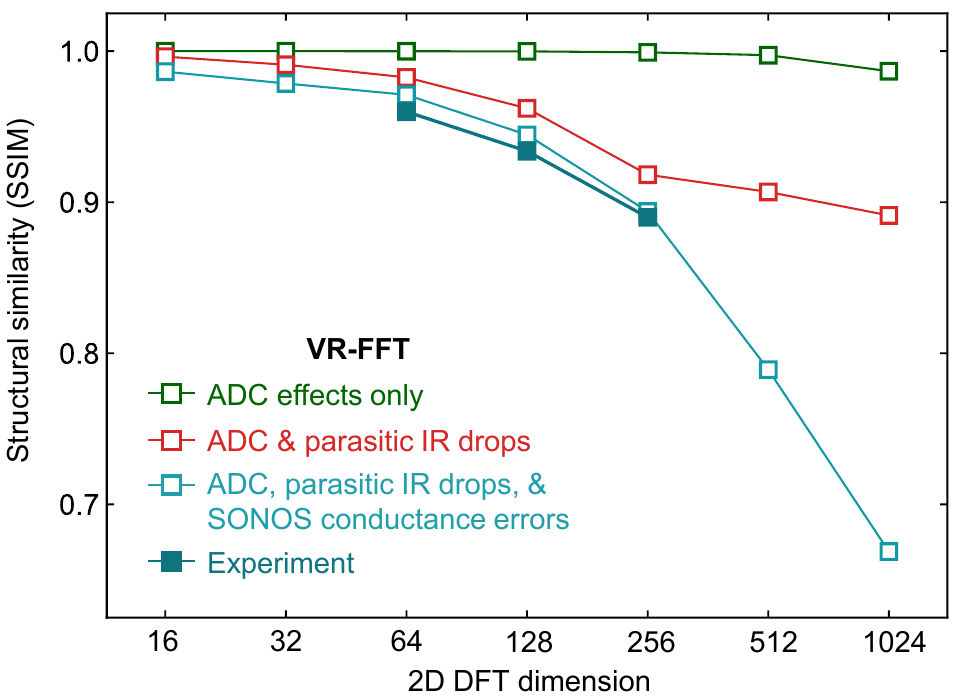}
\caption{Simulated SSIM of the reconstructed Rotterdam image for the analog VR-FFT, with selective enabling of different modeled error sources to show the breakdown of the error contributions to a loss of SSIM.
}
\label{fig:ssim_breakdown}
\end{figure}

Here, we attempt to quantify the relative contributions of different sources of error in the analog VR-FFTs that were conducted using the SONOS test chip.
While it is not possible to exactly determine these relative contributions from our experimental results, we can study this using simulations of the chip.
In Fig. \ref{fig:ssim_breakdown}, we show how the SSIM of the Rotterdam image reconstruction changes as we selectively enable the models for the major error sources: ADC quantization and clipping, parasitic $IR$ drops, and SONOS conductance errors (including programming errors, drift, and read noise).

The test chip's ADC has a minor effect on accuracy even at large transform sizes, because of its high resolution (12 bits).
Parasitic $IR$ drops have an overall larger effect that increases with array size.
The slowed rate of error growth due to $IR$ drops for $N=512$ and $N=1024$ is due to the fact that the SONOS conductances were reduced to optimize accuracy, as shown in Table \ref{tab:2d_dft_sonos}.
However, as shown in Table \ref{tab:G_snr}, the reduction of $G_\text{max}$ increases the error contribution of SONOS conductance errors by reducing the SNR in the DFT weights.
This can be seen directly in Fig. \ref{fig:ssim_breakdown} where the share of SSIM loss due to SONOS conductance errors increases significantly for the largest two sizes.

\subsection{Spectral error metrics}

In Table \ref{tab:spectrum_metrics}, we report quality metrics for the complex-valued frequency spectra (not the reconstructed images) that were experimentally computed by the SONOS array using the analog VR-FFT, for the three exemplar 256$\times$256 RGB images. Each SONOS-computed spectrum is compared to the exact spectrum that was computed using an FP32 digital 2D FFT. We report two metrics: (1) the spectral PSNR, which measures errors in the magnitude spectrum, and (2) the power-weighted mean absolute error (MAE) in the phase, defined in Equation \ref{eq:metric3}.

\def\arraystretch{1.3}
\begin{table}[h]
\centering
    \begin{tabularx}{0.9\textwidth}{|>{\hsize=1.3\hsize}X|>{\hsize=0.9\hsize}X|>{\hsize=0.9\hsize}X|>{\hsize=0.9\hsize}X|}
        \hline
     \textbf{Image} & ``Rotterdam'' & ``Sandia'' & ``Orchid''    \\
        \hline
     \textbf{Spectral PSNR} & 58.074 dB & 63.301 dB & 64.428 dB \\
        \hline
     \textbf{Weighted phase MAE} & 0.303$^\circ$ & 0.300$^\circ$ & 0.511$^\circ$ \\
        \hline
    \end{tabularx}
\caption{Metrics for the frequency spectra that are computed by the SONOS chip via the analog VR-FFT.}
\label{tab:spectrum_metrics}
\end{table}

\subsection{Simulated accuracy using test chip vs more efficient analog IMC core}

\begin{figure}[h]
\centering
\includegraphics[width=0.55\textwidth]{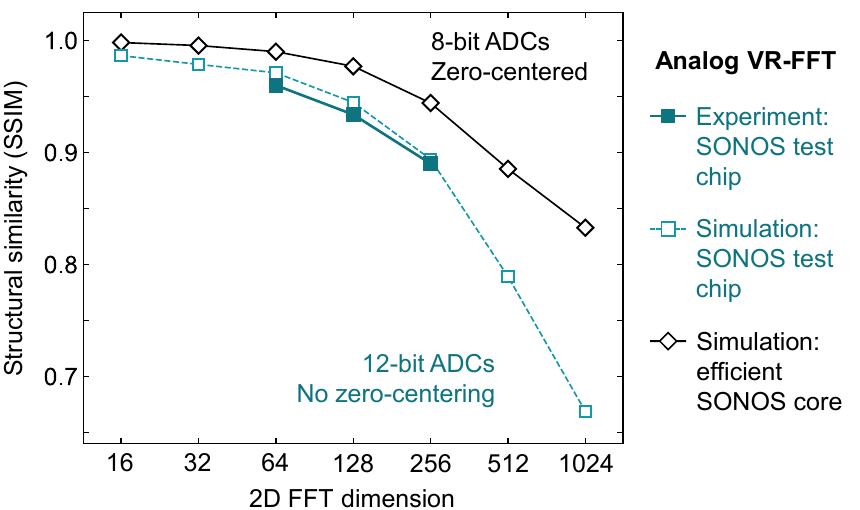}
\caption{Simulated SSIM of the reconstructed ``Rotterdam'' image vs analog VR-FFT transform size, comparing two hardware configurations: the 40-nm SONOS test chip, and a more optimized 40-nm SONOS core with 8-bit ADCs using zero-centered images. The teal points are the same as the points from Fig. \ref{fig:image_fft_dmvm}a.
}
\label{fig:ssim_12b_vs_8b}
\end{figure}

The previous simulation results in this section used an accuracy model of the fabricated prototype, while the results in Fig. \ref{fig:scaling} used a model of a more efficient SONOS IMC core, shown in Fig. \ref{fig:analog_core} and described in detail in Supplementary Section \ref{sec:energy_supp}.
The two accuracy simulation modes are described in more detail in the Methods section, but the most significant features of the optimized core are: (1) it uses 8-bit ADCs (rather than 12-bit), (2) it uses optimized input ranges for the ADCs, (3) it performs analog subtraction of currents from positive and negative DFT weights and inputs, and (4) it performs analog accumulation of results across input bits before the ADC.
The inputs to this core were also quantized to 8 bits.
Fig. \ref{fig:ssim_12b_vs_8b} compares the accuracy of the two hardware configurations on the Rotterdam image construction task.

One challenge with the lower-resolution 8-bit ADCs in the optimized core is that for the ``Rotterdam'' image in Fig. \ref{fig:vector_radix}, the zero-frequency component was orders-of-magnitude larger than all the other frequency components.
The large zero-frequency component is a natural consequence of the JPEG image format, where pixel values are stored as unsigned integers.
This led to a reduction in accuracy since the 8-bit ADCs did not quite have enough dynamic range to precisely represent all of these components.
However, we found that we could recover high reconstrution SSIM if each channel of the 2D image was zero-centered (by subtracting the pixel mean) prior to the analog VR-FFT to remove the zero-frequency component.
The digitally computed pixel means would then represent the zero-frequency component in the computed spectrum.
This simple pre-processing step greatly reduces the dynamic range requirement on the ADCs for this specific application, and led to higher accuracies across all transform sizes for the more efficient core, as shown in Fig. \ref{fig:ssim_12b_vs_8b}.
Importantly, we note that zero-centering is likely unnecessary for many practical sensing applications, where the signals are naturally close to zero-centered, or offsets would be normally removed by the analog front-end.
This was the case for the speech audio signals in Fig. \ref{fig:audio_spectrograms} and Fig. \ref{fig:scaling}b.
It was also true for the SAR phase history data used in Fig. \ref{fig:scaling}a, where no zero-centering was used.

%% file: _SI_sar.tex
\section{SAR image formation using analog FFTs}
\label{sec:sar_supp}

Here, we provide additional simulation results on the SAR image formation exemplar shown in Fig. \ref{fig:scaling}a.

\subsection{Analog FFT size scaling of SAR image formation accuracy}

We first show the simulated accuracy of polar-format SAR image formation for different image sizes when the analog VR-FFT is used for the 2D DFT step.
In these accuracy simulations, the model for the efficient SONOS IMC core design in Fig. \ref{fig:analog_core} is used, rather than the model for the fabricated chip.
To evaluate smaller image sizes than the original SAR image, we started with the full 2048$\times$2048 interpolated frequency-domain data, obtained using the polar-format algorithm up to the 2D DFT step.
This frequency-domain data was then center-cropped, reducing its size by 2$\times$ to 32$\times$ in both the range and azimuth dimensions.
This center crop reduces the range and azimuth resolution of the final formed image, and hence its pixel dimensions, without changing the spatial extent of the imaged scene.
At each transform size, a formed image is created using both a digital FP32 FFT and a simulated analog VR-FFT.
The analog VR-FFT simulations used the SONOS $G_\text{max}$ values listed in Table \ref{tab:2d_dft_sonos} for each elementary DFT size.
Specifically, for the full size 2048$\times$2048 image, we used $G_\text{max} = 5${~\textmu}S for the first stage (64-point DFT) and $G_\text{max} = 10${~\textmu}S for the second stage (32-point DFT).
The structural similarity is computed for the logarithm-scale analog formed image relative to the corresponding digital formed image.

\begin{figure}[h]
\centering
\includegraphics[width=0.95\textwidth]{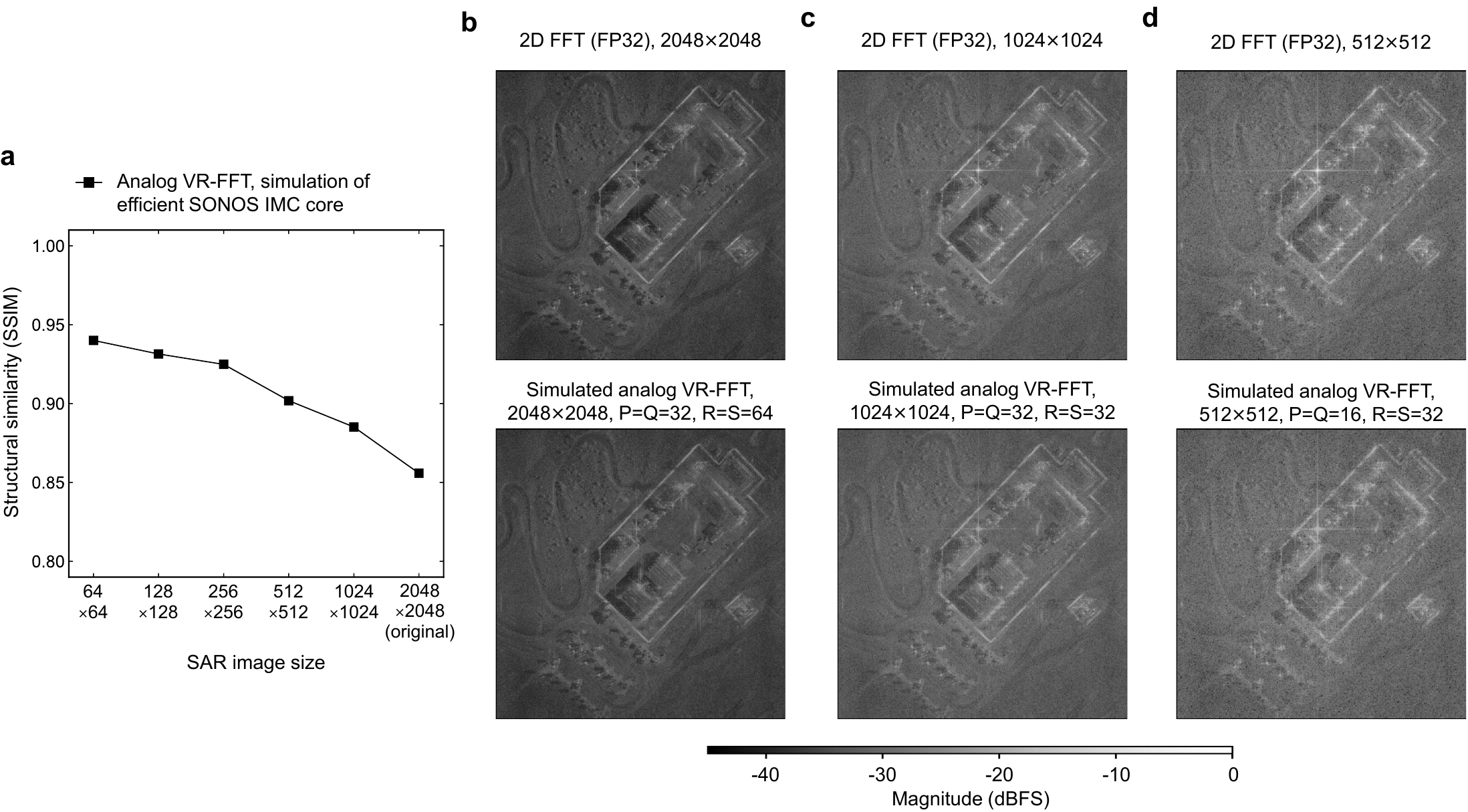}
\caption{(a) SSIM between the SAR images formed by a digital processor and the simulated SONOS IMC core, for different dimensions of the formed image. The same raw SAR sensor data is used as the starting point for all simulations. (b)-(d) Comparison of the SAR image formed using an FP32 2D FFT (left) and a simulated analog VR-FFT (right) as the final step of the polar-format algorithm, at three different 2D transform sizes.
}
\label{fig:sar_scaling}
\end{figure}

The results are shown in Fig. \ref{fig:sar_scaling}a, with some example formed images shown in Fig. \ref{fig:sar_scaling}b-d. 
The SSIM decreases gradually with the 2D transform size.
This is because a larger transform size leads to larger elementary analog DFT operations (represented by the values of $P$, $Q$, $R$ and $S$), which lead to a greater analog accumulation of errors originating from SONOS conductance errors and parasitic $IR$ drops.
This is the same effect that is response for the decline of SSIM with VR-FFT transform size in Fig. \ref{fig:image_fft_dmvm}a.
However, we note that the decline in SSIM with 2D transform size is significantly smaller for the SAR image compared to the Rotterdam image used in Fig. \ref{fig:image_fft_dmvm}a.
This is because of a significant difference in the distribution of input values for these different images.
As discussed in Supplementary Section \ref{sec:mvm_errors}, the natural images shown in Fig. \ref{fig:vector_radix}, including the Rotterdam image, have strictly positive pixel values and hence a large zero-frequency component, which is orders of magnitude larger than all other frequency components.
Accommodating this large component while maintaining resolution in all the other components stretches the limits of the dynamic range of 12-bit ADC, leading to some accuracy loss especially at large image sizes.
However, the raw RAW sensor data does not have this strong positive bias, and there is no single component of the input or output that dominates the signal to the same extent.
Therefore, a relatively high SSIM is possible at an image size of 2048$\times$2048, even when using 8-bit ADCs.

\subsection{Sensitivity analysis to parasitic resistance and conductance errors}

For the formed SAR images using analog VR-FFTs, we now perform a sensitivity analysis to two critical hardware parameters: the parasitic resistance of the array interconnects, and the magnitude of random device conductance errors.

\begin{figure}[b]
\centering
\includegraphics[width=0.95\textwidth]{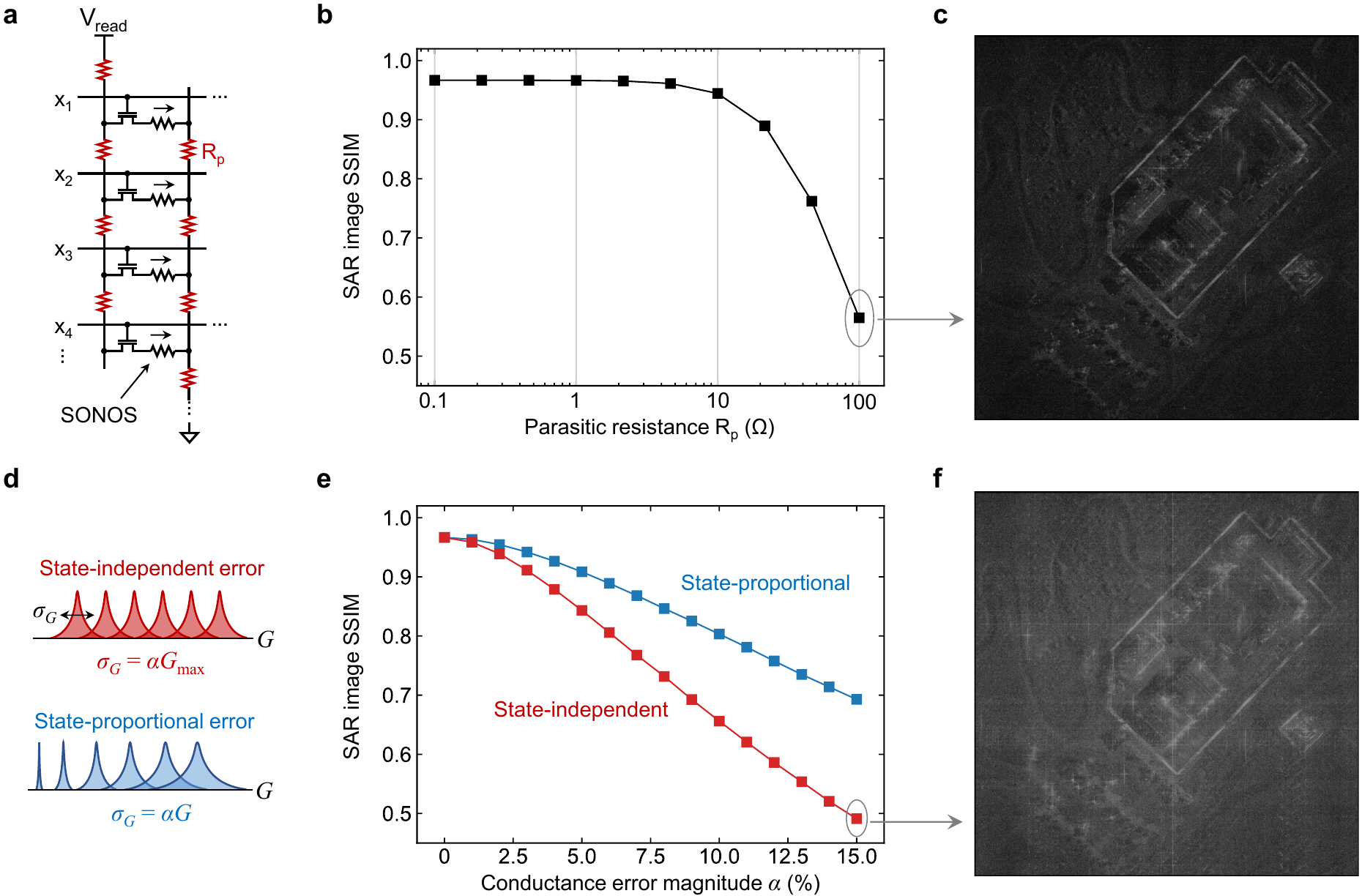}
\caption{Sensitivity analysis of the SSIM of formed SAR image using the analog VR-FFT. All results are on the original full-size 2048$\times$2048 SAR image. (a) Diagram showing the simulated array topology with parasitic wire resistance. (b) Sensitivity of the SSIM to unit cell parasitic resistance with zero conductance errors. (c) Example of a formed image at the indicated point in (b). (d) Definition of state-independent and state-proportional generic conductance errors. (e) Sensitivity of the SSIM to the magnitude of each type of generic conductance error. (f) Example of a formed image at the indicated point in (e).
}
\label{fig:sar_sensitivity}
\end{figure}

To study the effects of parasitic resistance, we simulate the array topology shown in Fig. \ref{fig:sar_sensitivity}a using the MVM circuit solver in CrossSim.
This topology accurately describes both the fabricated SONOS test chip and the SONOS IMC core design in Fig. \ref{fig:analog_core}.
The bit-wise inputs are applied as voltages on the select transistor gates (which do not draw current), the memory device is modeled as a linear resistor, and we define a parasitic metal resistance $R_p$ between every two unit cells.
For the memory devices, we use the conductance range of SONOS that was used in Fig. \ref{fig:scaling}a: $G_\text{max} = 5${~\textmu}S ($R_\text{min} = 200$k$\Omega$) for the first stage which used 128$\times$256 arrays, and $G_\text{max} = 10${~\textmu}S ($R_\text{min} = 100$ k$\Omega$) for the second stage which used 64$\times$128 arrays. 
We do not model any errors in the conductance to isolate the effect of $IR$ drops, but we do include the effect of 8-bit ADCs in all simulations, which leads to a baseline SSIM of 0.967 relative to floating-point digital.
Fig. \ref{fig:sar_sensitivity}b shows that the SSIM begins to decline sharply when $R_p > 10 \Omega$.
Fig. \ref{fig:sar_sensitivity}c shows an example of a formed image with large parasitic $IR$ drops ($R_p = 100 \Omega$).
Large $IR$ drops cause a large systematic reduction in summed currents, which in turn reduces the dot product values and leads to an overall darker image.
For the 40-nm test chip, we expect the effective $R_p$ (including the effects of series resistances at the array terminals) to be on the order of 1$\Omega$.
We note that the critical range of values for $R_p$ will scale with the memory device resistance; using higher resistances than the ones used above for SONOS would increase the tolerance for higher $R_p$.

To study the sensitivity of this SAR processing to random conductance errors, we simulate image formation using two generic conductance error profiles with a parameterizable error magnitude, rather than the conductance errors that are specific to SONOS devices.
In both cases, the conductance error in every device is randomly sampled from a normal distribution $\mathcal{N}(0,\sigma_G)$, where $\sigma_G$ can depend on the target conductance state $G$.
These two generic state dependences are illustrated in Fig. \ref{fig:sar_sensitivity}d: state-independent errors have an error magnitude that is flat across conductance, while state-proportional errors are proportional to the conductance.
As can be seen in Fig. \ref{fig:gradient}e, the combined effect of write errors and device-to-device variation in the SONOS devices has a conductance dependence that is closer to the state-proportional case, with: $\alpha \approx 11\%$ at $G=0${~\textmu}S, $\alpha \approx 5.5\%$ at $G=5${~\textmu}S, and $\alpha \approx 3.2\%$ at $G=10${~\textmu}S.

Fig. \ref{fig:sar_sensitivity}e shows how the SSIM depends on the error magnitude $\alpha$ for both types of error.
In general, the analog VR-FFT has similar sensitivity to state-proportional errors and state-independent errors.
This is in stark contrast to DNN inference workloads, where the accuracy tends to be much less sensitive to state-proportional errors than state-independent errors \cite{OnTheAccuracy}.
The reason for this difference was discussed in Supplementary Section \ref{sec:dft_weights}: when mapped to resistive crossbars, DFT matrices have conductance distributions that are close to uniform, while DNN weight matrices are highly concentrated around zero conductance.
Therefore, kernels that rely on the DFT matrix like the analog VR-FFT have errors that do not depend strongly on the state dependence of the conductance error.
This suggests that when considering random conductance errors alone, different memory devices should generally achieve similar accuracy for FFT applications if they have similar average conductance errors over their dynamic range.
In Fig. \ref{fig:sar_sensitivity}e, the difference between the two curves mainly arises from the definition of the two errors: for the same $\alpha$, the state-proportional case will have a lower average conductance error over the full range from 0 to $G_\text{max}$.
Fig. \ref{fig:sar_sensitivity}f shows an example of a formed SAR image with large state-proportional conductance errors, showing artifacts (mainly noise) that differ qualitatively from those in Fig. \ref{fig:sar_sensitivity}c despite the similar SSIM.

\subsection{Accuracy degradation due to conductance drift}

\begin{figure}[t]
\centering
\includegraphics[width=0.45\textwidth]{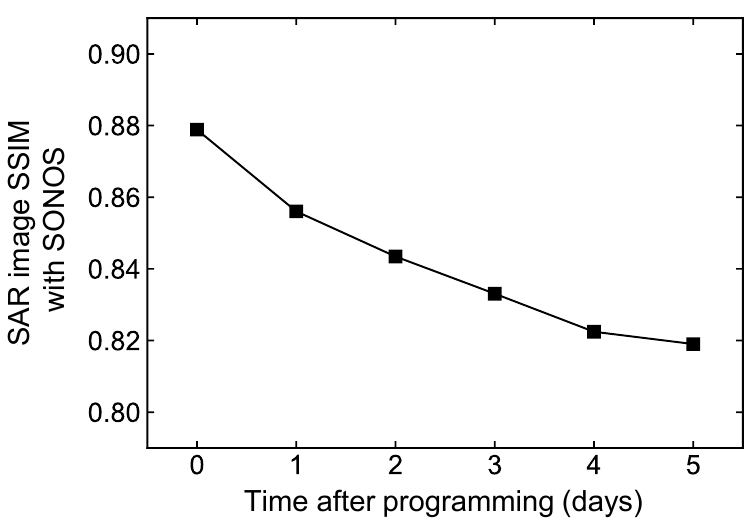}
\caption{Simulated SSIM of the formed SAR image using the analog VR-FFT, as a function of time after programming the SONOS devices.
}
\label{fig:sar_drift}
\end{figure}

Conductance drift in the SONOS devices also causes a degradation in the SSIM of the formed SAR image over time.
Here, drift refers to the combination of increasing random conductance variability over time (shown in Fig. \ref{fig:gradient}e) and increasing mean shift in conductance over time (shown in Fig. \ref{fig:gradient}f).
The simulated change in SSIM over the first five days after programming is shown in Fig. \ref{fig:sar_drift}, based on the SONOS drift characterization data in Fig. \ref{fig:gradient}.
The SSIM drops from 0.8789 to 0.8190 over this period, with a rate of decrease that slows over time, because the rate of conductance drift also slows over time.

%% file: _SI_energy_performance_area.tex
\section{Energy, performance, and area estimation of SONOS-based analog FFT core}
\label{sec:energy_supp}

\begin{figure}[b]
\centering
\includegraphics[width=\textwidth]{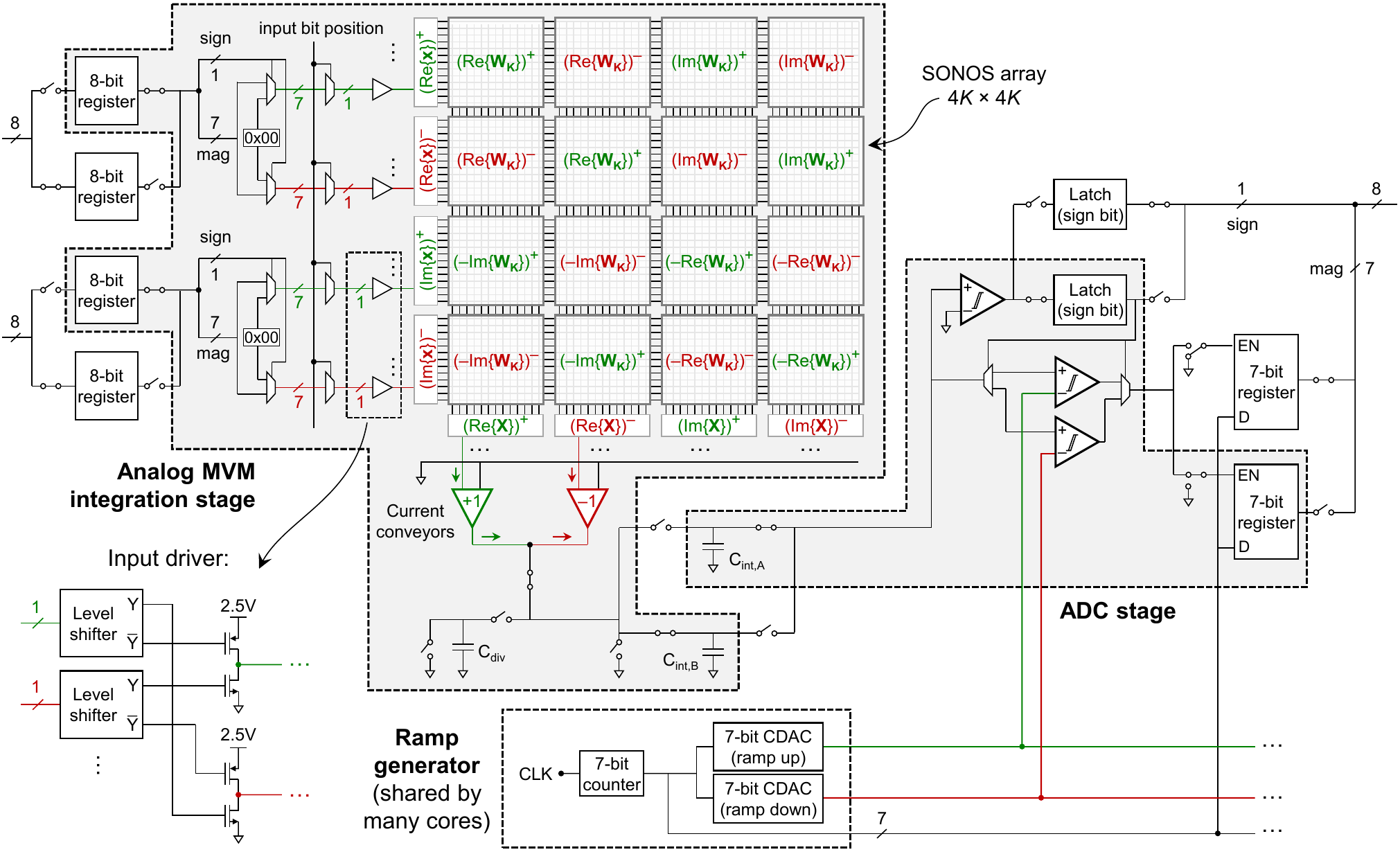}
\caption{High-level diagram of a SONOS analog IMC core for executing pipelined analog FFTs, shown for an input and output resolution of 8 bits. The core fully computes a $K$-point DFT using a single analog MVM operation. The different highlighted blocks represent pipelined stages that can be occupied by independent data batches. Different logical partitions of the SONOS array are shown, but there is no physical separation between these partitions.
}
\label{fig:analog_core}
\end{figure}

In this section, we describe the design of an energy-efficient analog IMC core and estimate its energy efficiency, area, and throughput for executing pipelined analog FFTs. This core uses the same 40-nm SONOS technology and uses the size of the SONOS array in the fabricated test chip (1024$\times$1024) as an upper bound, but with more optimal peripheral circuits, digital buffering, and dataflow. This design is roughly based on the 40-nm analog IMC core design in Ref.~\citenum{xiao2022accurate}, but with several modifications as noted below.

\subsection{Pipelined analog MVM core operation}

Fig. \ref{fig:analog_core} shows the design of the analog IMC core that can fully execute a DFT using a single analog MVM; i.e. there is a single ADC conversion for each real or imaginary output of the DFT.
The digital inputs and outputs are sign-magnitude integers.
The 2T SONOS cell and its operating bias are the same as for the array on the test chip, shown in Fig. \ref{fig:sonos_dft}a.
However, the DFT mapping in Fig. \ref{fig:sonos_dft}b is modified to allow for simultaneous processing of positive- and negative-valued inputs.
Since we operate the SONOS cells to conduct current only in one direction, this means that the array needs $4K \times 4K$ cells to execute a $K$-point complex DFT.
The maximum DFT size that can be computed by a single 1024$\times$1024 array is $K = 256$.
The columns of the array are held at virtual ground by current conveyors (CCs), which also pass the current from the column to an output node with a current gain of either $+1$ or $-1$.
Opposing gains from two CCs are used to subtract summed currents from positive and negative terms in the dot product, prior to digitization by the ADC.

The SONOS array processes the input vector bit-serially, but the bit-wise partial dot products are accumulated in the analog domain using the successive integration and rescaling technique to reduce the ADC energy consumption \cite{bavandpour2019efficient}.
For a given input bit, the differential current from the two CC's is integrated on an integration capacitor ($C_\text{int} = 150$ fF), which can be implemented as a metal-oxide-metal (MOM) capacitor using multiple metal layers.
After integration, the capacitor is disconnected from the array and connected to a second identical capacitor ($C_\text{div} = 150$ fF) to halve the charge and voltage on $C_\text{int}$.
For the next input bit, current is integrated on the accumulated charge on $C_\text{int}$ while $C_\text{div}$ is discharged.
Repeating this process for seven integration cycles (for the seven magnitude bits) implements an analog charge-domain shift-and-add accumulation.
The cycle time for each input bit is assumed to be 5 ns, which allows ample time for the array to settle \cite{marinella2018multiscale}.
This also allows a 150 fF capacitor to integrate a differential current of up to $\sim$30{~\textmu}A per input bit with a supply voltage of 1V.
Based on our accuracy simulations, this is more than enough to run all of the FFT workloads studied in this paper.

We implement double (ping-pong) buffering of the integration capacitor to allow the analog MVM integration stage to be pipelined with the subsequent ADC stage.
Each capacitor is used for integration in one cycle, then is used to hold the integrated voltage for the ADC stage in the next cycle.
We use a ramp ADC, which allows the analog dot products from all the columns to be digitized in parallel with a small footprint per column.
We find this to be an optimal choice at 8-bit resolution, though for higher precision, a successive-approximation-register (SAR) ADC may be optimal \cite{Lee2024}.
To increase speed, we use a first comparator stage to detect the sign bit.
Then, based on the sign bit, the input voltage is compared to one of two voltage ramp signals (ramp-up or ramp-down).
The ramp generation circuitry consists of a counter and two capacitive digital-to-analog converters (CDACs) that are driven by a 1 GHz clock.
We assume that each ramp generator can be shared by 1024 active columns, which can be distributed to one or more arrays depending on the DFT size.
When the comparator in this stage switches, the counter's value is latched to a 7-bit register which contains the magnitude bits.
The registers that store each 8-bit output are also double buffered to enable pipelining of the ADC operation with the subsequent stage, e.g. writing the intermediate digital outputs to SRAM.

\subsection{Energy consumption}
\label{sec:supp_energy}

\begin{figure}[t]
\centering
\includegraphics[width=0.5\textwidth]{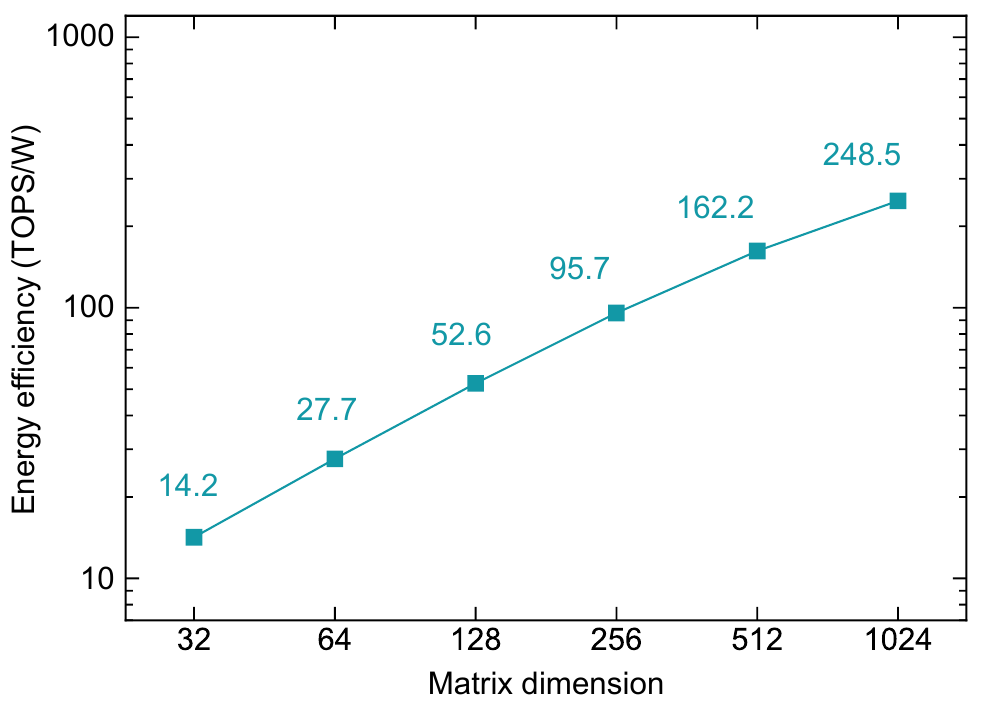}
\caption{Energy efficiency of MVMs using the analog IMC core in Fig. \ref{fig:analog_core} for a real-valued matrix. An operation is defined at 8-bit precision for real values.
}
\label{fig:TOPSW}
\end{figure}

We estimate the energy consumption of the analog FFT for various FFT sizes.
To simplify the analysis, we lump the energy costs into a total energy per 8-bit output value (real or imaginary) of an analog DFT operation.
As described and experimentally demonstrated in Supplementary Section \ref{sec:parallel_fft}, using the parallel analog FFT implementation enables the twiddle factor multiplications to be accurately folded into the analog MVMs, so that they do not incur any additional energy cost.
The dominant energy contributions are summarized below:

\begin{itemize}
	\item \textbf{SONOS array}: We include energy dissipation in the SONOS devices that draw current, and the $CV^2$ energy to charge the select gate lines from 0V to 2.5V.
	We assume $G_\text{max} = 16.7${~\textmu}S and $G_\text{min} = 10$ pS, based on the large on/off ratio of the SONOS device shown in Fig. \ref{fig:sonos_device}c. There are at most 4$K$ SONOS cells that contribute to a given digital output, since at least half the rows will be inactive due to input sign. The voltage across each cell's current-conducting terminals is 0.06V and power is dissipated for seven cycles. The average conductance for a DFT weight is $\approx$0.32$G_\text{max}$. We assume an input activity factor of 25\%, averaged over input bits. 
	For $CV^2$, we use values for the metal interconnect capacitance per length, SONOS cell dimensions, and the select gate capacitance from Ref.~\citenum{xiao2022accurate}, which is based on the same 40-nm process.
	The total energy consumption in the array is estimated as:
	\begin{itemize}
		\item[$\circ$] $IV$ energy: 0.011 pJ/output ($K=16$) to 0.17 pJ/output ($K=256$)
		\item[$\circ$] $CV^2$ energy: 0.11 pJ/output ($K=16$) to 1.8 pJ/output ($K=256$)
	\end{itemize}
   \item \textbf{Input drivers}: The input driver, shown on the bottom left of Fig. \ref{fig:analog_core}, converts each input bit into a voltage (0V or 2.5V) that can turn on or off the select transistors in a row of SONOS cells. These are implemented using 2.5V I/O transistors in the 40-nm process. The positive and negative input bit signals (which can be 1/0, 0/1, or 0/0) are stepped up in voltage using two level shifters, which then drive the gates of two pairs of transistors that connect the rows to the 2.5V or 0V voltage rails. The energy consumption of the transistors that drive the array is accounted for by the array $CV^2$ energy calculated above. These transistors are sized to be $4\times$ minimum width to enable fast charging of the select gate lines. For the level shifter, we use the design from Ref.~\citenum{marinella2018multiscale}, which can switch in 0.2 ns and consumes 29 fJ per transition (after accounting for our higher 2.5V voltage). Over seven cycles, the level shifters consume a total of 0.20 pJ/output.
	\item \textbf{Integrators}: We assume that two optimally designed CC's dissipate 21.6{~\textmu}W of power each, based on the fast CCII design that is used for analog IMC in Ref.~\citenum{marinella2018multiscale}. Integration over seven cycles leads to a total energy consumption of 1.5 pJ/output.
	\item \textbf{ADC}: The dominant energy contribution in the ADC step is the three comparators in Fig. \ref{fig:analog_core}. The energy consumption of the ramp generator is negligible since it is amortized across many digital outputs. Based on circuit simulations in a commercial 40-nm process for a similar design \cite{xiao2022accurate}, the total energy consumption for an 8-bit conversion is 2.1 pJ/output. This is about 8$\times$ more energy than the limit for 8-bit ADCs from a well-known ADC survey \cite{adc_survey}. (A SAR ADC can potentially operate closer to the limit, but consumes more area.)
	\item \textbf{SRAM access}: The digital outputs of the first DFT stage must be written to local SRAM buffers, then the values must be read out from SRAM for the second DFT stage. We assume that each analog IMC array can write to an independent SRAM bank with an 8-bit block size. We use CACTI 7.0 to estimate the access energy, latency, and area, using a custom technology file for a commercial 40-nm process \cite{Balasubramonian17}. The SRAM access energy, averaged over reads and writes, is 0.56 pJ/output. This energy is not included for small DFTs ($N \leq 256$) that can be implemented with a single MVM, and thus has no intermediate results.
\end{itemize}

The estimated total energy per digital output varies from 4.51 pJ to 5.75 pJ/output, depending on the matrix size.
Fig. \ref{fig:TOPSW} expresses the equivalent energy efficiency in TeraOperations/s/W (TOPS/W) of real-valued MVMs using the analog IMC core.
Since the energy consumption is dominated by peripheral circuits especially at smaller array sizes, the efficiency (in TOPS/W) increases directly with the size of the matrix and would have nearly the same value for FFTs and DNN layers.
To obtain the total DFT energy consumption, the per-output energy consumption is multiplied by the total number of intermediate and final digital outputs of the $N$-point DFT computation.
This is equal to $2N$ for small DFTs that can be implemented as a single MVM, and is equal to $4N_1N_2 = 4N$ for the analog FFT with a single Cooley-Tukey decomposition. 
These results are plotted in Fig. \ref{fig:scaling}c and compared to state-of-the-art digital FFT processors.

As a point of reference, we compare specifically to the efficiency of the AMD Versal AI core, which can accelerate both DSP and ML applications.
At the same 8-bit precision, the analog IMC core can compute FFTs with at least 59$\times$ higher efficiency for the FFT sizes from 64 to 2048 evaluated in Fig. \ref{fig:scaling}c.
For matrix multiplication, the analog IMC core is more efficient (in terms of TOPS/W) than the Versal by a factor of 34$\times$, 63$\times$, 106$\times$, and 163$\times$, for matrices containing 128, 256, 512, and 1024 rows, respectively \cite{versal_inference}.

\subsection{FFT throughput and latency}
\label{sec:supp_speed}

\begin{figure}[t]
\centering
\includegraphics[width=0.95\textwidth]{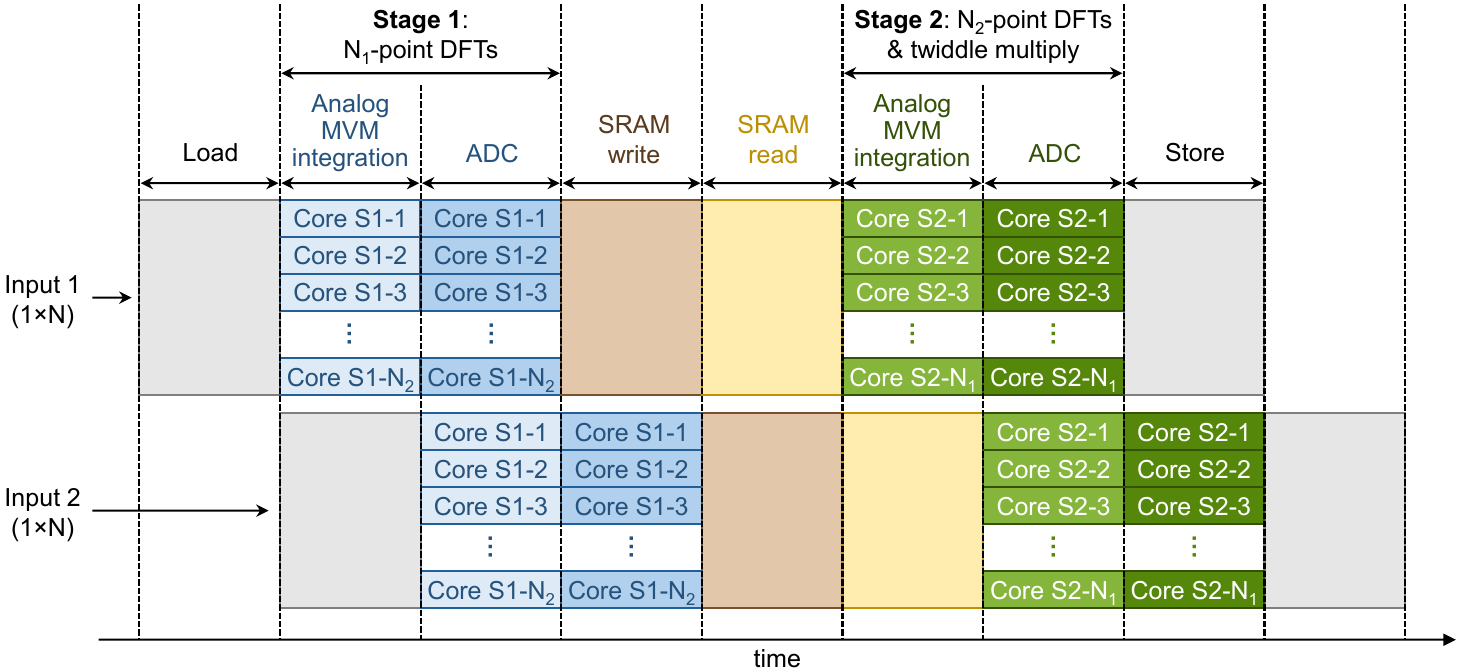}
\caption{Dataflow diagram showing the pipelined execution of the parallel analog $N$-point FFT in Fig. \ref{fig:parallel_fft}.
}
\label{fig:pipelining}
\end{figure}

For throughput estimation, we focus on the 4096-point FFT design point, which is a particularly relevant size for 5G wireless communications \cite{Zaidi2018,Guo2023}, though the cores can be dynamically re-configured to compute other FFT sizes.
This requires elementary DFTs of size $N_1 = N_2 = 64$.
To attain high throughput, we use the parallel FFT scheme in Fig. \ref{fig:parallel_fft}, where the independent analog DFTs within each stage of the FFT are computed in parallel on multiple arrays.
We further increase throughput by pipelining the stages of the analog FFT as shown in Fig. \ref{fig:pipelining}, which is supported by ping-pong buffers at the boundaries between stages.
Some of these buffers are shown in Fig. \ref{fig:analog_core}.
The SRAM memory is also double buffered. 
We also allocate a stage to load the input elements into the core's ping-pong input buffers in the correct order.
These buffers allow multiple independent FFTs to be processed concurrently in different stages of the pipeline.

The FFT throughput is set by the length of a single pipeline stage.
This length is set by the limiting step, which is the digitization of analog DFT outputs using the ramp ADC.
For an 8-bit ADC, the latency of this step is 130 ns, or 130 clock cycles at 1 GHz: two cycles for sign detection and 128 cycles for the 7-bit ramp to determine the magnitude.
Meanwhile, the analog MVM integration stage takes 42 ns to integrate all of the input bits.
The latency of SRAM access is 0.43 ns per 8-bit word (using CACTI), so the latency to read or write 128 words as needed for a 64-point DFT is at most $\sim$55 ns.
Therefore, with the analog IMC cores working in parallel, a 4096-point FFT can be completed once every 130 ns, and the throughput (in GigaSamples/s) is:
\begin{equation}
\nonumber
\text{FFT throughput} = 4096 \, \text{samples} \, / \, 130 \, \text{ns} = \textbf{31.5 GSamples/s}
\end{equation}

The latency of the 1D 4096-point FFT computation is the total time needed for a computation to traverse the full pipeline in Fig. \ref{fig:pipelining}. Since there are eight pipeline stages that all have the same length of 130 ns, the latency is:
\begin{equation}
\nonumber
\text{Latency of 4096-point FFT} = 8 \times 130 \, \text{ns} = \textbf{1040 ns}
\end{equation}

We can quantify the raw compute throughput of the accelerator in TOPS. The 4096-point analog FFT implementation involves 64 MVMs in the first stage and 64 MVMs in the second stage. Each MVM is a multiplication between a 64$\times$64 complex matrix and a 1$\times$64 complex vector, which has $2\times 128\times 128 = 32768$ real operations (multiply and add counted separately). A 4096-point FFT completes every 130 ns. Therefore, the compute throughput is:
\begin{equation}
\nonumber
\text{Compute throughput} = 2 \times 64 \times 32768 \, \text{operations} \, / \, 130 \, \text{ns} = \textbf{32.3 TOPS}
\end{equation}

\subsection{Area estimation}
\label{sec:supp_area}

To estimate the area-normalized performance of the analog FFT processor, we again focus on the 4096-point FFT design point described above.
This implementation requires 128 analog IMC cores that each contains a 256$\times$256 SONOS array.
It also requires 16KB of SRAM memory to buffer the intermediate results, and 16 ramp generators for the ADCs that are shared by 8 IMC cores each.
Since the IMC core design is similar to that in Ref.~\citenum{xiao2022accurate}, which was based on the same 40-nm process, we use the methods from that work to estimate the area of all components of the IMC core in Fig. \ref{fig:analog_core}.
We note that these are based on summed component areas rather than a full physical layout, so we have conservatively included an additional 25\% area overhead within the analog IMC core for wiring and a basic digital control unit.
We also use CACTI to estimate the area of the SRAM buffer.
SONOS flash arrays require charge pump circuits to generate the voltages needed for programming.
Since the SONOS arrays need only to be infrequently programmed (i.e. once every few days to reset the effects of drift), a set of charge pumps can be shared by multiple arrays.
To roughly estimate its area, we follow the 40-nm flash AIMC chip in Ref.~\cite{Fick2022}, where a single set of charge pumps with $\sim$2.0 mm$^2$ area was shared by 16 flash arrays that had a total of 33.5M devices.
The 4096-point FFT design needs a total of 8.4M devices, so we estimate the corresponding share of charge pump area required to be 0.5 mm$^2$.
The estimated total area for the 4096-point FFT design is 5.42 mm$^2$.
The area breakdown is summarized in Table \ref{tab:area}.

\def\arraystretch{1.2}
\begin{table}[t]
\centering
    \begin{tabularx}{0.95\textwidth}{|>{\centering\arraybackslash\hsize=1.4\hsize}X|>{\centering\arraybackslash\hsize=0.8\hsize}X|>{\centering\arraybackslash\hsize=0.8\hsize}X|}
        \hline
     \textbf{Contribution} & \textbf{40-nm design} & \textbf{22-nm design} \\
        \hline
     SONOS array           &  0.839 mm$^2$ & 0.254 mm$^2$ \\
     Input drivers and logic    &  0.461 mm$^2$ & 0.139 mm$^2$ \\
     \makecell{Column analog peripheral circuits\\(including integration capacitors)}  &  1.663 mm$^2$ & 0.973 mm$^2$ \\
     Column comparators       &  0.208 mm$^2$ & 0.063 mm$^2$ \\
     Output registers      &  0.629 mm$^2$ & 0.190 mm$^2$ \\
     SRAM               &  0.081 mm$^2$ & 0.019 mm$^2$ \\
     Ramp generators       &  0.056 mm$^2$ & 0.017 mm$^2$ \\
     Control and wiring layout overhead  &   0.987 mm$^2$ & 0.455 mm$^2$ \\
     Charge pumps  &    0.500 mm$^2$ & 0.339 mm$^2$ \\
     \hline
     \textbf{Total}  &  \textbf{5.424 mm}$^{\mathbf{2}}$ & \textbf{2.448 mm}$^{\mathbf{2}}$ \\
        \hline
    \end{tabularx}
\caption{Estimated area breakdown of the 4096-point analog FFT designs.}
\label{tab:area}
\end{table}

The area-normalized FFT throughput and compute throughput of this design are: 
\begin{align}
\nonumber
&\text{40-nm area-normalized FFT throughput} = (31.5 \, \text{GSamples/s}) \, / \, 5.42 \, \text{mm}^2 = \textbf{5.81 GSamples/s/mm}^{\mathbf{2}} \\
\nonumber
&\text{40-nm area-normalized compute throughput} = 32.3 \, \text{TOPS} \, / \, 5.42 \, \text{mm}^2 = \textbf{5.96 TOPS/mm}^{\mathbf{2}}
\end{align}

A 22-nm SONOS flash memory is being developed for analog IMC applications \cite{Agrawal2022}, so we also estimate the area of a 22-nm implementation of the analog IMC core.
We assume a simple quadratic scaling of the area of the transistors and the SONOS cells.
The density of the MOM capacitors increases by $\sim$47\% from the 40-nm node to the 22-nm node, from 5.9 fF/{\textmu}m$^2$ to 8.7 fF/{\textmu}m$^2$, due to the availability of additional metal layers \cite{Shi2018}.
We also use this capacitance density scaling to project the area reduction of the charge pumps.
We estimate that these changes reduce the total area to 2.45 mm$^2$, leading to an area-normalized FFT throughput of 12.9 GigaSamples/s/mm$^2$.
\begin{align}
\nonumber
&\text{22-nm area-normalized FFT throughput} = (31.5 \, \text{GSamples/s}) \, / \, 2.45 \, \text{mm}^2 = \textbf{12.9 GSamples/s/mm}^{\mathbf{2}} \\
\nonumber
&\text{22-nm area-normalized compute throughput} = 32.3 \, \text{TOPS} \, / \, 2.45 \, \text{mm}^2 = \textbf{13.2 TOPS/mm}^{\mathbf{2}}
\end{align}

%% file: _SI_vector_radix_overhead.tex
\section{Analog 2D FFT dataflow and overhead}
\label{sec:vector_radix_overhead}

In this section, we describe the dataflow and estimate the energy, throughput, and area of an analog 2D FFT accelerator based on the vector-radix FFT mapping in Fig. \ref{fig:vector_radix}a.
Estimates for the component-level overheads are based on the analysis in Supplementary Section \ref{sec:energy_supp}.
The 2D input and all intermediate data are assumed to be 16-bit complex values (8-bit real, 8-bit imaginary); this resolution is sufficient for accurate SAR image formation, as shown in Fig. \ref{fig:scaling}a.

\subsection{Analog VR-FFT dataflow}

\begin{figure}[b]
\centering
\includegraphics[width=\textwidth]{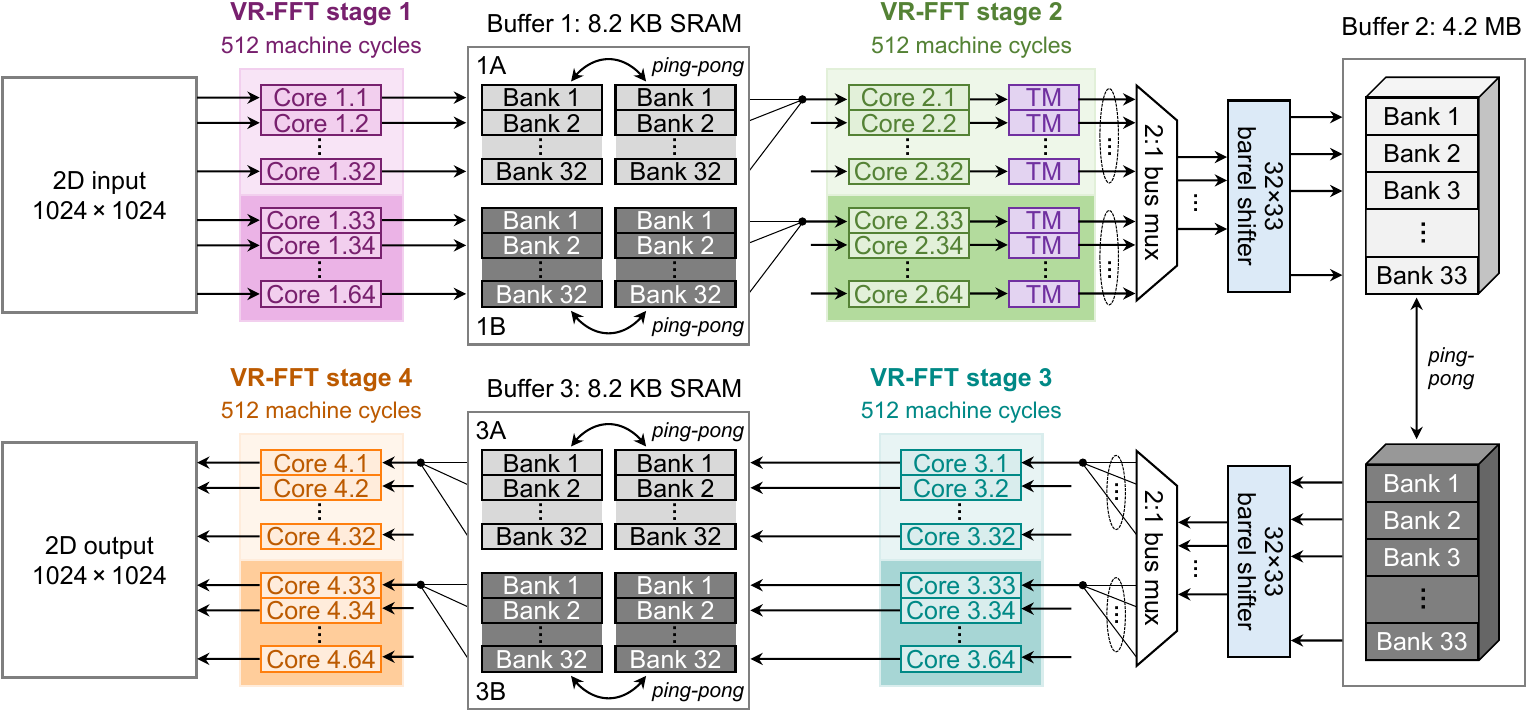}
\caption{High-level dataflow for an analog, 1024$\times$1024 vector-radix FFT. Buffer 1 and Buffer 3 are pipelined at the granularity of MVM cycles (sub-matrices), while Buffer 2 is pipelined across two streaming 2D images.
}
\label{fig:vector_radix_pipelining}
\end{figure}

One possible dataflow for an analog 2D VR-FFT is shown in Fig. \ref{fig:vector_radix_pipelining} and described below.
As an exemplar, we consider a configuration to process a large, 1024$\times$1024 2D VR-FFT using the radices $P=Q=R=S=32$.
We assume that the analog IMC cores use SONOS arrays with dimensions of 128$\times$128, which are each capable of executing 32-point analog DFTs.

A critical difference in the analog implementation of a large $N\times N$ 2D FFT vs. an $N$-point 1D FFT is that the analog MVMs in a given VR-FFT stage cannot practically be executed entirely in parallel.
Each VR-FFT stage in this example requires $32^3 = 32,768$ MVMs, compared to 32 MVMs per stage for the 1D 1024-point FFT.
We trade off throughput with area by allocating 64 analog IMC cores per stage to compute 64 MVMs in parallel.
Of these, cores 1-32 process a single sub-matrix at coordinate $(r, s)$ within the VR-FFT: see Equation \ref{eq:vrfft_1} in Methods.
Cores 33-64 concurrently process a second sub-matrix at $(r', s')$.
We can pipeline the analog MVM, ADC, and buffer access steps using a similar pipeline to Fig. \ref{fig:pipelining}, which will be described below.
Each DFT stage requires 512 machine cycles to process all 1024 sub-matrices, where one MVM is completed per IMC core per machine cycle. 
The machine cycle is the length of a pipeline stage and is set to 130 clock cycles (see Supplementary Section \ref{sec:supp_speed}).

\vspace{1ex}

\noindent \textbf{VR-FFT Stage 1 \& Buffer 1}: Every analog IMC core in Stage 1 simultaneously produces 32 complex-valued outputs, which are then written to Buffer 1.
This buffer is split into two independent buffers: Buffer 1A is connected to cores 1-32 in Stages 1 and 2, and Buffer 1B is connected to cores 33-64 in both stages.
Within each buffer, we allocate one SRAM bank per core so that data from all 64 cores can be written in parallel to Buffer 1A/1B within one machine cycle.
Buffers 1A and 1B are each double buffered (i.e. ping-pong buffers) so that writing Stage 1 outputs can occur in parallel with reading Stage 2 inputs.

An important property of the VR-FFT is that the computation in Stages 1 and 2 is local to a given sub-matrix, and different sub-matrices do not interact (see Equation \ref{eq:vrfft_1}).
This means that data in Buffer 1A is never needed by cores 33-64 in Stage 1 or 2, and data in Buffer 1B is never needed by cores 1-32 in these stages.
Additionally, the sub-matrix locality means that the first MVM in Stage 2 can start as soon as the first MVM in Stage 1 is completed, and the cores in the two stages can work in lockstep via sub-matrix-level pipelining.
Therefore, Buffer 1 only needs to hold the data for two sub-matrices at a time.

\vspace{1ex}

\noindent \textbf{Local transpose and VR-FFT Stage 2}: The output of a given sub-matrix in Stage 1 must be transposed before Stage 2.
Since each row of the sub-matrix is written to a unique bank, elements of a column are scattered across all 32 banks in Buffer 1A or 1B.
Therefore, the transpose can be implemented by loading the data in Buffer 1A/1B into Stage 2's cores in a transposed order without bank conflicts.
To load the input vector for a given Stage 2 core, one complex value is read out from all 32 banks in parallel within one clock cycle.
The inputs to two IMC cores can be loaded in the same clock cycle from the two buffers (1A and 1B).
On the next clock cycle, the input vectors for the next pair of cores is loaded.
In this manner, the inputs to all 64 cores in Stage 2 can be retrieved in one machine cycle.

\vspace{1ex}

\noindent \textbf{Twiddle multiplication}: Since processing all MVMs in Stage 2 in parallel is impractical at this FFT size, the twiddle factors cannot be folded into the device conductances within the analog IMC cores.
Therefore, we add a pipeline stage after Stage 2 to execute digital twiddle multiplication before storing the results to Buffer 2.
This can be done using a dedicated twiddle multiplication (TM) block per core, whose design is similar to that in Ref.~\citenum{Chen2017}.
The dimensions of the twiddle matrix $\mathbf{T}$ are 1024$\times$1024, but it has only 1024 unique values which can be stored in a lookup table, implemented as a read-only memory (ROM) inside each TM block.
The TM block performs the twiddle multiplications on one complex element at a time, using the following steps: (1) compute the ROM address of the desired twiddle factor by performing simple arithmetic on the current coordinate $(p, q, r, s)$, (2) fetch the value from the ROM, (3) multiply the output element by the twiddle factor, and (4) store the result in a double-buffered output register.
These steps can be pipelined inside each TM block to compute all 32 complex multiplications within a machine cycle.
Due to the periodicity of the DFT matrix that was described in Supplementary Section \ref{sec:reconfigurability}, the same ROM can be reconfigured after programming for multiple different FFT sizes (per dimension) by changing the address computation logic, as long as one size is a multiple of another.
Therefore, the ROM should be sized for the largest FFT size to be supported by the accelerator.

\vspace{1ex}

\noindent \textbf{Buffer 2}: Between Stages 2 and 3 of the VR-FFT, there is a global axis exchange operation between $(p, q)$ and $(r, s)$, which can be implemented by accessing Buffer 2 with a stride of 1024 elements when retrieving the MVM inputs for Stage 3.
This strided access should ensure: (1) The 32 inputs to a given Stage 3 core are retrieved from 32 different banks in the same cycle to maintain throughput, and (2) the 32 inputs arrive in the correct order to maintain algorithmic correctness.
The first requirement can be logically satisfied by allocating 33 banks and using a simple modulo-33 rule in the controller to assign a matrix address to a bank: \texttt{bank\_id = floor(address/32) \% 33}.
During buffer write, each IMC core in Stage 2 writes its outputs contiguously to one bank in Buffer 2.
This write occurs in two phases, controlled via a 2-to-1 bus multiplexer: the first phase writes data from cores 1-32 and the second phase writes data from cores 33-64.
The buffer read similarly occurs in two phases, and the modulo-33 rule guarantees that all 32 inputs to a given IMC core in Stage 3 are read from different banks without bank conflicts.
To satisfy the second requirement, a barrel shifter on both sides of Buffer 2 can correctly route data between the 33 banks and the 32 multiplexed IMC cores in Stages 2 and 3.
Each barrel shifter's connectivity is static over a given read/write phase, and rotates by one position twice per machine cycle.

The axis exchange implies that all of the outputs from Stage 2 must reside in Buffer 2 before the first MVM in Stage 3 can begin.
Therefore, Buffer 2 must be large enough to hold the global result of Stage 2 for a given image.
Buffer 2 is double buffered similarly to Buffer 1, but the ping-pong between its buffers occurs once every 521 machine cycles (see below), rather than on every machine cycle.
Effectively, double buffering of Buffer 2 enables pipelining the system's analog IMC cores across two images or frames, while double buffering of Buffers 1 and 3 enables more fine-grained pipelining of the cores over the sub-matrices of an image.

We note that the above only describes a high-level logical implementation of Buffer 2.
The large memory requirement for this global buffer is a consequence of the well-known ``corner turn'' issue in any hardware implementation of large 2D FFTs.
Due to its large size, significant optimization of Buffer 2's physical design (e.g. its temporal and spatial memory hierarchy) may be needed to jointly optimize its speed, access energy, and area.
We do not perform this optimization here, but note that hardware solutions to the corner turn have been an active research topic \cite{Dou2007,Kim2009,Chen2017,Garrido2020}, and we expect that these innovations can be leveraged together with analog FFT cores.

\vspace{1ex}

\noindent \textbf{Buffer 3 \& VR-FFT Stages 3-4}: Stages 3 and 4 of the VR-FFT operate locally on the same sub-matrix at a time (see Equation \ref{eq:vrfft_2} in Methods) and can therefore run in lockstep, similarly to Stages 1 and 2.
Buffer 3 is partitioned into Buffers 3A and 3B similarly to Buffer 1, and has the same total size as Buffer 1.

\subsection{Analog 2D FFT energy, throughput, and area}

We estimate the overhead of the 1024$\times$1024 analog VR-FFT using SONOS analog IMC cores, based on the analog IMC core design and component-level overhead estimates in Supplementary Section \ref{sec:energy_supp}.

\vspace{1ex}
\noindent \textbf{Energy}: Table \ref{tab:vrfft_energy} shows a breakdown of the estimated energy consumption for this 2D FFT using 40-nm analog IMC cores (same process as the fabricated chip).
The VR-FFT consists of $1.31\times10^5$ analog MVMs across four stages, for a total of $2.15\times10^9$ real-valued operations inside the IMC cores.
For Buffers 1 and 3, the per-byte SRAM access energy is about 0.56 pJ/byte.
For accesses to the larger Buffer 2, we assume a 10$\times$ larger energy of 5.6 pJ/byte, though optimal implementations may be able to achieve a lower energy.
Each complex twiddle multiplication is based on the sum of the energies of four INT8 multiplies (0.20 pJ each) and two INT8 adds (0.03 pJ each) from Ref.~\citenum{Horowitz2014} for a 45 nm node, and a conservative access energy of 0.25 pJ/byte for the local 2.0 KB ROM inside each TM block.
Due to the high energy efficiency of the analog FFT operations, reads and writes to the global buffer (Buffer 2) make up nearly half of the energy consumption.
\begin{table}[t]
\centering
\def\arraystretch{1.2}

\begin{minipage}[t]{0.4\textwidth}
\centering
\begin{tabularx}{\linewidth}{|>{\centering\arraybackslash\hsize=1.3\hsize}X|>{\centering\arraybackslash\hsize=0.7\hsize}X|}
    \hline
    \textbf{Contribution} & \textbf{Energy} \\
    \hline
    SONOS array                     & 2.02 {\textmu}J \\
    Column integration circuits     & 12.68 {\textmu}J \\
    Input level shifters            & 1.70 {\textmu}J \\
    ADCs                            & 17.70 {\textmu}J \\
    SRAM access                     & 30.70 {\textmu}J \\
    Twiddle multiplications 		& 1.43 {\textmu}J \\
    Barrel shifters                 & 0.005 {\textmu}J \\
    \hline
    \textbf{Total}                  & \textbf{66.23 {\textmu}J} \\
    \hline
\end{tabularx}
\caption{Estimated energy breakdown of the 40-nm design for the 1024$\times$1024 VR-FFT.}
\label{tab:vrfft_energy}
\end{minipage}
\hfill
\begin{minipage}[t]{0.56\textwidth}
\centering
\begin{tabularx}{\linewidth}{|>{\centering\arraybackslash\hsize=1.4\hsize}X|>{\centering\arraybackslash\hsize=0.8\hsize}X|>{\centering\arraybackslash\hsize=0.8\hsize}X|}
    \hline
    \textbf{Contribution} & \textbf{40-nm design} & \textbf{22-nm design} \\
    \hline
    Analog IMC cores               & 4.479 mm$^2$  & 1.981 mm$^2$ \\
    Charge pumps                   & 0.250 mm$^2$  & 0.170 mm$^2$ \\
    Buffer 1                       & 0.048 mm$^2$  & 0.013 mm$^2$ \\
    Buffer 2                       & 12.743 mm$^2$ & 4.515 mm$^2$ \\
    Buffer 3                       & 0.048 mm$^2$  & 0.013 mm$^2$ \\
    Twiddle multiply blocks  & 0.449 mm$^2$  & 0.127 mm$^2$ \\
    Multiplexers                   & 0.004 mm$^2$  & 0.001 mm$^2$ \\
    \hline
    \textbf{Total} & \textbf{18.021 mm}$^{\mathbf{2}}$ & \textbf{6.821 mm}$^{\mathbf{2}}$ \\
    \hline
\end{tabularx}
\caption{Estimated area breakdown of the 1024$\times$1024 VR-FFT designs in the 40-nm and 22-nm nodes.}
\label{tab:vrfft_area}
\end{minipage}

\end{table}

\vspace{1ex}

\noindent \textbf{Latency}: As described above, Stages 1 and 2 of the VR-FFT can work in lockstep.
The sub-matrix level pipeline for the first two stages is mostly similar to Fig. \ref{fig:pipelining}: (1) Load inputs to Stage 1, (2) Compute Stage 1 analog MVMs, (3) ADC conversions for Stage 1, (4) Write Stage 1 results to Buffer 1, (5) Load inputs to Stage 2 from Buffer 1, (6) Compute Stage 2 analog MVMs, (7) ADC conversions for Stage 2, (8) Digital twiddle multiplication, (9) Write results to Buffer 2.
The machine cycle's duration of 130 clock cycles (130 ns) is set by the ADC conversion steps (steps 3 and 7).
The buffer read/write steps can be executed in less than 130 ns by reading or writing one complex value per clock cycle to 64 banks in parallel over 32 clock cycles (steps 4, 5), or to 32 banks in parallel over 64 clock cycles (step 9).
The twiddle multiplications (step 8) can be completed in less than 130 ns by running each TM block at the system's clock speed of 1 GHz.
Since there are 512 sequential analog MVMs steps per stage, it takes 521 machine cycles (512+9) to write the final set of Stage 2 results to Buffer 2.
After Buffer 2 is fully populated with Stage 2's results, Stages 3 and 4 can work in lockstep with sub-matrix-level pipelining.
This pipeline is identical to that of Stages 1 and 2, but without the twiddle multiplication step.
Therefore, the latency of Stages 3 and 4 is 520 machine cycles.
With this configuration, the total latency of all four stages is:
\begin{equation}
\nonumber
\text{Latency of 1024$\times$1024 VR-FFT} = (521 + 520) \times 130 \, \text{ns} = \textbf{135.3 {\textmu}s}
\end{equation}

\vspace{1ex}

\noindent \textbf{Throughput}: The macro-pipeline described above can process two 1024$\times$1024 images concurrently: one is in the Stage 1-2 pipeline, and the other is in the Stage 3-4 pipeline.
A new 2D FFT can be computed every 521 machine cycles, which is the longer of the two macro-pipeline stages.
The FFT throughput is:
\begin{equation}
\nonumber
\text{FFT throughput} = 1024 \times 1024 \text{ samples} \, / \, (521 \times 130 \, \text{ns}) = \textbf{15.5 \text{GSamples/s}}
\end{equation}

\vspace{1ex}

\noindent \textbf{Area}: Table \ref{tab:vrfft_area} shows a breakdown of the estimated area for this 2D FFT design, implemented in both the 40-nm and 22-nm nodes as described in Supplementary Section \ref{sec:supp_area}.
The area breakdown of the components within the analog IMC cores is similar to that in Table \ref{tab:area}.
As with energy consumption, the area of both designs is dominated by the 4.2 MB global buffer (Buffer 2), which is needed due to the corner turn.
We have estimated the area of an SRAM implementation of this buffer using CACTI, but a more compact implementation might be feasible, possibly using denser on-chip memories like embedded DRAM (eDRAM).
The area-normalized FFT throughput values for these two designs are: 0.860 GSamples/s/mm$^2$ (40-nm) and 2.272 GSamples/s/mm$^2$ (22-nm).

%% file: _SI_scaling_laws.tex
\section{Asymptotic scaling laws for the analog FFT}
\label{sec:scaling_laws}

In this section, we derive mathematical scaling laws for the energy, area, and performance of the analog FFT and the analog direct MVM method, summarized in Table \ref{tab:scaling_laws}.
We note that these are theoretical, asymptotic scaling laws whose purpose is to illustrate the feasibility of scaling to very large DFT sizes, and do not account for the actual costs of the underlying hardware operations (e.g. ADC conversions) or effects at small DFT size, which were modeled in Supplementary Section \ref{sec:energy_supp}.
By comparison, digital FFT algorithms typically follow a $\mathcal{O}(N \text{log}_2 N)$ scaling for both time and energy \cite{Duhamel1990}, though the actual FFT processor may follow a different scaling law because the hardware is optimized to be most efficient at one or a few specific FFT sizes.
The analysis here considers 1D DFTs in the limit of large DFT size $N$, but can be straightforwardly generalized to multiple dimensions.

\def\arraystretch{1.5}
\begin{table}[h!]
    \begin{tabularx}{\textwidth}{|>{\centering\arraybackslash\hsize=0.85\hsize}X|>{\arraybackslash\hsize=0.8\hsize}X|>{\arraybackslash\hsize=1.05\hsize}X|>{\arraybackslash\hsize=1.3\hsize}X|}
        \hline
     \textbf{DFT processing method} & \textbf{Energy scaling} & \textbf{Area scaling} & \textbf{Time scaling} \\
        \hline
     Analog FFT  					& 	$\mathcal{O}(N \text{log}_K N)$ & Analog: $\mathcal{O}(K^3)$ or $\mathcal{O}(K^2)$* \newline Buffering: $\mathcal{O}(N \text{log}_K N)$ & MVMs: $\mathcal{O}(1)$ or $\mathcal{O}\left(\frac{N}{K^2} \text{log}_K N \right)\textsuperscript{\textdagger}$ \newline Buffering: $\mathcal{O}(N \text{log}_K N)$ \\
        \hline
     Analog direct MVM  	& 	$\mathcal{O}\left(\frac{N^2}{K}\right)$ & $\mathcal{O}(N^2)$ or $\mathcal{O}\left(\frac{N^2}{K}\right)$* & MVMs: $\mathcal{O}(1)$ \newline Digital adds: $\mathcal{O}\left(\text{log}_2 \frac{N}{K} \right)$  \\
        \hline
    \end{tabularx}
\vspace{8pt}
\caption{Asymptotic scaling laws for the energy, area, and time overheads of different analog DFT processing methods. *The area scaling depends on whether the memory elements or peripheral circuits dominate the area. \textsuperscript{\textdagger}If Cooley-Tukey decomposition is only applied once, the scaling is $\mathcal{O}(1)$.}
\label{tab:scaling_laws}
\end{table}

\vspace{-16pt}

\begin{figure}[h]
\centering
\includegraphics[width=0.67\textwidth]{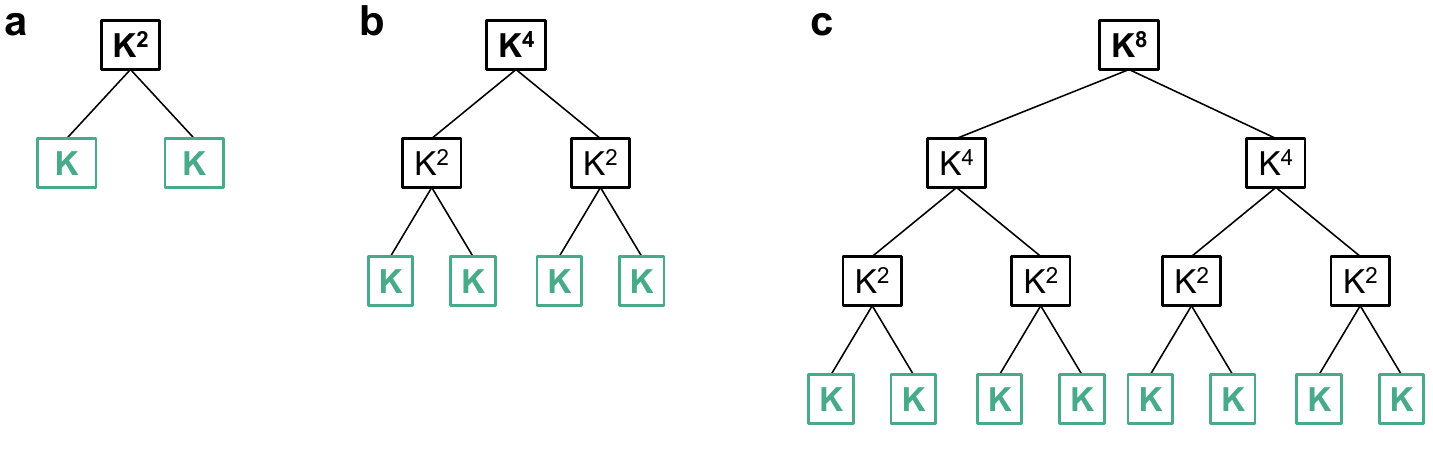}
\caption{Cooley-Tukey decompositions for three exemplary DFT sizes: (a) $N = K^2$, (b) $N = K^4$, and (c) $N = K^8$, where $K$ is the largest analog DFT that can fit in a single resistive crossbar array.
}
\label{fig:fft_tree}
\end{figure}

\vspace{-16pt}

\subsection{Energy scaling of the analog FFT}

To analyze the energy scaling properties, we use the basic assumption that the energy of an analog MVM is dominated by the output peripheral circuits such as TIAs, integrators, and ADCs. This has been true for published analog IMC accelerators that use 8-bit or higher-resolution ADCs \cite{HERMES2022,wan2022compute,xiao2020analog}, and for the analog core analyzed in Supplementary Section \ref{sec:energy_supp}. While DNNs can be trained to tolerate lower precision, we will assume that DSP applications require at least 8-bit ADC outputs. When peripheral circuits dominate the energy, the analog MVM's energy scales with the number of output values. For the full analog FFT, we account for the energy of analog MVMs, the energy of digital twiddle multiplications if needed, and the energy of intermediate reads and writes to digital buffers.

We consider the three exemplary DFT sizes in Fig. \ref{fig:fft_tree}. The $N = K^2$ case in Fig. \ref{fig:fft_tree}a is the largest DFT that can be computed using only one application of Cooley-Tukey decomposition. Following the procedure described in the main text, this involves two sets of $K$ analog DFTs of size $K$. Each analog DFT has $2K$ ADC conversions (for the real and imaginary outputs), so the total number of ADC conversions for the $K^2$-point analog FFT is: $n_\text{ADC}(K^2) = K\cdot 2K + K\cdot 2K = 4K^2$. We now consider the $N = K^4$ case in Fig. \ref{fig:fft_tree}b, which is the largest DFT that can be computed using a Cooley-Tukey decomposition tree of depth $\leq 2$. At the top level, this DFT is factored into two sets of $K^2$ DFTs of size $K^2$. Each $K^2$-point DFT can be computed using the analog FFT in Fig. \ref{fig:fft_tree}a, so the total number of ADC conversions is: $n_\text{ADC}(K^4) = 2 \cdot K^2\cdot n_\text{ADC}(K^2) = 8K^4$. Applying the same method, the total number of ADC conversions for the $N = K^8$ case in Fig. \ref{fig:fft_tree}c is found to be: $n_\text{ADC}(K^8) = 2 \cdot K^4 \cdot n_\text{ADC}(K^4) = 16 K^8$. For these three cases, the number of ADC conversions scales as $2N \text{log}_K N$, and this $\mathcal{O}(N \text{log}_K N)$ scaling law for the number of ADC conversions holds generally for increasing $N$. The sequential analog FFT in Fig. \ref{fig:cooley_tukey}b and the parallel analog FFT in Fig. \ref{fig:parallel_fft} have the same number of ADC conversions, so the same energy scaling law applies to both.

Since the twiddle factors are multiplied element-wise with the digitized first-stage DFT outputs, the number of multiplications scales proportionally with the number of ADC conversions. The same is true for the number of buffered intermediate results. We note that no digital twiddle multiplications are needed if using the parallel FFT scheme with only one Cooley-Tukey decomposition. Nonetheless, because all major contributions to the energy of the analog FFT follow the same worst-case asymptotic scaling behavior, the energy of the analog FFT scales as $\mathcal{O}(N \text{log}_K N)$.

\subsection{Time scaling of parallel analog FFTs}

In an analog MVM, all multiply-accumulate operations are processed in parallel across the full array.
We assume that the time to digitize analog MVM results does not scale with array size; this can be accomplished with a ramp ADC, or by increasing the number of ADCs together with the number of array columns. Under these assumptions, the latency of a single elementary analog DFT is $\mathcal{O}(1)$. 

Our analysis uses the parallel analog FFT scheme in Fig. \ref{fig:parallel_fft}. First, consider a single Cooley-Tukey decomposition ($N=K^2$). Because the MVMs occur in parallel and there are no digital twiddle multiplications, each DFT stages is $\mathcal{O}(1)$. Therefore, the time scaling for the full analog FFT is $\mathcal{O}(1)$.

When performing more than one level of Cooley-Tukey decomposition, we only assume that the lowest level of the decomposition is parallelized in order to maintain a relatively small footprint. Consider the case of two decompositions ($N=K^4$): this requires sequentially performing $2K^2$ DFTs of size $K^2$, each of which has a time cost of $\mathcal{O}(1)$. Similarly, the third level of decomposition ($N=K^8$) requires $4K^6$ DFTs of size $K^2$. Continuing this pattern, we find that the asymptotic time scaling associated with analog MVMs is $\mathcal{O}((N/K^2)\text{log}_K N)$. We note that if Cooley-Tukey decomposition is applied more than once, then digital twiddle multiplications are still needed in the higher-level decompositions. 

We assume that the time associated with buffering intermediate results scales with the number of results: $\mathcal{O}(N \text{log}_K N)$. While this scales faster asymptotically than the time cost of analog MVMs, it has a very different proportionality costant. Furthermore, the analog MVMs in the first stage, the storage of intermediate results, and the analog MVMs of the second stage can all be pipelined to increase throughput.

\subsection{Area scaling of parallel analog FFTs}

For a single resistive crossbar that implements a $K$-point analog DFT, the area scales as $\mathcal{O}(K^2)$ if the memory elements dominate the area. For highly scaled memory technologies, the area may instead be dominated by peripheral circuits, which would scale as $\mathcal{O}(K)$. For the parallel FFT implementation, we would allocate $K$ arrays of the above size to implement the first stage of the Cooley-Tukey decomposition, and another $K$ arrays to implement the second stage. The total area scales as either $\mathcal{O}(K^3)$ if the memory elements dominate, or $\mathcal{O}(K^2)$ if the peripheral circuits dominate. The scaling remains the same for multiple levels of Cooley-Tukey decomposition, since as mentioned above, the same arrays would be used to process the higher-level DFTs sequentially. The area overhead of intermediate digital storage scales as $\mathcal{O}(N \text{log}_K N)$, following the scaling of the number of intermediate results that need to be stored.

\subsection{Scaling laws for analog DFTs using direct MVMs}

A direct analog MVM mapping of a large DFT would require the DFT matrix to be split up across many arrays. In this case, $K$ is proportional to the maximum dimension of a given array in this partition. The number of outputs for an $N$-point DFT scales as $N$, and each of these outputs must be obtained by digitally summing the partial results of $N/K$ ADC conversions from different arrays. Therefore, the number of ADC conversions and the number of digital additions would both scale as $\mathcal{O}(N^2 / K)$, and this is the overall energy scaling of the direct MVM approach.

Since the direct MVM approach stores every element of the matrix $\mathbf{W}_N$ and does not require digital buffering, the area scaling is straightforward. The overall area scaling is $\mathcal{O}(N^2)$ if memory elements dominate the area, and is $\mathcal{O}(N^2/K)$ if the peripheral circuits dominate the area. For the digital additions, we assume that enough adders are allocated so that partial results for each DFT output can be added in parallel to increase throughput. This would require $\mathcal{O}(N^2/K)$ digital adders.

In terms of performance, all of the analog MVMs across all of the arrays implementing the DFT can in principle be performed simultaneously, so the time complexity of the analog processing is $\mathcal{O}(1)$. The digitized partial sums would then need to be accumulated. Assuming as above that partial sums are added in parallel, the optimal throughput can be obtained by using an adder tree for each DFT output. This would lead to a time complexity of $\mathcal{O}(\text{log}_2(N/K))$ for the partial sum accumulations.